%% file: jmlr-article.tex
\documentclass[twoside,11pt]{article}

\usepackage[abbrvbib, preprint]{jmlr2e}

\usepackage{algorithm,algpseudocode}
\usepackage{amsfonts,amsmath}
\usepackage{comment}
\usepackage{dsfont}
\usepackage{enumitem}
\usepackage{xcolor} 
\usepackage{url}
\usepackage{tikz}
\usetikzlibrary{positioning, calc, shapes.geometric, shapes.multipart, shapes,
arrows.meta, arrows, patterns, decorations.pathreplacing, decorations.markings,
external, trees, automata, backgrounds}
\usepackage[normalem]{ulem}
\usepackage{mathtools}
\usepackage{booktabs}
\usepackage{hyperref}
\usepackage{multirow}
\usepackage{array}
\usepackage[format=hang]{caption}
\usepackage{subcaption}
\usepackage[margin=1in]{geometry}
\usepackage[onehalfspacing]{setspace}

\newcommand{\proglang}[1]{\textsf{#1}}
\newcommand{\pkg}[1]{\textbf{#1}}

\newcommand{\RR}{\mathbb{R}}
\newcommand{\1}{\mathds{1}}

\algnewcommand\algorithmiccompute{\textbf{Compute }}
\algnewcommand\Compute{\State\algorithmiccompute}
\algnewcommand\algorithmicreturnb{\textbf{Return }}
\algnewcommand\Returnb{\State\algorithmicreturnb}
\algnewcommand\algorithmicsolve{\textbf{Solve }}
\algnewcommand\Solve{\State\algorithmicsolve}

\newtheorem{cor}{Corollary}
\newtheorem{defn}{Definition}
\newtheorem{lem}{Lemma}
\newtheorem{prop}{Proposition}
\newtheorem{thm}{Theorem}
\newtheorem{ex}{Example}

\newcommand{\di}[2]{d_{#1#2}^{\not\rightarrow}}
\newcommand{\lb}[2]{l_{#1#2}^\leftrightarrow}
\newcommand{\ld}[2]{l_{#1#2}^\rightarrow}
\newcommand{\lu}[2]{l_{\{#1,#2\}}^-}
\newcommand{\xb}[2]{x_{\{#1,#2\}}^\leftrightarrow}

\newcommand{\xd}[2]{x_{#1#2}^\rightarrow}

\newcommand{\starright}{{\ast\! \rightarrow}}
\newcommand{\xs}[2]{x_{#1#2}^{\starright}}
\newcommand{\lc}[3]{l_{\{#1,#2\}}^#3}

\newcommand{\lcb}[3]{l_{\{#1,#2\}}^{\leftrightarrow,#3}}
\newcommand{\zc}[3]{z_{\{#1,#2\}}^#3}
\newcommand{\zcb}[3]{z_{\{#1,#2\}}^{\leftrightarrow,#3}}
\newcommand{\zu}[2]{z_{\{#1,#2\}}^-}

\newcommand{\lcd}[3]{l_{#1#2}^{\rightarrow,#3}}
\newcommand{\Uc}[3]{U_{\{#1,#2\}}^#3}
\newcommand{\phic}[3]{\phi_{\{#1,#2\}}^{#3,G}}

\newcommand\an{\mathrm{an}}
\newcommand\bd{\mathrm{bd}}
\newcommand\ci{\perp\!\!\!\perp}

\newcommand{\dconD}[4]{#1\not\ci_d #2 \mid #3\ [#4]}
\newcommand{\msep}[3]{#1\ci_m #2 \mid #3}
\newcommand{\msepD}[4]{#1\ci_m #2 \mid #3\ [#4]}
\newcommand{\mconD}[4]{#1\not\ci_m #2 \mid #3\ [#4]}
\newcommand{\condIndep}[3]{#1\ci #2 \mid #3}
\newcommand{\sepD}[4]{#1\ci #2 \mid #3\ [#4]}
\newcommand{\aG}{a_{\mathbb{G}}}

\newcommand{\mathdash}{\relbar\mkern-9mu\relbar}

\usepackage{lastpage}
\jmlrheading{??}{2026}{1-\pageref{LastPage}}{?/??; Revised
?/??}{?/??}{21-0000}{Lucas Kook and S\o{}ren Wengel Mogensen}

\ShortHeadings{Exact Graph Learning via Integer Programming}{Kook and Mogensen}
\firstpageno{1}

\begin{document}

\title{Exact Graph Learning via Integer Programming}

\author{%
   \name Lucas Kook \email lucasheinrich.kook@gmail.com\\
   \addr Institute for Statistics and Mathematics\\
   Vienna University of Economics and Business\\
   Vienna, Austria%
   \AND%
   \name S\o{}ren Wengel Mogensen \email swm.fi@cbs.dk \\
   \addr Center for Statistics, Department of Finance \\
   Copenhagen Business School\\
   Copenhagen, Denmark%
}

\editor{TBD}

\maketitle

\begin{abstract}
Learning the dependence structure among variables in complex systems is a
central problem across medical, natural, and social sciences. These structures
can be naturally represented by graphs, and the task of inferring such graphs
from data is known as graph learning or causal discovery. Existing approaches
typically rely on restrictive assumptions about the data-generating process,
employ greedy oracle algorithms, or solve approximate formulations of the graph
learning problem. Therefore, they are either sensitive to violations of central
assumptions or fail to guarantee globally optimal solutions. We address these
limitations by introducing a nonparametric graph learning framework based on
conditional independence testing and integer programming. We reformulate the
graph learning problem as a mixed-integer program and prove that solving this
integer-programming problem provides a globally optimal solution to the original
graph learning problem. Our method leverages efficient encodings of graphical
separation criteria, enabling the exact recovery of larger graphs than was
previously feasible. We provide an open-source \proglang{R} package \pkg{glip}
which supports learning (acyclic) directed (mixed) graphs and chain graphs. We
demonstrate that our approach is often faster than existing exact graph learning
procedures and achieves state-of-the-art performance on simulated and benchmark
data across all aforementioned classes of graphs.
\end{abstract}


\section{Introduction}\label{sec:intro}

Graphs (or graphical models) can be used to encode the dependence structure
between the components of a random vector \citep{lauritzen1996graphical} and
estimating graphs from data is a central problem across various scientific
disciplines \citep{Glymour2019,brouillard}. In this paper, we study the task of
learning a graphical representation of the dependence structure of a random
vector from data, which we refer to as \emph{graph learning}, and specifically
how to learn the graph using tests of conditional independence. This is possible
because under a suitable Markov condition, graphical separation between nodes in
a graph implies conditional independence in the joint distribution
\citep{dawid1979,lauritzen1996graphical,Pearl:2009}. We study graph learning in
different classes of graphs, including \emph{directed acyclic graphs} (DAGs) and
\emph{acyclic directed mixed graphs} (ADMGs) that may represent DAGs in which
some nodes are not observed. In these classes of graphs, a causal interpretation
is often given to the output graph, in which case graph learning is also known
as \emph{causal discovery} \citep{spirtes2000causation}. Therefore, while we use
the term \emph{graph learning} in this paper, one may as well use the term
causal discovery when considering DAGs or ADMGs. 

Under a suitable Markov condition and faithfulness (such that graphical
separation implies conditional independence and vice versa) and without making
further assumptions, the graph can be learned only up to the set of graphs that
impose the same restrictions on the dependence relations between variables,
i.e.,~the graph's Markov equivalence class
\citep{verma1990causal,andersson1997markov}. In causal discovery, methods are
often classified as either \emph{score-based} or \emph{constraint-based}
\citep{spirtes2000causation,chickering2002optimal}. Constraint-based methods use
testable constraints implied by a graph, most often conditional independencies,
to select an output graph. Score-based methods seek to optimize some score that
measures the goodness-of-fit between a graph and the observed data. This
typically requires assumptions on the data-generating process,
e.g.,~distributional assumptions or assumptions on the functional form of the
relationship between variables.

Classical constraint-based causal discovery algorithms, such as the Peter--Clark
\citep[PC,][also referred to as the parent-children algorithm]{Spirtes1991} or
Fast Causal Inference \citep[FCI,][]{spirtes2000causation} algorithm, are
motivated by \emph{oracle} approaches: Given an oracle which provides the
correct answers to conditional independence queries, how can we output a correct
graph? While such algorithms can be shown to be sound and complete given an
independence oracle, their finite sample performance depends heavily on the
properties of the involved conditional independence tests. However, conditional
independence testing poses a challenge in its own right, as
\citet{shah2020hardness} have shown that there exists no uniformly valid
assumption-free conditional independence test that has non-trivial power against
any alternative. Yet, many oracle-based algorithms rely on the removal of edges
upon a non-rejection of a conditional independence test, which may lead to poor
finite-sample performance.

Classical score-based methods, such as greedy equivalence search
\citep[GES,][]{Meek1997}, treat the graph learning problem as an optimization
problem: How can we output a graph which is least inconsistent with the data?
Inconsistency can, for instance, be assessed through goodness-of-fit measures
such as the Bayesian Information Criterion under appropriate distributional
assumptions \citep{Cooper1995}. Score-based methods that do not rely on
restrictive distributional or functional assumptions are typically
\emph{greedy}, however, in that they solve an approximate formulation of the graph
learning problem and therefore do not guarantee globally optimal solutions
\citep[e.g.,][]{hu2024fast,lagrange2025efficient}. 

In this paper, we follow a line of research which uses the output of conditional
independence tests to score graphs without making restrictive distributional or
parametric assumptions and with guarantees to output globally optimal graphs
\citep{hyttinen2013discovering,hyttinen2014constraint,hyttinen2017core}. We
refer to methods that guarantee globally optimal solutions to the graph learning
problem they tackle as \emph{exact}, and \emph{approximate} otherwise. While
these existing works rely on answer set programming, we encode the graph
learning problem as a mixed-integer program and propose Graph Learning via
Integer Programming (GLIP). While exact graph learning based on integer
programming has been proposed for \emph{directed mixed graphs} (DMGs) with at
most six nodes \citep{eberhardt2025discovering}, GLIP is, to the best of our
knowledge, the first exact and nonparametric approach to graph learning based on
mixed-integer programming for directed mixed graphs (with more than six nodes),
chain graphs and for several subclasses thereof. 

\subsection{Globally optimal solutions to the graph learning problem}
\label{sec:intro:global}

We now give a high-level overview of the graph learning problem and showcase the
advantages of exact methods over greedy methods motivated through independence
oracles (Example~\ref{ex:oracle}). GLIP optimizes a score which is based on
conditional independence testing, and in this sense our method is both
score-based and constraint-based, creating a method which is as assumption-lean
as classical constraint-based methods and which simultaneously provides a
globally optimal solution.

Several papers
\citep{hyttinen2013discovering,hyttinen2014constraint,hyttinen2017core,
eberhardt2025discovering} provide algorithms that search for a global solution
to an
optimization problem of the form,
\begin{align}
    \min_{G\in \mathbb{G}_d} f(G, \mathcal{D}_n), \tag{GL} \label{tag:GL}
\end{align}
where $\mathbb{G}_d$ is a class of graphs, $\mathcal{D}_n$ is a collection of
$n$ i.i.d.\ observations of a random vector $X = (X_1,X_2,\ldots,X_d)$, and $f$
is a function which measures the degree of correspondence between its two
arguments. Every graph in $\mathbb{G}_d$ has $d$ nodes, and each node represents
a variable in the random vector, $X$. The following is a specific instance of
such a problem using $d$-separation in DAGs and where $f$ simply counts the
number of discrepancies between $d$-separation statements and $p$-values (using
0.05 as a threshold),
\begin{align*}
    \min_{D\in \mathbb{D}_d} \sum_{i,j,C}  \lvert \1_{\dconD{i}{j}{C}{D}}
    - \1_{p_{ijC} < 0.05} \rvert
\end{align*}
where $\mathbb{D}_d$ is the set of DAGs on $d$ nodes, 
$\1_{\dconD{i}{j}{C}{D}}$ is 1 if $i$ and $j$ are $d$-connected given
$C$ in $D$ and 0 otherwise, $\1$ is an indicator function, and $p_{ijC}$
is a $p$-value from testing the hypothesis that variable $X_i$ and variable $X_j$
are conditionally independent given the variables $X_C$, $i,j\in [d] \coloneqq
\{1,2,\ldots,d\}$, $C\subseteq [d] \setminus \{i,j\}$. The notion of a
$d$-connecting path can be found in Section \ref{sssec:dcsep}.

\begin{ex}[Exact and greedy approaches to graph learning]\label{ex:oracle}
Consider the following linear Gaussian structural causal model
\citep[e.g.,][Def.~6.2]{Peters2017} and induced graph:\\[6pt]
\begin{minipage}[c]{0.49\textwidth}
\begin{align*}
    A &\coloneqq N_A\\
    B &\coloneqq N_B\\
    C &\coloneqq A + B + N_C\\
    D &\coloneqq C + N_D
\end{align*}
\end{minipage}
\begin{minipage}[c]{0.49\textwidth}
\centering
\resizebox{0.8\linewidth}{!}{
\begin{tikzpicture}[node distance=0.7cm]
\node[circle, draw] (A) {$A$};
\node[below right = of A, circle, draw] (C) {$C$};
\node[above right = of C, circle, draw] (B) {$B$};
\node[below=of C, circle, draw] (D) {$D$};
\draw[-latex] (A) -- (C);
\draw[-latex] (B) -- (C);
\draw[-latex] (C) -- (D);
\draw[latex-latex, color=white] (A) -- node [midway, above=0.3cm, color=black] {$\rho = 0$} (B);
\end{tikzpicture}
\hspace{1cm}
\begin{tikzpicture}[node distance=0.7cm]
\node[circle, draw] (A) {$A$};
\node[below right = of A, circle, draw] (C) {$C$};
\node[above right = of C, circle, draw] (B) {$B$};
\node[below=of C, circle, draw] (D) {$D$};
\draw[-latex] (A) -- (C);
\draw[-latex] (B) -- (C);
\draw[-latex] (C) -- (D);
\draw[latex-latex] (A) -- node [midway, above=0.3cm, color=black] {$\rho > 0$} (B);
\end{tikzpicture}}
\end{minipage}\\[6pt]
Here, $(N_A, N_B, N_C, N_D) \sim \operatorname{N}(0, \Sigma)$ such that, for all
$i,j \in [4]$, $\Sigma_{ii} = 1$ and $\Sigma_{12} = \Sigma_{21} = \rho$, and
$\Sigma_{ij} = 0$ otherwise. We generate $n = 300$ independent observations from
this causal model for $\rho \in \{0, 0.9, 0.9999\}$ and apply the FCI
algorithm and GLIP using partial correlation tests. Repeating the simulation 50
times yields the most frequent (as indicated below each graph) \emph{partial
ancestral graphs} \citep[PAGs,][]{zhang2007characterization} shown in
Figure~\ref{fig:ex:global}.

\begin{figure}[H]
\centering
\resizebox{\textwidth}{!}{%
\begin{tabular}{cccccc}
\toprule
$\rho = 0$
&
$\rho = 0.9$
&
$\rho = 0.9999$
&
$\rho = 0$
&
$\rho = 0.9$
&
$\rho = 0.9999$
\\
\begin{tikzpicture}[node distance=0.7cm]
\node[circle, draw] (A) {$A$};
\node[below right = of A, circle, draw] (C) {$C$};
\node[above right = of C, circle, draw] (B) {$B$};
\node[below=of C, circle, draw] (D) {$D$};
\draw[o-latex] (A) -- (C);
\draw[o-latex] (B) -- (C);
\draw[-latex] (C) -- (D);
\end{tikzpicture}
&
\begin{tikzpicture}[node distance=0.7cm]
\node[circle, draw] (A) {$A$};
\node[below right = of A, circle, draw] (C) {$C$};
\node[above right = of C, circle, draw] (B) {$B$};
\node[below=of C, circle, draw] (D) {$D$};
\draw[o-o] (A) -- (C);
\draw[o-o] (A) -- (B);
\draw[o-o] (B) -- (C);
\draw[o-o] (C) -- (D);
\end{tikzpicture}
&
\begin{tikzpicture}[node distance=0.7cm]
\node[circle, draw] (A) {$A$};
\node[below right = of A, circle, draw] (C) {$C$};
\node[above right = of C, circle, draw] (B) {$B$};
\node[below=of C, circle, draw] (D) {$D$};
\draw[o-o] (A) -- (B);
\draw[o-o] (B) -- (C);
\draw[o-o] (C) -- (D);
\end{tikzpicture}
&
\begin{tikzpicture}[node distance=0.7cm]
\node[circle, draw] (A) {$A$};
\node[below right = of A, circle, draw] (C) {$C$};
\node[above right = of C, circle, draw] (B) {$B$};
\node[below=of C, circle, draw] (D) {$D$};
\draw[o-latex] (A) -- (C);
\draw[o-latex] (B) -- (C);
\draw[-latex] (C) -- (D);
\end{tikzpicture}
&
\begin{tikzpicture}[node distance=0.7cm]
\node[circle, draw] (A) {$A$};
\node[below right = of A, circle, draw] (C) {$C$};
\node[above right = of C, circle, draw] (B) {$B$};
\node[below=of C, circle, draw] (D) {$D$};
\draw[o-o] (A) -- (C);
\draw[o-o] (A) -- (B);
\draw[o-o] (B) -- (C);
\draw[o-o] (C) -- (D);
\end{tikzpicture}
&
\begin{tikzpicture}[node distance=0.7cm]
\node[circle, draw] (A) {$A$};
\node[below right = of A, circle, draw] (C) {$C$};
\node[above right = of C, circle, draw] (B) {$B$};
\node[below=of C, circle, draw] (D) {$D$};
\draw[o-o] (A) -- (B);
\draw[o-o] (C) -- (D);
\end{tikzpicture}
\\
GLIP [44/50]
& GLIP [49/50]
& GLIP [38/50]
& FCI [45/50]
& FCI [49/50]
& FCI [44/50]
\\
\bottomrule
\end{tabular}}
\caption{%
PAGs summarize the Markov equivalence class (MEC) of ADMGs by placing a tail or
an arrow head if and only if all members of the MEC agree on this tail or arrow
head, respectively. If there exist two ADMGs that disagree on an edge mark, a
circle edge mark is used in the PAG instead. Heads and tails are defined in
Section~\ref{sssec:dcsep}. See Example~\ref{ex:oracle}.
}\label{fig:ex:global}
\end{figure}
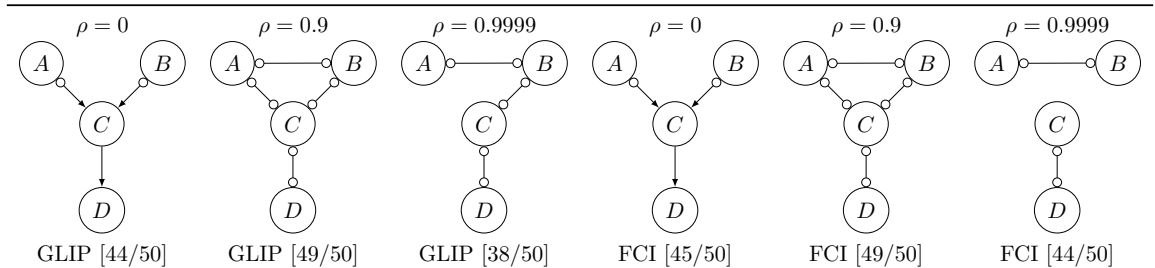

For the two smaller correlations, both algorithms produce the same (correct)
PAG. At $\rho = 0.9999$ the FCI algorithm falsely removes the edges from $A$ and
$B$ to $C$, due to the high correlation between $A$ and $B$. The proposed
optimization approach, however, does not rely on iteratively removing edges
given non-rejections of conditional independence tests and instead takes the
marginal correlation between $A$ and $C$ and $B$ and $C$ into account during
optimization. While such bivariate collinearities are easy to detect, the same
problem of falsely removing edges can appear under multicollinearity, which is
harder to detect, especially in larger systems. The problem illustrated here is
related to near-faithfulness violations
\citep{spirtes2000causation,colombo2014orderindependent}.
\end{ex}

\subsection{Related work}\label{sec:related}

There is a large literature on graph learning and causal discovery, and we have
given references to foundational works on causal discovery above. In the
following, we focus on prior work that is more directly related to ours in the
sense that it treats graph learning (or causal discovery) as an optimization
problem. While most prior work is approximate (in the sense defined above),
there is also some prior work on exact methods. In the following, we will first
cover other exact methods and then move to related but approximate methods based
on integer programming. We defer comparison of related work to our proposed
method to Section~\ref{sec:contrib}.

The work most closely related to ours is \citet{eberhardt2025discovering}, which
also uses an integer-programming approach to causal discovery in directed mixed
graphs and subclasses thereof. In the integer-programming problem formulation of
causal discovery proposed by \citet{eberhardt2025discovering}, there is a
parameter that controls the maximal length of paths that are taken into account
in the optimization. When all edges and all paths are used, their optimization
problem includes a variable for each path in the complete DMG (or in the
complete \emph{directed graph} (DG)), leading to a number of variables that
scales as $d!$. For each of these paths, $\rho$, say between $i$ and $j$, and
for each conditioning set, $C \subset [d]\setminus \{i,j\}$, there is a variable
which indicates whether $\rho$ is $d$-connecting given $C$. When using all
paths, \citet{eberhardt2025discovering} prove that their formulation will lead
to a global optimum, however, they also state that this will not work for $d >
6$ due to the sheer size of the integer-programming problem. In
\citet{eberhardt2025discovering}, a heuristic search approach among candidate
edges is proposed for use in larger graphs. This approach outputs a member of
the correct Markov equivalence class when oracle independence tests are used,
however, it is not guaranteed to find a global optimum to the finite-sample
optimization problem. 

There are other exact approaches to graph learning that do not use integer
programming. \citet{hyttinen2013discovering} formulate graph learning of DMGs
and subclasses thereof as a satisfiability problem to which off-the-shelf
Boolean satisfiability solvers (SAT solvers) could be applied. This does not
handle conflicts among input (in)dependences as they lead to unsatisfiability,
and \citet{hyttinen2014constraint} use answer set programming to allow conflicts
among input (in)dependences. \citet{hyttinen2017core} provide a similar, but
faster, method. \citet{rantanen2018learning, rantanen2020discovering} provide
branch-and-bound algorithms to find a global optimum to the causal discovery
problem.

Prior work based on integer programming, other than
\citet{eberhardt2025discovering}, imposes strong assumptions on the joint
distribution and some of those focus on specific subclasses of DMGs. For
instance, \citet{chen2021integer,dash2025integer} provide methods for causal
discovery of ADMGs, assuming that the true model is a linear Gaussian model.
\citet{pmlr-v213-cussens23a,yang2025inexact} consider only DAGs and assume a
decomposability property of the score. Other papers that use integer programming
or related methods for graph learning use so-called decomposable scores based
on, e.g., a likelihood and most only handle DAGs, examples include
\citet{jaakkola2010learning,studeny2014learning,cussens2017bayesian,kucukyavuz2023consistent}.

Finally, one can also learn weak equivalence classes instead of the more
expressive Markov equivalence classes which reduces the number of conditional
independence tests that are needed and the sizes of the conditioning sets which
is similar in spirit to recent advances in \emph{recursive} causal discovery
\citep{mokhtarian2025recursive} and DAG-learning based on dimension reduction
\citep{solea2025learning}. While we provide a nonparametric approach using only
conditional independencies induced by the causal graph, stronger assumptions may
facilitate considerably more scalable approaches to graph learning; see,
e.g.,~\citet{wienobst2025embracing}, which uses a decomposable score and the
assumption of a linear additive noise model to learn DAGs.
\citet{silander2006simple} provide an exact method in the context of DAGs and
under the assumption of a decomposable score such as BIC.

\subsection{Our contributions}\label{sec:contrib}

We propose a framework for Graph Learning via Integer Programming (GLIP) as a
versatile nonparametric and exact graph learning method for Markov equivalence
(or weak Markov equivalence, see Section \ref{subsec:weak}) classes of directed
(acyclic) graphs (Theorem~\ref{thm:DAGs}), (acyclic) directed mixed graphs
(Theorem~\ref{thm:DMGs}), and chain graphs (Theorem~\ref{thm:CGs}); see
Table~\ref{tab:graphclasses} for an overview of the supported graph classes and
graphical separation criteria. GLIP scores graphs based on the output of
arbitrary conditional independence tests, yields globally optimal solutions to
\eqref{tag:GL} (specifying $f$ as in \eqref{tag:GLG}, see
Section~\ref{sec:optim}), and can prove their optimality
(Theorem~\ref{thm:optim}). 

GLIP can be thought of as being score-based, i.e.,~selecting a graph by
optimizing a score. At the same time, the GLIP objective uses $p$-values from
tests of conditional independence which is most often used by constraint-based
methods. GLIP can be applied, with appropriate changes, to all classes of graphs
with a path-based (or walk-based) global Markov property (see
Section~\ref{ssec:DGs} for a detailed explanation) and makes no distributional
assumptions. 

Empirically we show that, compared to existing exact approaches based on answer
set programming \citep{hyttinen2014constraint,hyttinen2017core}, GLIP is faster
for graph learning in a large fraction of graphs with up to 9 nodes (after which
learning Markov equivalence classes becomes computationally burdensome and both
methods hit walltimes of 600~seconds). Similar conclusions hold for weak graph
learning in graphs with up to 14 nodes and considering conditioning sets of size
at most 1. 

In practical use, we recommend that one applies a computationally cheap graph
learning method first. The output is then used as a warmstart for our method.
This means that a computational budget can be allotted to GLIP for improving
upon the warmstart, even if no optimum can be found within the budget limit (or,
as may happen, an optimum is found but not proven to be optimal). In our
computational experiments, we show that even with a fairly small computational
budget, one can improve upon the warmstart graph. One could further improve the
performance of GLIP at the cost of additional computation by using a larger
walltime. For real-world applications of causal discovery and graph learning,
one may of course be willing to use a larger walltime to obtain the best
possible solution relative to the available computational power.

Empirically, we demonstrate that GLIP can yield exact solutions to graph
learning problems for graphs that are much larger than what
\citet[p.~3289]{eberhardt2025discovering} consider: \emph{``However, [our]
approach is not viable even for moderately sized graphs ($|V| > 6$), because
[our approach] searches over all possible paths''}. The reason for this
improvement is a more efficient minimal-length based encoding of constraints in
the integer program (Section~\ref{ssec:constraints}). It is sufficient to
consider the length of the shortest connecting path, rather than all possible
paths, since the existence of a connecting path trivially implies the existence
of a shortest such path. This leads to our encoding being more
parsimonious than that of \citeauthor{eberhardt2025discovering}, in the sense
that the number of variables in our integer program grows linearly with the
number of input $p$-values, whereas the number of variables in the encoding of
\citeauthor{eberhardt2025discovering} grows factorially.

GLIP, and the corresponding \proglang{R} package \pkg{glip}
(\url{https://github.com/LucasKook/glip}, see also Appendix~\ref{app:glipdemo}
for an illustration of the package) is highly modular regarding the
specification of the objective function. In this paper, we choose to minimize a
weighted sum of disagreements between graphical separations and thresholded
$p$-values representing the learned conditional independencies. Although this
choice is common in the literature
\citep{hyttinen2013discovering,hyttinen2014constraint,hyttinen2017core,
eberhardt2025discovering}, other choices may be equally sensible. In this vein,
GLIP supports any loss function that can be expressed as a linear function in
the variables of the underlying integer program, potentially through the use of
additional auxiliary variables. Similarly, since GLIP is based on integer
programming, it is straightforward to add domain-knowledge constraints, such as
``$v$ is a sink node in the graph''.

The rest of the paper is structured as follows. In Section~\ref{sec:ip}, we
provide an integer-programming formulation of graph learning and weak graph
learning. In Section~\ref{sec:theory}, we then derive explicit constraints for
learning directed (acyclic) graphs, (acyclic) directed mixed graphs (for chain
graphs, see Appendix~\ref{ssec:CGs}). Although the graph classes we consider
require different techniques, the results on the minimal-length encodings are
rather similar. Therefore, the reader is welcome to focus their reading on a
particular class of graphs, say DAGs, and skip the parallel sections for the
other classes of graphs. In Section~\ref{sec:optim}, we then introduce the
objective function and show in Theorem~\ref{thm:optim} that GLIP provides
globally optimal solutions. We then turn to the empirical evaluation of GLIP and
describe computational details in Section~\ref{sec:comp}, provide the empirical
results in Section~\ref{sec:results}, and close the paper with a discussion.

\section{Integer-programming formulation of graph learning}\label{sec:ip}

In this section, we will describe the fundamental idea behind an
integer-programming formulation of a graph learning problem. We consider three
settings depending on the type of graph we are trying to recover: Directed
graphs (DGs) and directed mixed graphs (DMGs). In the case of directed graphs
and directed mixed graphs, we provide one additional constraint which can be
used to restrict the search space to directed acyclic graphs (DAGs) and acyclic
directed mixed graphs (ADMGs), respectively. The ADMG (and chain graph) setting
both generalize the DAG setting. We include the DAG setting both for
illustration and to emphasize the fact that fewer constraints are needed in this
specialized setting. 

We wish to learn a graph based on tests of conditional independence, and an
integer-programming formulation of this problem must encode the graphical
separations represented by a graph. \citet{eberhardt2025discovering} use a
direct encoding for this problem where every path in any DMG (or DG) of size $d$
is represented by a variable which leads to a very large number of variables (in
the order of $d!$). We propose a different encoding which avoids keeping track
of individuals paths, i.e.,~in our formulation there are no variables that
represent individual paths. Instead we encode the lengths of the shortest
connecting path which is in the order $2^d$. This novel encoding is a central
contribution of this paper, and we call this technique a \emph{minimal-length
encoding}. We illustrate this in Example \ref{exmp:minlength}.

\begin{ex}[Minimal-length encoding]\label{exmp:minlength}
Section~\ref{sssec:dcsep} defines $d$-separation and\linebreak $d$-connecting
walks that
are used in this example. Other graphical terminology is defined in Section
\ref{ssec:prelims}. If we wish to decide if $i$ and $j$ are $d$-separated given
$C$ in a DAG on
nodes $[d] = \{1,2,\ldots,d\}$, $D$, it clearly suffices to compute the length
of the minimal $d$-connecting path between $i$ and $j$ given $C$ (by convention,
this length is $d$ if no such path exists). We let $\lc{i}{j}{C}$ denote the
length of a shortest $d$-connecting path between $i$ and $j$ given $C$. The
concatenation of two $d$-connecting
paths is not necessarily $d$-connecting, however, one can write down explicit
conditions such that the edges constrain the minimal-length
variables, and this forms the basis of a minimal-length encoding.
   
\begin{figure}[H]
\centering
\begin{tikzpicture}
\node[circle, draw] (A) {$1$};
\node[right = of A, circle, draw] (B) {$2$};
\node[right = of B, circle, draw] (C) {$3$};
\node[right = of C, circle, draw] (D) {$4$};
\node[above  = of C, circle, draw] (E) {$5$};
\node[above  = of D, circle, draw] (F) {$6$};
\draw[-latex] (A) -- (B);
\draw[-latex] (C) -- (B);
\draw[-latex] (B) -- (E);
\draw[-latex] (E) -- (F);
\draw[-latex] (D) -- (E);
\draw[-latex] (C) -- (D);
\end{tikzpicture}
\caption{See Example~\ref{exmp:minlength}.}
\label{fig:minlength}
\end{figure}
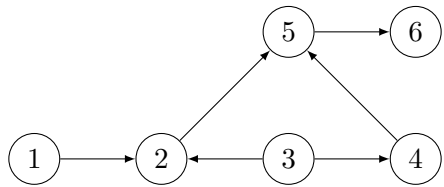

The DAG in Figure~\ref{fig:minlength} provides a concrete example of this. In
$D$, we see that $\lc{1}{3}{{\{2\}}} = 2$. The concatenation of any
$d$-connecting path between $1$ and $3$ given $\{2\}$ with the edge
$3\rightarrow 4$ is $d$-connecting between $1$ and $4$ given $\{2\}$, and we
must have $\lc{1}{4}{{\{2\}}}\leq 3$ (in fact, $\lc{1}{4}{{\{2\}}}= 3$ in this
graph). Similarly, we see that $\lc{1}{5}{{\{6\}}} = 2$ and that
$\lc{4}{5}{{\{6\}}} = 1$. We see that the concatenation of any $d$-connecting
path between $1$ and $5$ given $\{6\}$ and any $d$-connecting path between $4$
and $5$ given $\{6\}$ is $d$-connecting between $1$ and $4$ given $\{6\}$ since
$5$ is an ancestor of the conditioning set $\{6\}$, but not itself in the
conditioning set, leading to the constraint $\lc{1}{4}{{\{6\}}} \leq
\lc{1}{5}{{\{6\}}} + \lc{4}{5}{{\{6\}}}$. By considering a set of rules
including the above, one may give a complete description of how the minimal
lengths are related which leads to a parsimonious encoding of our learning
problem.
\end{ex}

In this paper, we consider DAGs, DGs, ADMGs, DMGs, and CGs. However, our
minimal-length encoding can be adapted to other classes of graphs and other
walk-based notions of graphical separation, including $\delta$-separation
\citep{Didelez:2008}, $\mu$-separation \citep{Mogensen:2020},
$\sigma$-separation \citep{forre2018constraint}, and E-separation
\citep{manten2025asymmetric}, and our work leads to a general and useful
integer-programming approach to causal discovery and graph learning in a broad
class of graphical models. We provide the intuition and concrete steps of
how to extend our minimal-length encoding to such cases in
Section~\ref{ssec:DGs}.

\subsection{Weak Markov equivalence}\label{subsec:weak}

Two graphs are said to be Markov equivalent if they encode the same set of
conditional independencies. In graph learning based on conditional independence,
one often tries to learn a Markov equivalence class of graphs, i.e., a set of
graphs that encode the same conditional independencies. It may not be feasible
to test all conditional independencies, even for node sets of a moderate size.
One can reduce the computational cost by learning a so-called \emph{$k$-weak
Markov equivalence class} of graphs instead of a Markov equivalence class of
graphs. This has the additional advantage that tests of conditional independence
will often have low power when the conditioning set is large. By reformulating
the learning target, one may avoid tests with large conditioning sets
altogether. This has led several authors to the notion of \emph{weak Markov
equivalence}. \citet{spirtes2001anytime} considered an \emph{anytime} version of
the FCI algorithm which would only consider small conditioning sets.
\citet{kocaoglu2024characterization} characterized $k$-weak equivalence in DAGs
while \citet{mogensen2025weak} characterized weak equivalence classes assuming a
so-called homogeneous equivalence in directed mixed graphs with $\mu$-separation
in the context of stochastic process models. 

The following definitions follow the terminology in \citet{mogensen2025weak}.
Let $\mathcal{J}$ be a collection of triples of the form $(i,j,C)$, i.e.,
$\mathcal{J} \subseteq \{(i,j,C): i,j\in [d], C\subseteq [d] \}$ (for notational
convenience we allow also triples $(i,j,C)$ such that $i \in C$ or $j \in C$. By
convention, we will simply say that $i$ and $j$ are separated given $C$ in any
graph if $i\in C$ or $j\in C$). Let $\mathcal{I}(G)$ denote the graphical
separations encoded by a graph, $G$ (e.g., the set of $d$-separations in $G$ if
$G$ is a DAG). Two graphs are said to be \emph{$\mathcal{J}$-weakly equivalent}
if $\mathcal{I}(G_1) \cap \mathcal{J} = \mathcal{I}(G_2) \cap \mathcal{J} $,
i.e., if they agree on all triples in the set $\mathcal{J}$. We say that
$\mathcal{J}$ is \emph{homogeneous} if there exists $\mathbb{C}\subseteq \{ C:
C\subseteq [d]\}$ such that $\mathcal{J}= \{(i,j,C): i,j\in [d], C\in
\mathbb{C}\}$. If $\mathcal{J}$ is homogeneous, the equivalence relation it
defines is said to also be \emph{homogeneous}. This can be further specialized
to \emph{$k$-weak equivalence}: For a class of graphs $\mathbb{G}$, two graphs,
$G_1 \in \mathbb{G}$ and $G_2 \in \mathbb{G}$, are said to be \emph{$k$-weakly
equivalent} if for all $i,j\in [d]$ and $C\subseteq [d]$ such that $\vert C\vert
\leq k$, $i$ and $j$ are separated given $C$ in $G_1$ if and only if $i$ and $j$
are separated given $C$ in $G_2$ (the notion of separation will naturally depend
on the class of graphs $\mathbb{G}$). Using $k$-weak equivalence with $k = d-2$
corresponds to Markov equivalence and we use the convention that if $i\in C$ or
$j\in C$, then there is no connecting walk between $i$ and $j$ given $C$ in any
graph.

The following theory applies equally well to graph learning of Markov
equivalence classes and of (homogeneous) weak equivalence classes. Some
constraints are indexed by a conditioning set $C$, and in case of a homogeneous
weak equivalence one should only use constraints such that $C \in \mathbb{C}$.
We use $[G]$ to denote the Markov equivalence class of $G$, and we use
$[G]_\mathbb{C}$ to denote the weak equivalence class of $G$ under a homogeneous
weak equivalence relation defined by $\mathbb{C}\subseteq \{C:C\subseteq
[d]\}$.

\subsection{Learning output}

We will reformulate the graph learning problem (see Section~\ref{sec:optim}) as
an optimization problem of the following form 
\begin{align}
\begin{split}
    \min_{\mathbf{y}\in \mathbb{R}^N}\quad & \mathbf{c}^\top \mathbf{y} \\
    \text{s.t.} \quad & \mathbf{A} \mathbf{y} \leq \mathbf{b}, \\
    & \underline{\mathbf{y}} \leq \mathbf{y} \leq \overline{\mathbf{y}},
\end{split} \tag{IP} \label{tag:IP}
\end{align}
where $\mathbf{y}$ is an $N$-vector of real and integer variables, and
$\mathbf{b}$, $\mathbf{c}$, $\mathbf{A}$ are coefficient matrices ($N\times 1$,
$N\times 1$, and
$N\times N$, respectively), $\underline{\mathbf{y}}$ and $\overline{\mathbf{y}}$
are vectors of lower and upper bounds on the variables, and inequalities should
be read entry-wise. Furthermore, there are \emph{integrality constraints}, i.e.,
some variables are restricted to be integers. This type of problem is known as a
mixed-integer linear programming problem. In our work, $\mathbf{y}$ will be
composed of 
\begin{itemize}
    \item \emph{edges variables}, $\mathbf{x}$, that define a graph,
\item \emph{length variables}, $\mathbf{l}$, that represent a certain type of minimal length, 
\item \emph{separation variables}, $\mathbf{z}$, that represent graphical separation/connection,
\item and auxiliary variables, $\mathbf{w}$,
\end{itemize}
such that $\mathbf{y} =
(\mathbf{x}^\top,\mathbf{l}^\top,\mathbf{z}^\top,\mathbf{w}^\top)^\top$. The integrality
constraints arise as, e.g., each edge variable is binary, representing either
the absence of the presence of a specific edge.

\begin{algorithm}
\caption{Graph Learning via Integer Programming}
\label{algo:cd}
\begin{algorithmic}[1]
\Require node set $V$, collection of conditioning sets $\mathbb{C}$, collection
of $p$-values corresponding to each triple $(i,j,C)$ such that $i,j\in [d]$, $C\in
\mathbb{C}$, graph class $\mathbb{G}$
\Statex
 \Compute coefficient matrices $\mathbf{b}$, $\mathbf{c}$, and $\mathbf{A}$ corresponding
 to $[d]$, $\mathbb{C}$, and $\mathbb{G}$ using Section \ref{sec:theory}
\Solve integer program in Equation \eqref{tag:IP} to obtain $\mathbf{y}^* =
((\mathbf{x}^\ast)^\top,(\mathbf{l}^\ast)^\top,(\mathbf{z}^\ast)^\top,(\mathbf{w}^\ast)^\top)^\top$
\label{algo:IP}
\Compute graph $G\in \mathbb{G}$ corresponding to $\mathbf{x}^\ast$
\Compute graphical representation of $[G]_\mathbb{C}$ using graphical theory (optional)
\Returnb $G$ (and optionally $[G]_\mathbb{C}$)
\label{algo:return}
\end{algorithmic}  
\end{algorithm}

One should note that the solution in Line~\ref{algo:IP} of
Algorithm~\ref{algo:cd} will not be unique in general as Markov equivalent
graphs will give the same objective value. The graph $G$ in
Line~\ref{algo:return} is therefore only one member of the optimal Markov
equivalence class (or more generally, weak equivalence class). However,
from this representative one may use graphical theory to compute a
representation of the entire equivalence class.

\begin{table}[!t]
\centering
\resizebox{\textwidth}{!}{%
\begin{tabular}{lrrrr}
\toprule
     \bf Class & \bf Subclass & \bf Separation criterion & 
     \bf Representation of $[G]$ & \bf Reference \\
     \midrule
     \multirow{3}{*}{DMG}
     & DMG  & $m_c$- or $m$-separation &  -                  &                                 - \\
     & ADMG & $m_c$- or $m$-separation & PAG                 & \citet{zhang2007characterization} \\
     & DG   & $d_c$- or $d$-separation &  -                  &                                 - \\
     \midrule
     \multirow{3}{*}{Chain}
     & Chain & $c$-separation   & Largest chain graph & \citet{frydenberg1990chain}         \\
     & DAG   & $d_c$- or $d$-separation & Essential graph     & \citet{Andersson:1997}              \\
     \bottomrule
\end{tabular}}
\caption{%
Classes of graphs considered for graph learning in this work.
An ADMG can be represented by a Markov equivalent 
\emph{maximal ancestral graph} (MAG), and PAGs can be used as
representations of Markov equivalence classes of MAGs. DAGs can be considered a
subclass of chain graphs or of DMGs. Chain graphs are used with the
LWF-interpretation. This table only covers the case of learning
Markov equivalence classes. Our methods can also be applied to learn a member of
the globally optimal weak equivalence class in DMGs and chain graphs (see
Section~\ref{subsec:weak}). However, there are no graphical representations of
those equivalence classes, to our knowledge.
}\label{tab:graphclasses} 
\end{table}

\section{Minimal-length encodings and constraints}\label{sec:theory}

This section describes how to define a set of variables, $\mathbf{y}$, and a set
of constraints that implicitly define the matrix $\mathbf{A}$ and the vectors
$\mathbf{b}$, $\underline{\mathbf{y}}$, and $\overline{\mathbf{y}}$
in~\eqref{tag:IP}. Our approach uses a novel \emph{minimal-length encoding} of
the conditional independence constraints encoded by a graph. This encoding is
parsimonious in the sense that the number of variables in the
integer-programming problem is linear in the number of input $p$-values. 

\begin{figure}[!ht]
    \centering
\begin{tikzpicture}
 \node (DMGs) at (0,0) {DMGs};
     \node [below of=DMGs] (DGs)  {DGs};
    \node [right of=DGs, xshift = 2cm] (ADMGs)  {ADMGs};
    \node [left of=DGs, xshift = -2cm] (CGs)  {CGs};
    \node [below of=DGs] (DAGs)  {DAGs};  
        \node [above of=CGs] (HGs)  {HGs};  
    \draw [thick] (DMGs) -- (DGs);
    \draw [thick] (DMGs) -- (ADMGs);
    \draw [thick] (DGs) -- (DAGs);
    \draw [thick] (ADMGs) -- (DAGs);
    \draw [thick] (CGs) -- (DAGs);
    \draw [thick] (HGs) -- (CGs);
\end{tikzpicture}
    \caption{Hasse diagram of classes of graphs. A line between classes of
    graphs indicates that the lower class is a subclass of the upper class. DMG:
    Directed mixed graph, ADMG: Acyclic directed mixed graph, DG: Directed
    graph, DAG: Directed acyclic graph, CG: Chain graph, HG: Hybrid graph. The
    hybrid graphs are only used as a convenient superclass of chain graphs, and
    we do not consider hybrid graph learning in this paper.}
    \label{fig:hasse}
\end{figure}
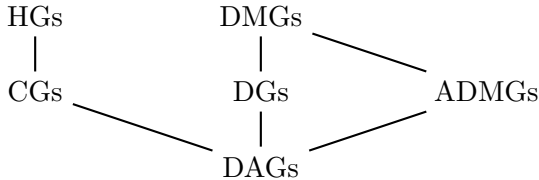

Our results cover several classes of graphs (see Figure~\ref{fig:hasse}). We
first introduce the relevant graphical concepts (Section~\ref{ssec:prelims}) and
describe the variables that will go into the vector $\mathbf{y}$
(Section~\ref{ssec:var}, which the reader may skip at first and consult when
needed). We then start defining the constraints that we need for the
integer-programming formulation (Section~\ref{ssec:constraints}), some of which
are shared between different classes of graphs. Section~\ref{ssec:DGs},
Section~\ref{ssec:DMGs}, and Appendix~\ref{ssec:CGs} describe the minimal-length
constraints which are specific to each class, and these sections formulate the
relevant mathematical results for directed graphs (DGs), directed mixed graphs
(DMGs), and chain graphs (CG), respectively. The intuition behind the
minimal-length encoding is illustrated in Example~\ref{ex:constraints} in
Section~\ref{ssec:DGs}.

In the acyclic case (DAGs, ADMGs, CGs), we make no parametric or distributional
restrictions on the model class as we only match $p$-values from tests of
conditional independence to a graphical structure. On the other hand, when we
consider general directed (mixed) graphs and allow directed cycles, we only
consider the case of linear Gaussian models, similarly to
\citet{eberhardt2025discovering}. In this case, $d$-separation is valid and the
model can be given a causal interpretation \citep{spirtes1995directed,
hyttinen2012learning, eberhardt2025discovering}. 

\subsection{Preliminaries} \label{ssec:prelims}

A graph is a pair $(V,E)$ where $V$ is a finite set of nodes and $E$ is a finite
set of edges. Each edge is between a distinct pair of nodes, and we say that
these nodes are the \emph{endpoints} of the edge. The graphs that we consider do
not have any edges for which the two endpoints are equal. We will consider three
different types of edges: Undirected edges, $-$, directed edges, $\rightarrow$,
and bidirected edges, $\leftrightarrow$. For nodes $i,j\in V$, we use the
notation $i - j$, $i \rightarrow j$, and $i \leftrightarrow j$ to indicate the
existence of an undirected, directed, and bidirected edge, respectively, between
$i$ and $j$. Undirected edges and bidirected edges are symmetric in the sense
that, e.g., $i - j$ if and only if $j - i$. On the other hand, $i \rightarrow j$
and $j \rightarrow i$ are different edges. 

A \emph{walk}, $\omega$, in a graph is an alternating sequence of nodes,
$i_0,\ldots,i_{n}$, and edges, $e_1,\ldots,e_n$,
\begin{align*}
    (i_0, e_1, i_1, e_2, \ldots, e_n, i_{n})
\end{align*}
such that each edge $e_k$ is between $i_{k-1}$ and $i_k$ for each $k\in
\{1,2,\ldots,n\}$. A \emph{path} is a walk such that no
node is repeated. We say that $i_0$ and $i_{n}$ are the \emph{endpoints} of
$\omega$, and we say that $\omega$ is \emph{between} its endpoints $i_0$ and
$i_{n}$. The \emph{length} of a walk, $\omega$, is the number of edges on the
walk, and we denote this length by $\vert \omega \vert$. For $0\leq
n_1\leq n_2 \leq n$, we define the \emph{subwalk of $\omega$ between the
$n_1$'th node and the $n_2$'th node} as the walk $(i_{n_1}, e_{n_1+1}, i_{n_1+1},
e_{n_1+2}, \ldots, e_{n_2-1}, i_{n_2})$, and we denote it by $\omega(n_1,n_2)$.
When $i_{n_1} = k$ and $i_{n_2}=  l$, we will also use the notation
$\omega(l,k)$ to denote the subwalk of $\omega$ between the $n_1$'th node and
the $n_2$'th node. When $\omega$ is a walk, but not a path, this may not define
a subwalk uniquely as $l$ or $k$ may occur more than once on $\omega$. In our
usage, this will not lead to ambiguity. When $\omega$ is a path, we also refer
to $\omega(l,k)$ as a \emph{subpath}. 

Let $e_1,\ldots,e_n,f_1,\ldots,f_m \in E$ and
$i_0,\ldots,i_{n},j_0,\ldots,j_{m}\in V$. Let $\omega_1$ be a walk between $i$
and $j$, and let $\omega_2$ be a walk between $j$ and $l$,
\begin{align*}
    \omega_1 &= (i = i_0, e_1,
i_1, e_2, \ldots, e_n, i_{n} = j), \\
\omega_2 &= (j = j_0, f_1, j_1, f_2, \ldots, f_m, j_{m} = l).
\end{align*}
The \emph{concatenation} of $\omega_1$ and $\omega_2$ is the
walk between $i$ and $l$ which first traverses the nodes and edges of $\omega_1$
and then the nodes and edges of $\omega_2$,
\begin{align*}
   (i = i_0, e_1, i_1, e_2, \ldots, e_n,  j , f_1, j_1, f_2, \ldots,
   f_m, j_{m} = l).
\end{align*}

If $G_1 = (V,E_1)$ and $G_2= (V,E_2)$ are two graphs, we define their
\emph{union}, $G_1\cup G_2$, as the graph $(V,E_1\cup E_2)$. Throughout the
paper, we assume that the nodes are numbered 1 through $d$ such that
$V=\{1,2,\ldots,d\}$, and we denote this set by $[d]$.

\subsubsection{Classes of graphs}\label{sssec:classes}

We will consider graph learning in five of the six different classes of graphs
that are represented in Figure~\ref{fig:hasse}. Each of these five classes is a
subclass of either \emph{directed mixed graphs} (Definition~\ref{def:dmg}) or of
\emph{chain graphs} (Definition~\ref{def:cg}). This section defines the classes
of graphs that we will need. We let $S_{ij}$ denote the set of edges that are
between nodes $i$ and $j$. If $S_{ij}\neq\emptyset$, we say that $i$ and $j$ are
\emph{adjacent}.

\begin{defn}[Directed mixed graph]\label{def:dmg}
A graph, $G = (V,E)$, is a \emph{directed mixed graph} (DMG) if for every pair
of distinct nodes $i$ and $j$, the set $S_{ij}$ is a subset of $\{i\rightarrow
j, i\leftarrow j, i \leftrightarrow j\}$.
\end{defn}

\begin{defn}[Directed graph]
    A directed mixed graph is a \emph{directed graph} (DG) if it
    contains no bidirected edges.
\end{defn}

In a DMG, $G=([d],E)$, a path between $i$ and $j$,
\begin{align*}
    (i = i_0, e_1, i_1, \ldots, e_n, i_{n} = j),
\end{align*}
is \emph{directed} from $i$ to $j$ if for each $k = 1,\ldots,n$ the edge $e_k$
is directed and points towards $i_{k}$, i.e., if $i = i_0 \rightarrow i_1
\rightarrow \ldots \rightarrow i_n \rightarrow i_{n} = j$. A \emph{trivial
path} is a path with no edges and only a single node, and such a path is
directed by convention. A \emph{directed cycle} is the concatenation of a
directed path from $i$ to $j$ and the edge $j\rightarrow i$, $i \neq j$.

\begin{defn}[Acyclic directed mixed graph]
    A directed mixed graph is an \emph{acyclic directed mixed graph} (ADMG) if
    it contains no directed cycles.
\end{defn}

\begin{defn}[Directed acyclic graph]
    A directed graph is a \emph{directed acyclic graph} (DAG) if it contains no
    directed cycles.
\end{defn}

We say that a graph is \emph{simple} if there is at most one edge between any
pair of distinct nodes (i.e., $\vert S_{ij} \vert \leq 1$ for all $i$ and $j$
such that $i\neq j$) and there is no edge for which the endpoints are equal.
Using the
terminology in \citet{studeny1998chain}, we define the following superclass of
chain graphs.

\begin{defn}[Hybrid graph]\label{def:hybrid}
    A simple graph is a \emph{hybrid graph} if every edge is either directed or
    undirected.
\end{defn}

In hybrid graphs, a \emph{partially directed cycle} consists of $n + 1$ distinct
nodes $i_0,i_1,\ldots,i_n$, $n \geq 2$, such that for all $k$, $0 \leq k \leq
n$, and using $i_{n+1} = i_0$, we have $i_k \mathdash i_{k+1}$ or $i_k
\rightarrow i_{k+1}$ and such that there exists a $k_0$ such that $i_{k_0}
\rightarrow i_{k_0+1}$.

\begin{defn}[Chain graph]\label{def:cg}
Let $G$ be a hybrid graph. We say that $G$ is a \emph{chain graph} if it
contains no partially directed cycles.
\end{defn}

A hybrid graph with no directed edges is an \emph{undirected graph}, and a
hybrid graph with no undirected edges is a directed graph. A chain graph with no
undirected edges is a directed acyclic graph. We will only use the hybrid graphs
as a convenient superclass of the chain graphs when formulating our results, and
we will not consider graph learning of hybrid graphs.

Hybrid graphs and DAGs are simple graphs.
This means that a sequence of nodes, $i_0,i_1,\ldots,i_n$ such that $i_k$ and
$i_{k+1}$ are adjacent for $k = 0,\ldots,n-1$ defines a unique walk. On the
other hand, DMGs and DGs are not necessarily simple and a walk should be
specified by a sequence of nodes, $i_0,i_1,\ldots,i_n$ such that $i_k$ and
$i_{k+1}$ are adjacent for $k = 0,\ldots,n-1$ along with a choice of edge
between each pair $(i_k,i_{k+1})$ when several edges are between $i_k$ and
$i_{k+1}$. For brevity, we do not always make this choice explicit and simply
write a walk as a sequence of nodes $i_0,i_1,\ldots,i_n$, also in DMGs and DGs.
We use the symbol $\sim$ to represent a generic edge of any type, and we will at
times write a generic walk as $i_0 \sim i_1 \sim \ldots \sim i_n$.

We use the notation $i\rightarrow j$, $i\leftrightarrow j$, and $i - j$ as
shorthand for stating that an edge of the specified type exists between $i$ and
$j$. As hybrid graphs are simple, the existence of the edge $i \rightarrow j$ in
a hybrid graph implies that there is no other edge between $i$ and $j$. On the
other hand, in a DMG, the statement $i \rightarrow j$ does not imply  absence of
the directed edge from $j$ to $i$ nor does it imply absence of the bidirected
edge between $i$ and $j$.

We will provide additional theory on chain graphs in Appendix~\ref{ssec:CGs}. The
next section introduces some of the graphical separation criteria that we will
use to connect graphs to conditional independence statements. The paper uses
different graphical separation criteria that are based on a notion of connecting
walks or paths, e.g.,~so-called $m$-connecting walks. When we write `a
connecting walk' without specifying the type of graphs, we refer to a connecting
walk of any type in a relevant class of graphs.

\subsubsection{\texorpdfstring{$m$}{d}-separation and
\texorpdfstring{$m_c$}{dc}-separation}\label{sssec:dcsep}

Let $i,j\in [d]$ be nodes in a graph, $G = ([d], E)$. We say that edges $i
\rightarrow j$ and $i\leftrightarrow j$ have a \emph{head} at $j$, and that the
edge $i \rightarrow j$ has a \emph{tail} at $i$. Let $\omega$ be a walk in $G$,
$i_0 \sim i_1 \sim \ldots \sim i_n$. For $k \in \{1, \ldots,
n-1\}$, we say that $i_k$ is a \emph{collider} on $\omega$ if both adjacent
edges have heads at $i_k$ (that is, $i_{k-1}\rightarrow i_k \leftarrow i_{k+1}$,
$i_{k-1}\leftrightarrow i_k \leftarrow i_{k+1}$, $i_{k-1}\rightarrow i_k
\leftrightarrow i_{k+1}$, or $i_{k-1}\leftrightarrow i_k \leftrightarrow
i_{k+1}$), and otherwise we say that $i_k$ is a 
\emph{noncollider}. Note that the endpoints of the walk are neither
colliders nor noncolliders. A node may appear more than once on a walk, and it
may appear as both a collider and a noncollider. The property of being a
collider/noncollider is therefore a property of each \emph{instance} of a node
on a walk.

If there is a directed path from $i$ to $j$, $i \rightarrow \ldots \rightarrow
j$, then we say that $i$ is an \emph{ancestor} of $j$ (each $i$ is an ancestor
of itself as the trivial path is directed). When $G$ is a DMG, we let
$\an_G({i})$ denote the set of ancestors of $i$ in $G$, and at times we use
$\an({i})$ when omitting $G$ does not lead to ambiguity. For $C \subseteq [d]$,
we define $\an_G(C) = \cup_{i\in C} \an_G({i})$. 

\begin{defn}[$m$-connecting walk]\label{def:mconn}
    Let $G = ([d], E)$ be a DMG. Let $i,j \in [d]$, $i\neq j$, $C\subseteq [d] \setminus
    \{i,j\}$, and let $\omega$ be a walk between $i$ and $j$ in $G$. We say that
    $\omega$ is \emph{$m$-connecting between $i$ and $j$ given $C$ in $G$} if every
    collider on $\omega$ is in $\an_G(C)$ and no noncollider on $\omega$ is in $C$.
\end{defn}

Every path is a walk, and an $m$-connecting path is simply a path which
satisfies Definition \ref{def:mconn}.

\begin{defn}[$m$-separation]\label{def:msep}
    Let $G = ([d], E)$ be a DMG, let $i,j\in [d]$, and let $C \subseteq [d]
    \setminus \{i,j\}$. We say that \emph{$i$ and $j$ are $m$-separated given
    $C$ in $G$} if there is no $m$-connecting path between $i$ and  $j$ given
    $C$ in $G$.
    
    Let $A,B,C \subseteq [d]$ be disjoint node sets. We say that \emph{$A$ and
    $B$ are $m$-separated given $C$ in $G$} if for every pair $i,j$ such that
    $i\in A$ and $j\in B$, nodes $i$ and $j$ are $m$-separated given $C$ in $G$,
    and we denote this by $\msep{A}{B}{C}$ or by $\msepD{A}{B}{C}{G}$.
    \end{defn}

The above definition uses $m$-connecting paths. One would obtain an equivalent
definition by using $m$-connecting walks. When $G$ is a DG in
Definition~\ref{def:mconn}, we will also use the term \emph{$d$-connecting walk}, and when
$G$ is a DG in Definition~\ref{def:msep}, we will also use the term \emph{$d$-separation.}

\begin{defn}[Global Markov property, DMGs]\label{def:globalDMG}
    Let $X = (X_1,\ldots,X_d)$ be a random vector, and let $G = ([d], E)$ be a
    DMG. We say that the distribution of $X$ satisfies \emph{the global
    $m$-separation Markov property} with respect to $G$ if for all disjoint sets
    $A,B,C \subseteq [d]$
    \begin{align*}
       \msepD{A}{B}{C}{G} \implies \condIndep{X_A}{X_B}{X_C},
    \end{align*}
    where $(\condIndep{\bullet}{\bullet}{\bullet})$ denotes conditional
    independence, and $X_A$ denotes the subvector $(X_i)_{i\in
    A}$.
\end{defn}

Definition~\ref{def:globalDMG} formulates the global $m$-separation Markov
property for DMGs. However, it is more often used in the smaller class of ADMGs
\citep{Richardson:2003}. When further restricting the class of graphs to DAGs,
it is known as the global $d$-separation Markov property
\citep{Pearl:2009}. When considering
cyclic graphs, we restrict to the linear Gaussian setting in which case
$m$-separation is still valid
\citep{spirtes1995directed,hyttinen2012learning,eberhardt2025discovering}.

While $d$-separation and $m$-separation are classical notions of graphical
separation in DAG-based models, we will also use a related notion of graphical
separation which is based on \emph{$d_c$-connecting walks} and
\emph{$m_c$-connecting walks}. 

\begin{defn}[$m_c$-connecting walk, $m_c$-separation]\label{def:mcsep}
Let $G = ([d], E)$ be a DMG. Let $i,j\in [d]$, $i\neq j$, $C\subseteq [d]
\setminus \{i,j\}$. A walk between $i$ and $j$ is \emph{$m_c$-connecting
given $C$} if it is $m$-connecting and all colliders on the walk are in
$C$. 

We say that \emph{$i$ and $j$ are $m_c$-separated given $C$ in $G$} if
there is no $m_c$-connecting walk between $i$ and  $j$ given $C$ in $G$.

Let $A,B,C \subseteq [d]$ be disjoint node sets. We say that \emph{$A$ and
$B$ are $m_c$-separated given $C$ in $G$} if for every pair $i,j$ such that
$i\in A$ and $j\in B$, nodes $i$ and $j$ are $m_c$-separated given $C$ in
$G$.
\end{defn}


\begin{ex}\label{ex:graphs}
Consider the graph $D$ in Figure~\ref{fig:graphs}. The path $3 \rightarrow
1 \rightarrow 2$ is $m$-connecting
(and $m_c$-connecting)
between $3$ and $2$ given $\emptyset$. On
the other hand, it is not $m$-connecting (nor $m_c$-connecting) between $3$ and
$2$ given $\{1\}$
since $1$ is a noncollider on the path and $1$ is in the conditioning set $C
= \{1\}$. The path $3 \rightarrow 4 \leftarrow 2$ is also not $m$-connecting 
(nor $m_c$-connecting)
given $\{1\}$ as $4$ is a collider on this path and $4\notin \an(\{1\})$. In
fact, there is no $m$-connecting path between $2$ and $3$ given $\{1\}$, and
therefore $\msepD{2}{3}{\{1\}}{D}$. 

\begin{figure}[H]
\centering
\setlength{\tabcolsep}{50pt}
\resizebox{\textwidth}{!}{%
\begin{tabular}{cc}
\begin{tikzpicture}
\node[circle, draw] (1) {$1$};
\node[below = of 1, circle, draw] (3) {$3$};
\node[right = of 1, circle, draw] (2) {$2$};
\node[below=of 2, circle, draw] (4) {$4$};
\node[right=of 4, circle, draw] (5) {$5$};
\node[above=of 5, circle, draw] (6) {$6$};
\draw[-latex] (3) -- (1);
\draw[-latex] (1) -- (2);
\draw[-latex] (2) -- (4);
\draw[-latex] (3) -- (4);
\draw[-latex] (4) -- (5);
\draw[-latex] (6) -- (2);
\draw[-latex] (6) -- (4);
\end{tikzpicture}
&
\begin{tikzpicture}
\node[circle, draw] (1) {1};
\node[below = of 1, circle, draw] (3) {3};
\node[right = of 1, circle, draw] (2) {2};
\node[below=of 2, circle, draw] (4) {4};
\node[right = of 4, circle, draw] (5) {5};
\draw[-latex] (3) -- (1);
\draw[-latex] (1) -- (2);
\draw[latex-latex] (2) to [bend left] (4);
\draw[-latex] (2) -- (4);
\draw[-latex] (3) -- (4);
\draw[-latex] (4) -- (5);
\end{tikzpicture}\\
$D$
& $G$
\end{tabular}}
\caption{
Example graphs: $D$ is a DAG, and $G$ is an ADMG. ADMGs may be
constructed as so-called \emph{latent projections} of DAGs
\citep{verma1991equivalence, richardson2023nested}: $G$ is a latent projection
of $D$, and therefore $\msepD{A}{B}{C}{D} \iff \msepD{A}{B}{C}{G}$
for all disjoint $A,B,C \subseteq \{1,2,3,4,5\}$. In this sense, $G$ is a
graphical marginal of $D$. See also Example~\ref{ex:graphs}.}
\label{fig:graphs}
\end{figure}
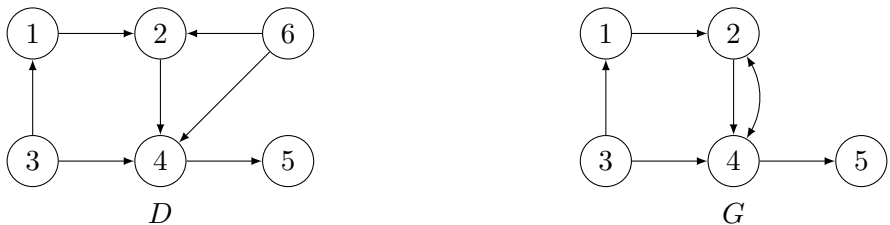

In $G$, we see that the path $3 \rightarrow 4 \leftarrow 2$ is
$m$-connecting between $3$ and $2$ given $\{5\}$ since the collider 4 is in
$\an(\{5\})$. On the other hand, this path is not $m_c$-connecting as
$4\notin \{5\}$. Whenever there is an $m$-connecting path between $i$ and
$j$ given $C$, we can also find an $m_c$-connecting walk between $i$ and $j$
given $C$ (Theorem \ref{thm:dc}). In this example, $3 \rightarrow 4
\rightarrow 5 \leftarrow 4 \leftarrow 2$ is $m_c$-connecting between $3$ and
$2$ given $\{5\}$.
\end{ex}

The notion of $m_c$-connecting walks is related to so-called
\emph{paths-with-tails} in DAGs \citep{matuvs1997conditional}. When we apply
$m_c$-separation to DGs, we will use the term \emph{$d_c$-separation}. The
$d_c$-connecting walks characterize $d$-separation in the sense that there is a
$d$-connecting path between $i$ and $j$ given $C$ if and only there is a
$d_c$-connecting walk between $i$ and $j$ given $C$. Analogously,
$m_c$-connecting walks characterize $m$-separation (Theorem~\ref{thm:dc}).

We defined $m$-separation in terms of \emph{paths}, i.e.,~walks where each node
appears at most once. In $G=([d], E)$, a path is of length at most $d-1$, and
this means that $d$-separation is characterized by walks of length at most $d -
1$. For technical reasons, we need a similar bound on the length of
$m_c$-connecting walks, and this bound is given by $\tilde{n}$,
\begin{align*}
    \tilde{n} = \begin{cases}
        d - 1, & d < 4, \\
        2d - 4, & d \geq 4,
    \end{cases}
\end{align*}
as formalized in Theorem \ref{thm:dc}. From Corollary \ref{cor:dConndcConn}, we
see that $m$-separation and $m_c$-separation lead to the same independence
models. We defer all proofs to Appendix~\ref{app:proofs}.

\begin{thm}\label{thm:dc}
    Let $G = ([d],E)$ be a directed mixed graph, and let $i,j\in [d]$,
    $C\subseteq [d] \setminus \{i,j\}$. If there is an $m$-connecting walk
    between $i$ and $j$ given $C$, then there is an $m_c$-connecting walk
    between $i$ and $j$ given $C$ of length at most $\tilde{n}$.
\end{thm}

\begin{lem}\label{lem:mconn}
    Let $G = ([d],E)$ be a DMG, and let $i,j\in [d]$, $C\subseteq [d]\setminus
    \{i,j\}$. If there is an $m$-connecting walk between $i$ and $j$ given $C$,
    then there is also an $m$-connecting path between $i$ and $j$ given $C$.
\end{lem}

\begin{cor}\label{cor:dConndcConn}
    Let $G = ([d],E)$ be a DMG, and let $i,j\in [d]$, $C\subseteq [d]\setminus \{i,j\}$.
    There is an $m$-connecting path between $i$ and $j$ given $C$ in $G$ if and
    only if there is an $m_c$-connecting walk between $i$ and $j$ given $C$ in
    $G$.
\end{cor}

\subsection{Variables}\label{ssec:var}

We will now describe some of the variables that will go into the vector
$\mathbf{y}$ in the integer-programming problem~\eqref{tag:IP}. A central
contribution of this paper is the concept of a minimal-length encoding. This
allows us to represent the graph learning problem as an integer-programming
problem where the number of variables in $\mathbf{y}$ is linear in the number
of $p$-values. The $p$-values are the input to the learning algorithm, and this
is therefore a parsimonious encoding of the problem.

\subsubsection{Edge variables} \label{sssec:edges}

We define a collection of variables $\mathbf{x}^\rightarrow=\{\xd{i}{j}\}_{i\neq j}$ where
$\xd{i}{j}$ is a binary variable such that $\xd{i}{j} = 1$ represents the
presence of the directed edge from $i$ to $j$. We also
define a collection variables $\mathbf{x}^\leftrightarrow = \{\xb{i}{j}\}_{i\neq
j}$ where $\xb{i}{j}$ is a binary variable such that $\xb{i}{j} = 1$ represents
the existence of the bidirected edge between $i$ and $j$. As indicated by the
notation, $\xd{i}{j}$ and $\xd{j}{i}$ are different variables while $\xb{i}{j}$
and $\xb{j}{i}$ are the same variable. We let $G_\mathbf{x}$ be the
graph corresponding to $\mathbf{x} = (\mathbf{x}^\rightarrow,
\mathbf{x}^\leftrightarrow)$.

In DMGs, we also define binary variables $\xs{i}{j}$ such that $\xs{i}{j} =
\max(\xd{i}{j}, \xb{i}{j})$, i.e,~$\xs{i}{j} = 1$ if and only if there is an
edge between $i$ and $j$ with a head at $j$, and we denote $\{\xs{i}{j}\}_{i\neq
j}$ by $\mathbf{x}^*$. In chain graphs, the variables $\mathbf{x}^\ast$ will
have a different meaning, see Appendix~\ref{ssec:CGs}.

When representing chain graphs, we use the set
$\mathbf{x}^\rightarrow=\{\xd{i}{j}\}_{i\neq j}$ while $\xb{i}{j} = 0$ for all
$i,j$, $i\neq j$. In chain graphs, we use the convention that $\xd{i}{j} = \xd{j}{i} = 1$
 represents the existence of an undirected edge. This is an important
distinction from DMGs, as $\xd{i}{j} = \xd{j}{i} = 1$ represents a directed
cycle $i \rightarrow j \rightarrow i$ in an DMG.

\subsubsection{Minimal-length variables}

We will describe different encodings of the graph learning problem, and each of
them use a notion of connecting paths or connecting walks of a minimal length.
For each encoding, we use the constant $\aG$ to represent the maximal possible
length (in any graph in the relevant class) of a shortest possible connecting
walk,
\begin{align*}
    \aG = \sup_{G\in \mathbb{G},i,j, C} \inf \{ \lvert \omega \rvert : \omega
    \text{ is a connecting walk between $i$ and $j$ given $C$} \},
\end{align*}
where $\mathbb{G}$ is a
collection of graphs each with node set $[d]$, $\lvert \omega\rvert$
denotes the length of the path, and using the convention $\inf\{\emptyset\}=
0$. The constant $\aG$ depends on the class of graphs and the notion of
separation that we use which is why we simply write `a connecting walk' in the
definition of $\aG$. As an example, when using $d$-separation in DAGs, the
relevant connecting walks would be the set of $d$-connecting paths. The length
of a path is at most $d - 1$, and we see that $\aG = d - 1$ in this case.

For each $i,j \in [d]$, $i\neq j$, and $C \subseteq [d]\setminus \{i,j\}$, we
define a variable $\lc{i}{j}{C} \in \{1,2,\ldots, \aG + 1\}$. We use the set
$\{i,j\}$ to index the variable, i.e., $\lc{i}{j}{C}$ and $\lc{j}{i}{C}$ refer
to the same variable. We will use the collection $\mathbf{l}^{m} =
\{\lc{i}{j}{C}\}_{i\neq j, C\subseteq [d] \setminus \{i,j\}}$ to represent the
length of the shortest connecting walk from $i$ to $j$ given $C$ if such a walk
exists (the exact meaning of `connecting walk' will depend on the specific
choice of graphs and separation criteria). The case $\lc{i}{j}{C} = \aG+1$ will
represent separation, i.e.,~that there is no connecting walk between $i$ and $j$
given $C$. 

For each $i,j \in [d]$, $i\neq j$, and $C \subseteq [d]\setminus \{i,j\}$, we
also define the binary variable $\zc{i}{j}{C} \in \{0, 1\}$. This variable will
represent the existence of a connecting walk between $i$ and $j$ given $C$ such
that $\zc{i}{j}{C} = 0$ if and only if $\lc{i}{j}{C} = \aG+1$.

\subsubsection{Directed length variables}

We say that a walk, $i_0,i_1,\ldots,i_n$, is \emph{descending} from $i_0$ to
$i_n$ if for all $k=0,\ldots,n-1$ we have $i_k \rightarrow i_{k+1}$ or $i_k -
i_{k+1}$ \citep{studeny1998chain}, and by convention the trivial walk is
descending. We say that $i$ is \emph{anterior} to $j$ if there exists a
descending path from $i$ to $j$. In the context of hybrid graphs, $\an(j)$
denotes the set of nodes that are anterior to $j$. Similarly, $\an(C)$ denotes
the set of nodes that are anterior to at least one node in $C$. In a DMG, a path
is descending if and only if it is directed. When $G$ is a DMG, this means that
$i$ is an ancestor of $j$ in $G$ if and only if it is anterior to $j$ in $G$.

For $i \neq j$, we define a variable $\ld{i}{j} \in \{1,2,\ldots,d\}$ which we
will use to represent the length of a shortest descending path from $i$ to $j$
if such a path exists, and otherwise $\ld{i}{j} = d$. We will denote
$\{\ld{i}{j}\}_{i\neq j}$ by $\mathbf{l}^\rightarrow$. We let $\di{i}{j}$ be a
binary variable such that $\di{i}{j} = 1$ if and only if there is no descending
path from $i$ to $j$, and we denote $\{\di{i}{j}\}_{i\neq j}$ by
$\mathbf{d}^{\not\rightarrow}$. For $C\subseteq [d]$ and $i\notin C$, we also
define the variable $\di{i}{C}$ such that $\di{i}{C} = 1$ if and only if there
is no descending path from $i$ to any node in $C$, and we denote
$\{\di{i}{C}\}_{i\notin C}$ by $\mathbf{d}_C^{\not\rightarrow}$.

\subsubsection{Bidirected length variables}\label{sssec:bidir}

We say that a walk is \emph{bidirected} if every edge on the walk is bidirected.
In DMGs and for $i \neq j$, we let $\lcb{i}{j}{C}\in \{1,2,\ldots,d\}$ represent
the length of the shortest bidirected path between $i$ and $j$ such that all
nodes on the path are in $C$ if such a path exists, and otherwise
$\lcb{i}{j}{C} = d$. Note that the endpoint nodes are also required to be in $C$
which means, e.g., that $\lcb{i}{j}{\emptyset} = d$ in any DMG. On the
other hand, if $i \leftrightarrow j$ is in the graph, then $\lcb{i}{j}{\{i,j\}}
= 1$. We denote the collection $\{\lcb{i}{j}{C}\}_{i\neq j}$ by
$\mathbf{l}^\leftrightarrow$.

We say that a walk is \emph{semi-bidirected} from $i$ to $j$ if it is bidirected
or if it is the concatenation of an edge $i\rightarrow k$ and a bidirected walk
between $k$ and $j$ (the trivial walk with no edges is bidirected). We will let
$\lcd{i}{j}{C} \in \{1,2,\ldots, d\}$ denote the length of a shortest
semi-bidirected path from $i$ to $j$ such that $i\notin C$ and all other nodes
on the path are in $C$ if such a path exists, and otherwise $\lcd{i}{j}{C} = d$.
We denote the collection $\{\lcd{i}{j}{C}\}$ by $\mathbf{l}^\twoheadrightarrow$.

\subsection{Constraints}\label{ssec:constraints}

Some of the constraints that go into the integer-programming formulation are
shared between the three settings (DGs, DMGs, CGs), while other constraints are
class-specific. We start by describing those that are shared. The theorems in
Section~\ref{ssec:DGs}, Section~\ref{ssec:DMGs}, and Appendix~\ref{ssec:CGs} list
the necessary and sufficient set of constraints for each class of graphs. The
minimal-length constraints are class-specific, and they are listed in
Section~\ref{ssec:DGs}, Section~\ref{ssec:DMGs}, and Appendix~\ref{ssec:CGs}.

\subsubsection{Consistency constraints}

Consistency constraints connect related sets of variables. As an example,
$\di{i}{j}$ is a binary version (indicating the existence
or non-existence of a descending path)
of $\ld{i}{j}$ (the length of a shortest descending path), and
Constraints \eqref{tag:C2} and \eqref{tag:C3} enforce consistency between
$\ld{i}{j}$ and $\di{i}{j}$. Similarly, Constraints \eqref{tag:C4} and
\eqref{tag:C5} enforce the equivalence
\begin{align*}
    \zc{i}{j}{C} = 0 \iff
\lc{i}{j}{C} = \aG + 1,
\end{align*}
representing that  $i$ and $j$ are separated given $C$ if and only if there is
no shortest connecting walk between $i$ and $j$ given $C$.
\begin{align}
& \forall i , j: i\neq j : &  1-\di{i}{j} &\leq d - l_{ij}^{\rightarrow}
\tag{C2}\label{tag:C2} \\
& \forall i , j: i\neq j : &  d - l_{ij}^{\rightarrow} &\leq (d-1)(1 -
\di{i}{j}) \tag{C3}\label{tag:C3} \\
&    \forall i, j: i < j \forall C :  &  \zc{i}{j}{C} & \leq a_\mathbb{G} + 1 -
\lc{i}{j}{C} \tag{C4} \label{tag:C4}\\
&    \forall i, j: i < j \forall C : & a_\mathbb{G} + 1 - \lc{i}{j}{C} & \leq
a_\mathbb{G} \zc{i}{j}{C}  \tag{C5}\label{tag:C5} \\
 & \forall i \forall C:   i\notin C : &  \di{i}{C} & = \min_{k\in C}(\di{i}{k},
 1)  \tag{Ra}\label{tag:R1a} \\
   &  \forall i \forall C:  i\in C: &   \di{i}{C} & = 0,
   \tag{Rb}\label{tag:R1b}
\end{align}
The constant 1 in~\eqref{tag:R1a} handles the case where $C = \emptyset$. The
proof of the following lemma follows immediately from the constraints.

\begin{lem}\label{lem:consist}
    Let $\mathbf{l}^m = \{ \lc{i}{j}{C}\}$. We define a collection of variables
    $\{\zc{i}{j}{C}\}$ such that $\zc{i}{j}{C} = 0$ if $\lc{i}{j}{C} = \aG + 1$,
    and $\zc{i}{j}{C} = 1$ otherwise. The collection $\{\zc{i}{j}{C}\}$ is the
    only solution to Constraints \eqref{tag:C4}--\eqref{tag:C5} when
    $\mathbf{l}^m$ is fixed.
\end{lem}

\subsubsection{Anterior distance constraints}\label{sssec:ant}

In a DMG or in a hybrid graph, we define the \emph{anterior distance from $i$ to
$j$} to be the length of a shortest descending path from $i$ to $j$ if such a
path exists, and otherwise we define it to be $d$. We use the following
\emph{anterior distance constraints} to ensure that $\ld{i}{j}$ equals the
anterior distance from $i$ to $j$ in $G_\mathbf{x}$ for each $i$ and $j$, $i
\neq j$. Constraints~\eqref{tag:D1}--\eqref{tag:D2} below define variables
$u_{ij}^{\text{D1}}$ and $u_{ijk}^{\text{D2}}$ that are used in
Constraint~\eqref{tag:N1}. Constraint~\eqref{tag:N1} implies
Constraints~\eqref{tag:D1}--\eqref{tag:D2}, and \eqref{tag:D1}--\eqref{tag:D2} are only
listed to provide an interpretation of the $u_{ij}^{\text{D1}}$- and
$u_{ijk}^{\text{D2}}$-variables.
\begin{align}
    & \forall i \forall j: i\neq j & & \ld{i}{j} \leq u_{ij}^{\text{D1}} \coloneqq 1 -
    (d-1)(\xd{i}{j} - 1)  \tag{D1}\label{tag:D1} \\
    & \forall i \forall j: i\neq j \forall k: k\neq i,j & & \ld{i}{j} \leq
    u_{ijk}^{\text{D2}} \coloneqq 1 + \ld{i}{k} - (d-2)(\xd{k}{j} - 1)
    \tag{D2}\label{tag:D2} \\
    & \forall i \forall j: i\neq j  & & \ld{i}{j} = \min_k(u_{ij}^{\text{D1}},
    u_{ijk}^{\text{D2}})  \tag{N1}\label{tag:N1}
\end{align}

\begin{lem}\label{lem:antDist}
    Let $G_\mathbf{x}$ be a fixed DMG or hybrid graph corresponding to
    $\mathbf{x}$.
    \begin{enumerate}[label=(\roman*)]
        \item The collection of variables $\mathbf{l}^\rightarrow =
        \{\ld{i}{j}\}$ solve \eqref{tag:N1} if and only if $\ld{i}{j}$ equals
        the anterior distance from $i$ to $j$ in $G_\mathbf{x}$ for each $i,j$,
        $i\neq j$. \label{bul:lDirdDir1}
        \item Define $\mathbf{d}^{\not\rightarrow} = \{\di{i}{j}\}$ for $i\neq j$
        such that $\di{i}{j} = 1$ if there is no descending path from
        $i$ to $j$ in $G_\mathbf{x}$, and $\di{i}{j} = 0$ otherwise. The set
        $\mathbf{d}^{\not\rightarrow}$ is the
        unique solution to \eqref{tag:C2}--\eqref{tag:C3}, when
        $\mathbf{l}^\rightarrow$ is the collection of anterior distances in
        $G_\mathbf{x}$. 
        \label{bul:lDirdDir2}
        \item Assume that \eqref{tag:C2}--\eqref{tag:C3} hold. Define $\di{i}{C}
        = 1$ if there is a descending path from $i$ to $C$ in
        $G_\mathbf{x}$, and $\di{i}{C} = 0$ otherwise. The set $\{\di{i}{C}\}$
        is the unique solution to
        \eqref{tag:R1a}--\eqref{tag:R1b}. Moreover, $i \in
        \an_{G_\mathbf{x}}(C)$ if and only if $\di{i}{C} = 0$ when
        $\mathbf{l}^\rightarrow$ is the collection of anterior distances in
        $G_\mathbf{x}$. \label{bul:lDirdDir3}
    \end{enumerate}
\end{lem}

\subsubsection{Acyclicity}\label{sssec:acyc}

We will use the following constraint to enforce acyclicity in ADMGs and DAGs
(Theorems~\ref{thm:DAGs} and~\ref{thm:DMGs}). The constraint ensures that for
$i\neq j$, there is not both a directed path from $i$ to $j$ and a directed path
from $j$ to $i$,
\begin{align}
    & & \forall i \forall j: i< j: & & 1 \leq \di{i}{j} + \di{j}{i}.
    \tag{AC}\label{tag:AC} 
\end{align}

\subsection{Directed graphs}\label{ssec:DGs}

For a DG, $G = ([d],E)$, and $i,j\in [d]$, $C\subseteq [d] \setminus \{i,j\}$,
we define the \emph{$d$-distance between $i$ and $j$ given $C$ in $G$} to
be the length of a shortest $d$-connecting path between $i$ and $j$ given
$C$ if such a path exists, and otherwise to be $\aG + 1$. We use $\aG = d-1$ in
this context as this is the maximal length of a path in graphs with node set
$[d]$. In this section, we list a set of \emph{minimal-length constraints} in
directed graphs (\eqref{tag:L1a}--\eqref{tag:L4}, \eqref{tag:M1}), that is, a
set of constraints that ensure that the $\lc{i}{j}{C}$-variables equal the
$d$-distances in $G_\mathbf{x}$. The constraints are listed below, and
Example~\ref{ex:constraints} explains the technique that they employ. This
technique can be used in different classes of graphs, including DMGs and chain
graphs as shown in Section~\ref{ssec:DMGs} and Appendix~\ref{ssec:CGs}. For DGs,
the minimal-length constraints are:
\begin{align}
      \forall i , j: i < j & \forall C: i,j\notin C:    \notag \\  &
      \lc{i}{j}{C}  \leq u_{ijC}^{\text{L1a}} \coloneqq 1 - (d-1)(\xd{i}{j} - 1)
      \tag{L1a}\label{tag:L1a} \\
      &   \lc{i}{j}{C}  \leq u_{ijC}^{\text{L1b}} \coloneqq 1 - (d-1)(\xd{j}{i} - 1)
      \tag{L1b}\label{tag:L1b} \\
 \forall i , j: i < j  &   \forall k: i,j,k \text{ distinct }  \forall C: i,j,k
 \notin C:  \notag \\ &     \lc{i}{j}{C} \leq u_{ijkC}^{\text{L2a}} \coloneqq 1 +
 \lc{i}{k}{C}  - (d - 2)(\xd{k}{j} - 1)   \tag{L2a}\label{tag:L2a} \\
  &    \lc{i}{j}{C} \leq u_{ijkC}^{\text{L2b}} \coloneqq 1 + \lc{j}{k}{C}  - (d -
  2)(\xd{k}{i} - 1)   \tag{L2b}\label{tag:L2b} \\
    \forall i, j: i < j &  \forall k: i,j,k \text{ distinct }  \forall C:
    i,j\notin C, k\in C:   \notag \\  &   \lc{i}{j}{C}  \leq
    u_{ijkC}^{\text{L3}} \coloneqq 2 - (d - 2 )(\xd{i}{k} - 1)- (d-2
    )(\xd{j}{k} - 1)  \tag{L3}\label{tag:L3} \\
 \forall i, j: i< j &  \forall k: i,j,k \text{ distinct }  \forall C:
 i,j,k\notin C:   \notag \\  &    \lc{i}{j}{C}   \leq u_{ijkC}^{\text{L4}} \coloneqq
 \lc{i}{k}{C} + \lc{k}{j}{C} + (d-2)\di{k}{C}   \tag{L4}\label{tag:L4} \\
   \forall i,  j: i< j  & \forall C:i,j\notin C  :  \notag \\   &
   \lc{i}{j}{C}   =
   \min_{k}(u_{ijC}^{\text{L1a}},u_{ijC}^{\text{L1b}},u_{ijkC}^{\text{L2a}},
   u_{ijkC}^{\text{L2b}},u_{ijkC}^{\text{L3}},u_{ijkC}^{\text{L4}}).
   \tag{M1}\label{tag:M1}
\end{align}
For each $i$ and $j$, $i<j$, Constraints~\eqref{tag:L1a}--\eqref{tag:L4} define
the variables $u_{ijC}^{\text{L1a}}$, $u_{ijC}^{\text{L1b}}$,
$u_{ijkC}^{\text{L2a}}$, $ u_{ijkC}^{\text{L2b}}$, $u_{ijkC}^{\text{L3}}$, and
$u_{ijkC}^{\text{L4}}$ for the appropriate values of $k$ and $C$ as stated
above. We refer to these variables as \emph{$u$-variables}. For each $i,j,C$, we
let $\Uc{i}{j}{C}$ denote the set of $u$-variables that go into the minimum
which defines $\lc{i}{j}C$ in \eqref{tag:M1}. Constraint~\eqref{tag:M1} implies
Constraints~\eqref{tag:L1a}--\eqref{tag:L4}.

The Constraints~\eqref{tag:L1a}--\eqref{tag:L4} implement the four rules
below. The validity of these rules is an immediate consequence of the definition
of a $d$-connecting walk (using the convention that if $i\in C$ or $j\in C$,
then there is no $d$-connecting walk between $i$ and $j$ given $C$ in any
graph). We let $\lc{i}{j}{C}$ denote the $d$-distance between $i$ and $j$ given
$C$.
\begin{itemize}
    \item \eqref{tag:L1a}, \eqref{tag:L1b}: If $i$ and $j$ are adjacent, then
    $\lc{i}{j}{C} =1$.
    \item \eqref{tag:L2a}, \eqref{tag:L2b}: If there is a $d$-connecting walk
    between $i$ and $k$ given $C$ of length $\lc{i}{k}{C}$ and $k\rightarrow j$,
    then there is a $d$-connecting walk between $i$ and $j$ given $C$ of length
    $\lc{i}{k}{C} + 1$.
    \item \eqref{tag:L3}: If $i\rightarrow k \leftarrow j$ and $k \in C$, then
    there is a $d$-connecting walk of length $2$ between $i$ and $j$ given $C$.
    \item \eqref{tag:L4}: If there is a $d$-connecting path between $i$ and $k$
    given $C$ of length $\lc{i}{k}{C}$ and a $d$-connecting path between $k$ and
    $j$ given $C$ of length $\lc{k}{j}{C}$ such that $k \in \an(C)\setminus C$,
    then there is a $d$-connecting walk between $i$ and $j$ given $C$ of length
    $\lc{i}{k}{C}+\lc{k}{j}{C}$.
\end{itemize}
These rules describe how $d$-distances and edges between certain pairs
of variables constrain the $d$-distances between other pairs of variables. In
order to obtain an encoding of the graph learning problem, we need to select a
system of such rules which is \emph{complete} in the sense that for any DG it
specifies the $d$-distance for any triple $(i,j,C)$, and the completeness of
(\eqref{tag:L1a}--\eqref{tag:L4}, \eqref{tag:M1}) is proven in Theorem
\ref{thm:DAGs}. Constraints \eqref{tag:K1a}--\eqref{tag:K4} and Constraints
\eqref{tag:I1a}--\eqref{tag:I4} formulate analogous, complete systems of rules for
DMGs with $m$-separation (Section~\ref{ssec:DMGs}) and CGs
(Appendix~\ref{ssec:CGs}), respectively. Appendix~\ref{app:encodings} provides
analogous, complete systems of rules for DGs with $d_c$-separation and for DMGs
with $m_c$-separation. The system (\eqref{tag:L1a}--\eqref{tag:L4},
\eqref{tag:M1}) is \emph{cubic} in the sense that it uses at most triples of
nodes $(i,j,k)$ (conditionally on the set $C$). The $d_c$- and $m_c$-encodings
in Appendix~\ref{app:encodings} are of higher order.
Appendix~\ref{app:encodings} also includes an complete system of rules which
avoids using Constraints~\eqref{tag:L2a}--\eqref{tag:L2b} at the cost of using
more variables.

Example~\ref{ex:constraints} gives concrete examples of the above rules in the
context of DGs and DAGs.

\begin{ex}\label{ex:constraints}    
Constraints~\eqref{tag:L1a}--\eqref{tag:L4} all describe how the existence of a
path along with certain edge configurations restrict other $d$-distances
as illustrated in this example. We consider first the DAG in the topleft corner
of Figure \ref{fig:constraints}. When $C = \emptyset$, the walk $2 \leftarrow
3 \leftarrow 4$ is $d$-connecting between $2$ and $4$ given $C$. This
walk is of length two, and the edge $1 \leftarrow 2$ is also
present in the
graph. Concatenating the walk with this edge gives a $d$-connecting walk
between $1$ and $4$ given $C$ of length 3 (the concatenation is
$d$-connecting since $2\notin C$ and the edge $1 \leftarrow 2$ has a tail at $2$). This is an
example of a general rule: If $i,j,k\notin C$, the $d$-distance between
$i$ and $k$ given $C$ is $\lc{i}{k}{C}$ and the edge $\xd{k}{j}$ is present,
then we can concatenate these to obtain a $d$-connecting path between $i$ and
$j$ given $C$ of length $1 + \lc{i}{k}{C}$. The concatenation is always
$d$-connecting since $k \notin C$ and $k$ is necessarily a noncollider on this
new walk. This means that the shortest $d$-connecting walk between $i$ and $j$
given $C$ is at most $1 + \lc{i}{k}{C}$. This rule is encoded by
Constraint~\eqref{tag:L2a} as 
\begin{align*}
\lc{i}{j}{C} \leq 1 + \lc{i}{k}{C}  - (d - 2)(\xd{k}{j} - 1),
\end{align*}
using that $\xd{k}{j} = 1$ if and only if $k \rightarrow j$ is in $G_\mathbf{x}$
(see also the illustration in Figure~\ref{fig:constraints} on the bottom left).
When $k\rightarrow j$ is not in the graph, we see that $d-2$ is added to the
right-hand side above, making the inequality trivial due to the range of the
$\lc{i}{j}{C}$-variables.

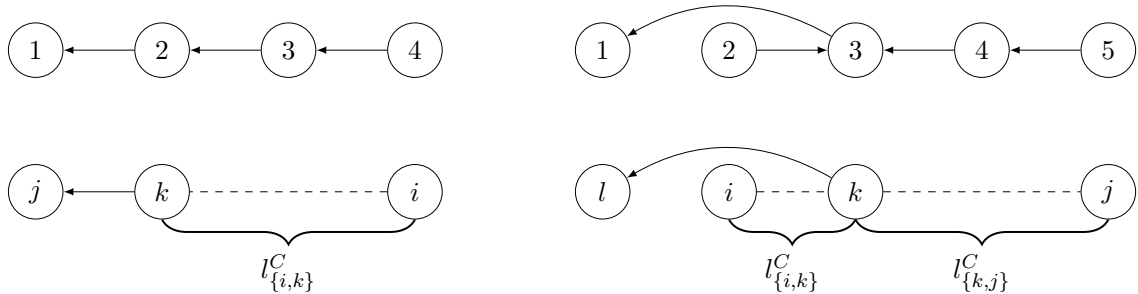
\begin{figure}[H]
 \resizebox{\textwidth}{!}{
    \begin{tikzpicture}
\tikzset{every node/.style={minimum size=2em}}
\node[circle, draw] (i1) {$1$};
\node[circle, draw, right = of i1] (i2) {$2$};
\node[circle, draw, right = of i2] (i3) {$3$};
\node[circle, draw, right = of i3] (i4) {$4$};
\draw[-latex] (i2) -- (i1);
\draw[latex-] (i2) -- (i3);
\draw[latex-] (i3) -- (i4);

\begin{scope}[shift={(8,0)}]
\node[circle, draw] (i1) {$1$};
\node[circle, draw, right = of i1] (i2) {$2$};
\node[circle, draw, right = of i2] (i3) {$3$};
\node[circle, draw, right = of i3] (i4) {$4$};
\node[circle, draw, right = of i4] (i5) {$5$};
\draw[latex-] (i1) to [bend left] (i3);
\draw[latex-] (i3) -- (i2);
\draw[latex-] (i3) -- (i4);
\draw[latex-] (i4) -- (i5);
\end{scope}

\begin{scope}[shift={(0,-2)}]
   \tikzset{every node/.style={minimum size=2em}}
\node[circle, draw] (j) {$j$};
\node[right = of j, circle, draw] (k) {$k$};
\node[right = of k, circle] (u1) {};
\node[right = of u1, circle, draw] (i) {$i$};
\draw[-latex] (k) -- (j);
\draw[dashed] (i) -- (k);
\draw[decorate,thick, decoration={brace, amplitude=10pt,mirror}] (k.south) -- (i.south)
          node [midway,below=10pt] {$\lc{i}{k}{C}$};

\begin{scope}[shift={(8,0)}]
    \node[circle, draw] (l) {$l$};
\node[right = of l, circle, draw] (i) {$i$};
\node[right = of i, circle, draw] (k) {$k$};
\node[right = of k, circle] (u1) {};
\node[right=of u1, circle, draw] (j) {$j$};
\draw[-latex] (k) to [bend right] (l);
\draw[dashed] (i) -- (k);
\draw[dashed] (k) -- (j);
\draw[decorate,thick, decoration={brace, amplitude=10pt,mirror}] (i.south) -- (k.south)
          node [midway,below=10pt] {$\lc{i}{k}{C}$};
\draw[decorate,thick, decoration={brace, amplitude=10pt,mirror}] (k.south) -- (j.south)
          node [midway,below=10pt] {$\lc{k}{j}{C}$};
\end{scope}
\end{scope}
\end{tikzpicture}
}
\caption{Illustration of minimal-length rules and Constraints~\eqref{tag:L2a}
and~\eqref{tag:L4}, see Example~\ref{ex:constraints}.}
\label{fig:constraints}
\end{figure}

Constraint~\eqref{tag:L4} is illustrated similarly in the DAG on the top right
of Figure~\ref{fig:constraints}: The path $2 \rightarrow 3$ is
$d$-connecting given $C = \{1\}$, and the path $3 \leftarrow 4 \leftarrow
5$ is also $d$-connecting given $C = \{ 1 \}$. The node $3$ is an ancestor
of the conditioning set $\{1\}$, but not an element of
$\{1\}$, and the
concatenation of the two paths is $d$-connecting between $2$ and $5$ given
$\{1\}$. Again, this an example of a general rule: If $i,j,k\notin C$, $k\in
\an(C)$, and there is a $d$-connecting walk between $i$ and $k$ given $C$ of
length $\lc{i}{k}{C}$ as well as a $d$-connecting walk between $k$ and $j$
given $C$ of length $\lc{k}{j}{C}$, then there is a $d$-connecting walk
between $i$
and $j$ given $C$ of length $\lc{i}{k}{C} + \lc{k}{j}{C}$ (see the illustration
on the bottom right of Figure~\ref{fig:constraints}).
Constraint~\eqref{tag:L4} enforces this rule by requiring
\begin{align*}
    \lc{i}{j}{C} \leq \lc{i}{k}{C} + \lc{k}{j}{C} + (d -2 )\di{i}{C},
\end{align*}
where $\di{k}{C} = 0$ if $k\in \an(C)$, and otherwise $\di{k}{C} = 1$. When
$k\notin \an(C)$, we see that the inequality is always satisfied.
\end{ex}

On the basis of Constraints~\eqref{tag:L1a}--\eqref{tag:L4} and
Example~\ref{ex:constraints}, it is natural to ask if analogous constraints
would work using the binary variables $\zc{i}{j}{C}$ instead of the
integer-valued variables $\lc{i}{j}{C}$. In short, this does not work as there
could be solutions with $\zc{i}{j}{C} = 1$ even if there is no $d$-connecting
path between $i$ and $j$ given $C$ in $G_\mathbf{x}$.


The following theorem describes how the relevant set of constraints encodes the
$d$-distances corresponding to a DG or a DAG, $G_\mathbf{x}$ by showing that the
$\lc{i}{j}{C}$-variables will equal the $d$-distances when
Constraint~\eqref{tag:M1} is satisfied.


Recall that $\mathbf{x} = (\mathbf{x}^\rightarrow, \mathbf{x}^\leftrightarrow)$.
When $G_\mathbf{x}$ is a DG, we have $\mathbf{x}^\leftrightarrow = 0$.

\begin{thm}[DGs and DAGs]\label{thm:DAGs}
    Let $G_\mathbf{x}$ be a fixed DG, and assume that $\mathbf{x}$ corresponds
    to $G_\mathbf{x}$. Assume that $\mathbf{x}^\rightarrow$,
    $\mathbf{l}^\rightarrow$, $\mathbf{d}^{\not\rightarrow}$, and
    $\mathbf{d}_C^{\not\rightarrow}$ satisfy \eqref{tag:C2}--\eqref{tag:C3},
    \eqref{tag:N1}, and \eqref{tag:R1a}--\eqref{tag:R1b}, and let $a_\mathbb{G}
    = d-1$. Then, the following holds:
    \begin{enumerate}[label=(\roman*)]
        \item Constraint \eqref{tag:AC} is satisfied if and only if
        $G_\mathbf{x}$ is a directed acyclic graph. \label{bul:thmDAG1}
        \item The set
        \begin{align*}
            \{\lc{i}{j}{C}: i,j\in [d], C\subseteq [d]\setminus \{i,j\}, i\neq j\}
        \end{align*}
        is a solution to \eqref{tag:M1} if and only if $\lc{i}{j}{C}$ equals the
        $d$-distance between $i$ and $j$ given $C$ in $G_\mathbf{x}$ for each
        triple $(i,j,C)$ such that $i\neq j$ and $i,j\notin C$. That is, the 
        $d$-distances in $G_\mathbf{x}$ are the unique solution of
        \eqref{tag:M1} when $\mathbf{x}$ corresponds to $G_\mathbf{x}$.
        \label{bul:thmDAG2}
    \end{enumerate}
\end{thm}

\subsection{Directed mixed graphs}\label{ssec:DMGs}

The approach from the previous section can be generalized to directed mixed
graphs (DMGs). For this, we need additional variables as well as additional consistency
constraints and distance constraints. Towards the end of the section, we list
the minimal-length constraints in the class of directed mixed graphs.

\subsubsection{Consistency constraints}

We let $\mathbf{x}^\leftrightarrow$ denote the collection of
$\xb{i}{j}$-variables and we let $\mathbf{x}^\ast$ denote the collection of
$\xs{i}{j}$-variables (see Section \ref{sssec:edges}), and we let
$\mathbf{l}^\leftrightarrow$ denote the collection of $\lb{i}{j}$-variables (see
Section \ref{sssec:bidir}). The proof of Lemma~\ref{lem:consistDMG} follows
immediately from the constraints below.
\begin{align}
& \forall i , j: i\neq j : &  \xs{i}{j} & \leq  \xd{i}{j}+ \xb{i}{j} \tag{C6}\label{tag:C6}  \\
& \forall i , j: i\neq j : &  \xd{i}{j} &\leq  \xs{i}{j} \tag{C7}\label{tag:C7} \\
& \forall i , j: i < j : &  \xb{i}{j} &\leq  \xs{i}{j} \tag{C8}\label{tag:C8} \\
& \forall i , j: i < j : &  \xb{i}{j} &\leq  \xs{j}{i} \tag{C9}\label{tag:C9} \\
& \forall i, j: i < j \forall C :  &  \zcb{i}{j}{C} & \leq d - \lcb{i}{j}{C} \tag{C10} \label{tag:C10}\\
& \forall i, j: i < j \forall C : &  d - \lcb{i}{j}{C} & \leq (d-1)\zcb{i}{j}{C}  \tag{C11}\label{tag:C11} 
\end{align}

\begin{lem}\label{lem:consistDMG}
Let $\mathbf{x}^\rightarrow$, $\mathbf{x}^\leftrightarrow$, and
$\mathbf{l}^\leftrightarrow$ be fixed. Define $\mathbf{x}^\starright
=\{\xs{i}{j} \}$ such that $\xs{i}{j} = \max(\xd{i}{j}, \xb{i}{j})$, and define
$\mathbf{z}^\leftrightarrow =\{ \zcb{i}{j}{C} \}$ such that $\zcb{i}{j}{C} = 0$
if and only if $\lcb{i}{j}{C} = d$. 
    \begin{enumerate}[label=(\roman*)]
        \item The set $\mathbf{x}^\starright$ is the
        only solution to \eqref{tag:C6}--\eqref{tag:C9}.
        \item The set
        $\mathbf{z}^\leftrightarrow$ is the only solution to
        \eqref{tag:C10}--\eqref{tag:C11}.
    \end{enumerate}
\end{lem}

\subsubsection{Bidirected distance constraints}

The following set of distance constraints were not needed in directed graphs. 
We say that a walk is \emph{bidirected} if it consists of bidirected edges only.
For $i,j \in [d]$, $i\neq j$, and $C\subseteq [d]$, the \emph{bidirected
distance between $i$ and $j$ relative to $C$} is the minimal length of a
bidirected path between $i$ and $j$ such that all nodes on the path are in $C$
if such a path exists, and otherwise it equals $d$. In this definition all nodes
on the path are required to be in $C$, including the endpoints $i$ and $j$.
\begin{align*}
     \forall i \forall j: i < j & \forall C: i,j\in C:  \\
    &   \lcb{i}{j}{C} \leq
    u_{ijC}^{\text{F1}} \coloneqq 1 - (d-1)(\xb{i}{j} - 1)  \tag{F1}\label{tag:F1} \\
     \forall i , j: i< j & \forall k: k\neq i,j \forall C: i,j,k \in C: \\
    & \lcb{i}{j}{C} \leq u_{ijkC}^{\text{F2}} \coloneqq 1 + \lcb{i}{k}{C} -
    (d-2)(\xb{j}{k}- 1)  \tag{F2}\label{tag:F2} \\
     \forall i \forall j: i< j & \forall C : \\ &  \lcb{i}{j}{C}  =
    \min_k(u_{ijC}^{\text{F1}}, u_{ijkC}^{\text{F2}}, d)  \tag{O1}\label{tag:O1}
\end{align*}
The minimum in \eqref{tag:O1} is only over variables $u_{ijC}^{F1}$ and
$u_{ijkC}^{F2}$ that are defined in \eqref{tag:F1}--\eqref{tag:F2}, and it
includes the constant $d$ such that $\lcb{i}{j}{C} = d$ when no such variables
are defined, e.g.,~when $i\notin C$. We let $\mathbf{l}^\twoheadrightarrow$ denote the
collection of $\lcd{i}{j}{C}$-variables (see Section \ref{sssec:bidir}). For $i,j
\in [d]$, $i\neq j$, and $C\subseteq [d]$, the \emph{semi-bidirected distance
from $i$ to $j$ relative to $C$} is the minimal length of a semi-bidirected path
from $i$ to $j$ such that $i \notin C$ and all other nodes on the path are in
$C$ if such a path exists, and otherwise it equals $d$.
\begin{align*}
     \forall i \forall j: i \neq j &\forall C: i\notin C, j \in C :  \\ &
     \lcd{i}{j}{C} \leq u_{ij}^{\text{E1}} \coloneqq
    1 - (d-1)(\xs{i}{j} - 1) \tag{E1}\label{tag:E1} \\
     \forall i \forall j: i\neq j & \forall k: i,j,k \forall C: i\notin C,
     j,k\in C, j\neq k: \\ &
    \lcd{i}{j}{C}  \leq  u_{ijk}^{\text{E2}} \coloneqq 1 + \lcb{k}{j}{C} -
    (d-2)(\xs{i}{k} - 1)
    \tag{E2}\label{tag:E2} \\
    \forall i \forall j: i\neq j & \forall C: i\notin C, j \in C:  \\ &
    \lcd{i}{j}{C}  = \min_k(u_{ij}^{\text{E1}},
    u_{ijk}^{\text{E2}})  \tag{G1}\label{tag:G1}
\end{align*}

\begin{lem}\label{lem:biDist}
    Let $G_\mathbf{x}$ be a fixed DMG. The collection of variables $\{
    \lcb{i}{j}{C} \}$ solves \eqref{tag:O1} if and only if $\lcb{i}{j}{C}$ equals
    the  bidirected distance between $i$ and $j$ relative to $C$ for each
    $i,j,C$.
\end{lem}

\begin{lem}\label{lem:semibiDist}
    Assume that \eqref{tag:C6}--\eqref{tag:C9} are satisfied. Let $G_\mathbf{x}$
    be a fixed DMG, and assume that $\mathbf{l}^\leftrightarrow$ solve
    \eqref{tag:O1}. The collection of variables $\{
    \lcd{i}{j}{C} \}$ solve \eqref{tag:G1} if and only if $\lcd{i}{j}{C}$ equals
    the  semi-bidirected distance from $i$ to $j$ relative to $C$ for each
    $i,j,C$ such that $i \neq j$.
\end{lem}

\subsubsection{Minimal-length constraints}

For a DMG, $G = ([d],E)$, and $i,j\in [d]$, $C\subseteq [d] \setminus \{i,j\}$,
we define the \emph{$m$-distance between $i$ and $j$ given $C$ in $G$} to
be the length of a shortest $m$-connecting path between $i$ and $j$ given
$C$ if such a path exists, and otherwise to be $\aG + 1$. The following are the
minimal-length constraints for directed mixed graphs and acyclic directed mixed
graphs using $m$-separation.
\begin{align}
  \forall i , j: i < j & \forall C: i,j\notin C:     \notag \\  &  \lc{i}{j}{C}
  \leq u_{ijC}^{\text{K1a}} \coloneqq 1 - (d-1)(\xs{i}{j} - 1) \tag{K1a}\label{tag:K1a}
  \\
 & \lc{i}{j}{C}  \leq u_{ijC}^{\text{K1b}} \coloneqq 1 - (d-1)(\xs{j}{i} - 1)
 \tag{K1b}\label{tag:K1b} \\
 \forall i , j: i < j   & \forall k: i,j,k \text{ distinct }  \forall C: i,j,k
 \notin C:  \notag \\  
 &     \lc{i}{j}{C}  \leq u_{ijkC}^{\text{K2a}} \coloneqq  1 + \lc{i}{k}{C}  -
 (d-2)(\xd{k}{j}  - 1)  \tag{K2a}\label{tag:K2a} \\
  &  \lc{i}{j}{C}  \leq u_{ijkC}^{\text{K2b}} \coloneqq  1 + \lc{j}{k}{C}  -
  (d-2)(\xd{k}{i}  - 1)   \tag{K2b}\label{tag:K2b} \\
     \forall i, j: i < j &  \forall k: i,j,k \text{ distinct } \forall C:
     i,j\notin C, k \in C:   \notag \\  &   \lc{i}{j}{C}  \leq
     u_{ijkC}^{\text{K3}} \coloneqq 
     \lcd{i}{k}{C} + \lcd{j}{k}{C} \tag{K3}\label{tag:K3} \\
  \forall i, j: i< j &  \forall k: i,j,k \text{ distinct }  \forall C:
  i,j,k\notin C:   \notag \\   &  \lc{i}{j}{C}   \leq u_{ijkC}^{\text{K4}} \coloneqq
  \lc{i}{k}{C} + \lc{k}{j}{C} + (d-2)\di{k}{C}   \tag{K4}\label{tag:K4} \\
   \forall i,  j: i< j &  \forall C:i,j\notin C :    \notag \\   &
   \lc{i}{j}{C}   =
   \min_{k}(u_{ijC}^{\text{K1a}},u_{ijC}^{\text{K1b}},u_{ijkC}^{\text{K2a}},u_{ijkC}^{\text{K2b}},
   u_{ijkC}^{\text{K3}},u_{ijkC}^{\text{K4}})  
   \tag{P1}\label{tag:P1}
 \end{align}

\begin{thm}[DMGs and ADMGs]\label{thm:DMGs}
Let $G_\mathbf{x}$ be a fixed DMG such that  $\mathbf{x} =
(\mathbf{x}^\rightarrow, \mathbf{x}^\leftrightarrow)$ corresponds to
$G_\mathbf{x}$. Assume that $\mathbf{x}^\rightarrow$,
$\mathbf{x}^\leftrightarrow$, $\mathbf{x}^\ast$, $\mathbf{l}^\rightarrow$,
$\mathbf{d}^{\not\rightarrow}$, $\mathbf{d}_C^{\not\rightarrow}$, and $\mathbf{l}^\twoheadrightarrow$ satisfy
\eqref{tag:C2}--\eqref{tag:C3}, \eqref{tag:C6}--\eqref{tag:C9}, \eqref{tag:O1},
\eqref{tag:G1}, \eqref{tag:N1}, and \eqref{tag:R1a}--\eqref{tag:R1b}. Then, the
following holds:
\begin{enumerate}[label=(\roman*)]
    \item Constraint \eqref{tag:AC} is satisfied if and only if $G_\mathbf{x}$
    is an acyclic directed mixed graph. \label{bul:thmDMG1}
    \item The set
    \begin{align*}
        \{\lc{i}{j}{C}:i,j\in [d],C\subseteq [d]\setminus \{i,j\}, i\neq j\}
    \end{align*}
    is a solution to \eqref{tag:P1} if and only if $\lc{i}{j}{C}$ equals the
    $m$-distance between $i$ and $j$ given $C$ in $G_\mathbf{x}$ for each triple
    $(i,j,C)$ such that $i\neq j$ and $i,j\notin C$. That is, the $m$-distances
    in $G_\mathbf{x}$ are the unique solution of \eqref{tag:P1} when
    $\mathbf{x}$  corresponds to $G_\mathbf{x}$. \label{bul:thmDMG2}
\end{enumerate}
\end{thm}

\section{Optimization and learning output}\label{sec:optim}

In this section, we show how to formulate the graph learning problem as an
integer-programming problem using the theory in Section~\ref{sec:theory}. We let
$\mathbb{G}_d$ be the set of graphs over which we wish to minimize the
objective, i.e., $\mathbb{G}_d$ is the set of DAGs, DGs, ADMGs, DMGs, or CGs on
$d$ nodes. For $i,j\in [d]$ and $C\subseteq [d] \setminus \{i,j\}$, we let
$w_{ijC}$ denote a fixed weight, we let $f_{ijC}$ denote a known transformation,
and we let $p_{ijC}$ denote a $p$-value from a test of conditional independence 
between $X_i$ and $X_j$ given
$X_C$ based on $n$ i.i.d.\ observations of the random
vector $X \in \RR^d$. We now consider the objective
\begin{align}\label{tag:GLG}
    \min_{G\in \mathbb{G}_d} \sum_{i,j \in [d]}  \sum_{C\in \mathbb{C}:
    i,j\notin C} w_{ijC} \vert \1_{ijCG} - f_{ijC}(p_{ijC}) \vert
\end{align}
where $\1_{ijCG}$ indicates whether $i$ and $j$ are $m$-separated given $C$
in $G$ in the case of DMGs or whether $i$ and $j$ are separated given $C$ in the
moral graph of $G_{\an(\{i,j\}\cup C)}$ in the case of CGs. In our
experiments, for example, we use $w_{ijC} = 1$ for all
$i,j,C$, and for all $i,j,C$ we let
$f_{ijC}(p_{ijC}) = 1$ if $p_{ijC} > \alpha$, and otherwise $f_{ijC}(p_{ijC}) =
0$, where $\alpha \in (0, 1)$ is a fixed threshold parameter.
As has been pointed out before \citep{colombo2014orderindependent},
without multiple testing corrections $\alpha$ cannot be interpreted as a
type~I error rate. Instead $\alpha$ should be seen as a tuning parameter
controlling the sparsity of the graph and can, for instance, be varied with
sample size.
GLIP can
also accommodate
objectives that are not of the form \eqref{tag:GLG} by using auxiliary variables
to linearize objectives. We discuss the choice of objective further in
Section~\ref{ssec:discussTheo}.

We now state the integer program,
\begin{align}
\begin{split}
    \min_{\mathbf{y}_{\mathbb{G}_d}\in \mathbb{Z}^N}\quad &
    \mathbf{c}_{\mathbb{G}_d}^\top \mathbf{y}_{\mathbb{G}_d}  \\
    \text{s.t.} \quad & A_{\mathbb{G}_d} \mathbf{y}_{\mathbb{G}_d} \leq
    \mathbf{b}_{\mathbb{G}_d} , \\
    & \underline{\mathbf{y}}_{\mathbb{G}_d}  \leq \mathbf{y}_{\mathbb{G}_d} \leq
    \overline{\mathbf{y}}_{\mathbb{G}_d}.
\end{split} \tag{IP} \label{tag:IPG}
\end{align}
In~\eqref{tag:IPG}, the
vector $\mathbf{y}_{\mathbb{G}_d}$ is the concatenation
\begin{align*}
    \mathbf{y}_{\mathbb{G}_d}^\top = (\mathbf{x}_{\mathbb{G}_d}^\top,
    \mathbf{l}_{\mathbb{G}_d}^\top, \mathbf{z}_{\mathbb{G}_d}^\top,
    \mathbf{w}_{\mathbb{G}_d}^\top)
\end{align*}
where $\mathbf{x}_{\mathbb{G}_d}$ is a binary vector representing
presence/absence of edges, $\mathbf{l}_{\mathbb{G}_d}$ is a vector of
minimal-length variables, $\mathbf{z}_{\mathbb{G}_d}$ is a vector of separation
variables (e.g., representing $d$-separation), and $\mathbf{w}_{\mathbb{G}_d}$
is a vector of auxiliary variables. The vector of auxiliary variables includes
the $u$-variables (for DAGs with $d$-separation these are defined in
\eqref{tag:L1a}--\eqref{tag:L4}) and variables to linearize the absolute value
in the objective.

In \eqref{tag:IPG}, the subscript ${\mathbb{G}_d}$ emphasizes that both the
coefficients (e.g.,~$\mathbf{A}_{\mathbb{G}_d}$) and the variable
$\mathbf{y}_{\mathbb{G}_d}$ will depend on both $d$ and the chosen class of
graphs. Table \ref{tab:constraints} shows which constraints to use for each
class of graphs. From these constraints, a collection of $p$-values, and choices
of weights, $w_{ijC}$, and functions $f_{ijC}$, one can construct the matrix and
the vectors needed in \eqref{tag:IPG} in a straightforward manner. 

The main result for each class of graphs is summarized in the following theorem.
The theorem implicitly uses a collection of conditioning sets, $\mathbb{C}$, in
\eqref{tag:GLG}. Therefore, the theorem applies to graph learning of equivalence
classes under a homogeneous equivalence relation (see Section
\ref{subsec:weak}), and this includes, in particular, Markov equivalence classes
and $k$-weak equivalence classes.

\begin{thm}\label{thm:optim}
Let $\mathbb{G}_d$ be the set of either DAGs, DGs, ADMGs, DMGs, or chain graphs
on $d$ nodes. Let $\hat{\mathbf{y}}$ be an optimal solution of \eqref{tag:IPG}
with the appropriate set of constraints (corresponding to
$\mathbb{G}_d$ and an appropriate notion of graphical separation) as defined in
Table \ref{tab:constraints}, and let
$\hat{\mathbf{x}}$ be the subvector corresponding to edge variables. Let
$G_{\hat{\mathbf{x}}}$ be the graph corresponding to $\hat{\mathbf{x}}$. In this
case, $G_{\hat{\mathbf{x}}}$ is a global minimum of
\eqref{tag:GLG}. 
\end{thm}

Theorem~\ref{thm:optim} shows how to find a globally optimal graph. Such a graph
will generally not be unique as Markov equivalent graphs will attain the same
objective value. However, from a learned graph, one can compute a representative
of its equivalence class using graphical theory as described in the next
section.

\begin{table}[!t]
\centering
\resizebox{\textwidth}{!}{%
\begin{tabular}{lrrrrrrrrr}
\toprule
     \bf Class & \bf Subclass & \bf Separation  & &  &\bf Constraints & & & \\
     \midrule
     \multirow{3}{*}{DMG}
     & DMG & $m$-separation & \eqref{tag:C2}--\eqref{tag:C9} & \eqref{tag:N1} &
     \eqref{tag:R1a}--\eqref{tag:R1b} &  & \eqref{tag:O1} & \eqref{tag:G1} &
     \eqref{tag:P1}     \\
     & DMG  & $m_c$-separation & \eqref{tag:C4}--\eqref{tag:C11} &  &  & &
     \eqref{tag:O1} &  &   \eqref{tag:P1c}  \\
     & ADMG & $m$-separation & \eqref{tag:C2}--\eqref{tag:C9} & \eqref{tag:N1} &
     \eqref{tag:R1a}--\eqref{tag:R1b} & \eqref{tag:AC} & \eqref{tag:O1} &
     \eqref{tag:G1} &   \eqref{tag:P1}     \\
     & ADMG & $m_c$-separation & \eqref{tag:C2}--\eqref{tag:C11} &
     \eqref{tag:N1} & & \eqref{tag:AC} & \eqref{tag:O1}  & &  \eqref{tag:P1c}
     \\
     & DG   & $d$-separation & \eqref{tag:C2}--\eqref{tag:C5} & \eqref{tag:N1} &
     \eqref{tag:R1a}--\eqref{tag:R1b} &  &  &  & \eqref{tag:M1}       \\
     & DG   & $d_c$-separation & \eqref{tag:C4}--\eqref{tag:C5} &  &  &  &  & &
     \eqref{tag:M1c}      \\
     \midrule
     \multirow{3}{*}{Chain}
     & Chain & $c$-separation   &
     \begin{tabular}{@{}c@{}}\eqref{tag:C2}--\eqref{tag:C5}, \\
     \eqref{tag:C12}--\eqref{tag:C13}\end{tabular}  & \eqref{tag:N1} &
     \eqref{tag:R1a}--\eqref{tag:R1b} &
     \begin{tabular}{@{}c@{}}\eqref{tag:CH1a} \\ \eqref{tag:CH1b}\end{tabular} &
     \eqref{tag:W1} & \eqref{tag:Z1} & \eqref{tag:Q1}  \\
     & DAG   & $d$-separation   & \eqref{tag:C2}--\eqref{tag:C5} &
     \eqref{tag:N1} & \eqref{tag:R1a}--\eqref{tag:R1b} & \eqref{tag:AC} & & &
     \eqref{tag:M1}       \\
     & DAG   & $d_c$-separation & \eqref{tag:C2}--\eqref{tag:C5} &
     \eqref{tag:N1} & & \eqref{tag:AC} & & & \eqref{tag:M1c}   \\
     \bottomrule
\end{tabular}}
\caption{%
Overview of the set of constraints needed for each combination of separation
criterion and class of graphs. The minimal-length encodings using $m_c$-
and $d_c$-separation can be found in Appendix~\ref{app:encodings}. 
}\label{tab:constraints}
\end{table}

\subsection{Representatives of equivalence classes}

The integer-programming problem~\eqref{tag:IP} has a solution. This solution
need not be unique, and we let $\mathbf{y}_\text{opt}$ denote the set of global
minimizers of~\eqref{tag:IP}. From solving the integer-programming
problem~\eqref{tag:IP}, we obtain a vector in $\mathbf{y}_\text{opt}$, and
Theorem~\ref{thm:optim} proves that we can construct a graph, $G$, which is in
the optimal equivalence class. This equivalence class
is either a Markov equivalence class or a weak equivalence class (or more
generally, a homogeneous equivalence class (Section \ref{subsec:weak})). Using
$G$, we can use graphical theory to construct a graphical representation of the
entire equivalence class, e.g.,~when learning a Markov equivalence class of
DAGs, we can compute the essential graph (see Example~\ref{ex:ess}) as a
representative of the Markov equivalence class \citep{Andersson:1997}.

In the following, we describe how to compute the relevant equivalence class
representative from the output of our graph learning algorithm in different
classes of graphs. One should note that the graphical theory does not exist for
every case that our theory covers. For example, there is no characterization of
weak equivalence of ADMGs, to our knowledge. However, our method is agnostic to
this missing graphical theory, and once such a characterization is available,
one may apply it to the output of our graph learning procedure.

\subsubsection{DAGs}

A Markov equivalence class of DAGs is often represented using an \emph{essential
graph}. Essential graphs are a subclass of chain graphs \citep{Andersson:1997}.
In short, essential graphs provide a concise representation of a Markov
equivalence class of DAGs: If $1 \rightarrow 2$ is in the essential graph, then
this edge is in every DAG in the Markov equivalence class. On the other hand,
if $1 - 2$ is in the essential graph, then there exists a DAG in the Markov
equivalence class such that $1\rightarrow 2$ and there exists a DAG in the
Markov equivalence class such that $1 \leftarrow 2$. See Example~\ref{ex:ess} below.

\begin{ex}\label{ex:ess}
DAGs $D_1$, $D_2$, and $D_3$ in Figure~\ref{fig:ess} constitute a Markov
equivalence class: They agree on all $d$-separations, and there is no other
Markov equivalent DAG. The chain graph $G$ is their \emph{essential graph} which
summarizes the Markov equivalence class: We see that $3 \rightarrow 4$ and
$2\rightarrow 4$ in the essential graph as these edges are present in all three
DAGs. We see that $1 - 2$ and $1 - 3$ in $G$ as the orientation of the edges between
nodes $1$ and $2$ and between nodes $1$ and $3$ differ between the three DAGs.

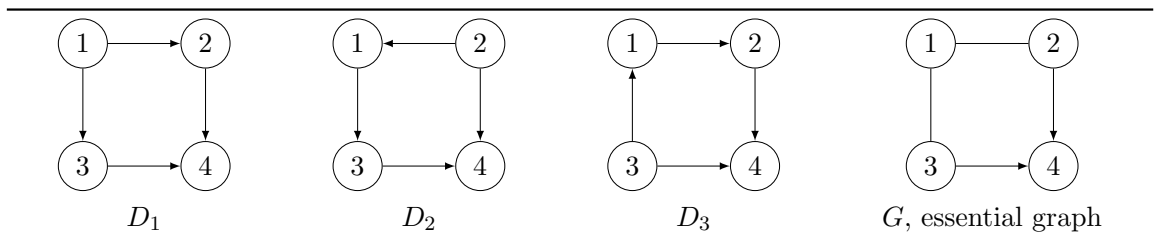
\begin{figure}[H]
\setlength{\tabcolsep}{20pt}
\resizebox{\textwidth}{!}{%
\begin{tabular}{cccc}
\toprule
\begin{tikzpicture}
\node[circle, draw] (A) {$1$};
\node[below = of A, circle, draw] (C) {$3$};
\node[right = of A, circle, draw] (B) {$2$};
\node[below=of B, circle, draw] (D) {$4$};
\draw[-latex] (A) -- (C);
\draw[-latex] (A) -- (B);
\draw[-latex] (B) -- (D);
\draw[-latex] (C) -- (D);
\end{tikzpicture}
&
\begin{tikzpicture}
\node[circle, draw] (A) {$1$};
\node[below = of A, circle, draw] (C) {$3$};
\node[right = of A, circle, draw] (B) {$2$};
\node[below=of B, circle, draw] (D) {$4$};
\draw[-latex] (A) -- (C);
\draw[-latex] (B) -- (A);
\draw[-latex] (B) -- (D);
\draw[-latex] (C) -- (D);
\end{tikzpicture}
&
\begin{tikzpicture}
\node[circle, draw] (A) {$1$};
\node[below = of A, circle, draw] (C) {$3$};
\node[right = of A, circle, draw] (B) {$2$};
\node[below=of B, circle, draw] (D) {$4$};
\draw[-latex] (C) -- (A);
\draw[-latex] (A) -- (B);
\draw[-latex] (B) -- (D);
\draw[-latex] (C) -- (D);
\end{tikzpicture}
&
\begin{tikzpicture}
\node[circle, draw] (A) {$1$};
\node[below = of A, circle, draw] (C) {$3$};
\node[right = of A, circle, draw] (B) {$2$};
\node[below=of B, circle, draw] (D) {$4$};
\draw[-] (A) -- (C);
\draw[-] (A) -- (B);
\draw[-latex] (B) -- (D);
\draw[-latex] (C) -- (D);
\end{tikzpicture}\\
$D_1$
& $D_2$
& $D_3$
& $G$, essential graph
\\
\bottomrule
\end{tabular}}
\caption{A Markov equivalence class of DAGs and their essential graph. See
Example~\ref{ex:ess}.
}\label{fig:ess}
\end{figure}

Consider now the learning problem in \eqref{tag:GLG} where $\mathbb{G}_d$ is the
set of DAGs on four nodes and $\mathbb{C}$ is the power set of $\{1,2,3,4\}$.
The DAGs $D_1$, $D_2$, and $D_3$ will be scored equivalently, i.e., if $D_1$ is
a global minimizer, so are $D_2$ and $D_3$. If the learning algorithm outputs
$D_1$, this graph is a global minimizer. From this graph, one can compute the
essential graph, $G$, to obtain a representation of the entire Markov
equivalence class.
\end{ex}

\subsubsection{ADMGs}

The independence model of an ADMG can also be represented by a maximal ancestral
graph \citep{richardson2002ancestral}, and partial ancestral graphs can be used
as representatives of Markov equivalence classes of maximal ancestral graphs
\citep{zhang2007characterization}. This means that one can compute a Markov
equivalent PAG from a learned ADMG as a unique representation of its Markov
equivalence class. An example of a PAG with an explanation of the different edge
types is given in Example~\ref{ex:oracle} in Section~\ref{sec:intro:global}.

\subsubsection{Weak DAGs}

A weak equivalence class of a DAG may be larger than its Markov equivalence
class, and a classical essential graph is therefore not a valid representation
of weak equivalence class. \citet{kocaoglu2024characterization} provides a
characterization of $k$-weakly equivalent DAGs, leading to a so-called
\emph{$k$-essential graph}, i.e., a graphical representation of the $k$-weak
equivalence class of a DAG. \citet{kocaoglu2024characterization} also defines a
learning algorithm, \emph{$k$-PC}. This algorithm is not complete, i.e.,~it will
not necessarily output the correct weak essential graph with oracle tests. Our
algorithm is complete in this sense as it outputs a DAG in the correct weak
equivalence class.

\subsubsection{Chain graphs}

\citet{frydenberg1990chain} showed that any Markov equivalence class contains a
\emph{largest chain graph}, and this graph can be used as a unique
representative of its Markov equivalence class. \citet{studeny1997recovery}
provides an algorithm to compute the largest chain graph of a Markov equivalence
class starting from a member of the Markov equivalence class.
\citet{roverato2005unified} provides an algorithm which uses the same idea, but
is conceptually more straightforward (see also \citealp{studenvr2009two}).

\section{Computational details}
\label{sec:comp}

We evaluate GLIP against other exact and approximate graph learning methods
using both simulated and available benchmark datasets in terms of DAG and ADMG
learning with both discrete and continuous variables. Numerical experiments for
chain graph learning are given in Appendix~\ref{app:empirical:chain}. Below, we
describe how the simulation studies and benchmark experiments were carried out
and list implementation details at the end.

\subsection{Timing comparisons against other exact methods}
\label{sec:comp:exact}

We compare GLIP against the Answer Set Programming (ASP) approach in
\citet{hyttinen2014constraint} in terms of optimization time only, as both
methods are exact and yield the same output for the same input $p$-values in
both weak ($k < d - 2$) and non-weak ($k = d - 2$) DAG and ADMG learning. For
that, we take the same data generating mechanism as in
\citet{hyttinen2014constraint} based on Gaussian graphical models each with a
randomly generated graph using a uniform edge probability of $1 / (d - 1)$,
where $d$ is the number of nodes in the graph, uniform edge weights between 1
and 4, and unequal error variances of $1 + 0.1 Z$, where $Z$ is standard normal.
ADMGs are generated using correlated error terms. We consider DAGs and ADMGs of
dimensions 3 to 9 and 3 to 14 for graph learning ($k = d-2$) and weak graph
learning ($k = 1$), respectively. For each iteration, we sample 10'000 data
points from the Gaussian graphical model and repeat each setting 20 times. We
apply partial correlation tests and use $\alpha = 0.001$. Both ASP and GLIP are
run with a walltime of 600~seconds. For DAG learning and ADMG learning, we
warmstart GLIP with a DAG or an ADMG computed from the output of PC and FCI,
respectively. The time this warmstart takes is negligible compared to the
optimization time, especially for medium- to large-sized graphs. 

\subsection{Performance comparisons against approximate methods}
\label{sec:comp:sim}

\subsubsection{Simulated data}

For our empirical evaluation of the proposed graph learning framework, we use
the following procedure: We first generate a random graph (DAG or ADMG) with
randomly generated edge weights. We then generate data according to the implied
linear structural causal model. We then apply GLIP and competing methods on the
dataset and record the performance. Each step is outlined in more detail below.

For DAGs, we generate random Erd\H{o}s-R\'enyi graphs with $d \in \{3, \ldots,
9\}$ nodes for non-weak and $d \in \{3, \ldots, 12\}$  nodes for weak learning
and expected node degree $e = 2$, and we ensure that they are acyclic. For
ADMGs, we generate a DAG with $d + 3$ nodes and apply latent projection over $3$
nodes chosen at random. We  generate edge weights uniformly between~0 and~1.

Given a DAG and edge weights, we sample from a linear additive noise model with
jointly independent Gaussian noise with unequal error variances. For ADMGs, the
data is generated in the same way but for the larger underlying DAG that has
been marginalized to the given ADMG. In turn, only data for the observed nodes
is used in the subsequent graph learning algorithms.

\subsubsection{Benchmark datasets}\label{sec:comp:bench}

Besides simulated data, we assess the performance of GLIP on several synthetic
benchmark datasets for which the ground-truth graph is available. We use ALARM
($n = 10'000$, $d = 37$, 46 edges), ASIA ($n = 10'000$, $d = 8$, 8 edges), CHILD
($n = 10'000$, $d = 20$, 25 edges, synthetic), HEPAR2 ($n = 10'000$, $d = 70$,
123 edges), and SACHS ($n = 10'000$, $d = 11$, 17 edges) from the
\pkg{py-why} repository (\url{https://github.com/py-why/causal-learn}) based on
the \pkg{bnlearn} repository (\url{https://www.bnlearn.com/bnrepository/}), in
which all variables are discrete. We perform an asymptotic $\chi^2$-test as a
conditional independence test since all variables are discrete.

\subsubsection{Competitors}

For both simulation and benchmark datasets, we compare GLIP (with warmstarts as
described in Section~\ref{sec:comp:exact} and a walltime of 300~seconds) to two
constraint-based algorithms, PC and FCI \citep{spirtes2000causation}, and
NOTEARS \citep{zheng2018notears} as a popular score-based method. We use PC as a
competitor in DAG learning and FCI as a competitor in ADMG learning,
exclusively, as PC assumes causal sufficiency. We run NOTEARS with linear models
and regularization strength $\lambda = 0.02$ for all experiments. As a naive
baseline for causal discovery in linear Gaussian SCMs, we also compare against
$R^2$-sort-and-regress (R2SORT) \citep{reisach2021varsort,reisach2023r2sort}.
For all constraint-based methods, since the data are generated according to a
linear Gaussian SCM, we use partial correlation tests to test conditional
independence. All tests use $\alpha = 0.001$ at sample sizes $n \in \{400,
10'000\}$. Since PC (and theoretically also FCI, although this was rarely
observed in our experiments) can return outputs with cycles, we additional
compare against an ``honest'' version of PC (HPC) in which we compute an
arbitrary DAG that is consistent with the PC output and, in turn, compute the
essential graph corresponding to this DAG.

If GLIP does not run to optimality, we return the current best graph.
Given the short walltime in our experiments, this leads to a conservative
evaluation of GLIP's performance.

\subsubsection{Evaluation metrics}

For learning DAGs and ADMGs, we compute the corresponding essential graph and
PAG, respectively, for all learned outputs. In the following, let $E$ and $O$
denote $d\times d$ adjacency matrices 
corresponding to the computed
representative of the Markov equivalence class of the learned graph and the
ground-truth graph, respectively. 
For PAGs, this adjacency matrix is not necessarily binary:
The relationship between nodes \(i\) and \(j\) is defined by the pair
\((E_{ji},E_{ij})\). For example, a directed edge \(i\rightarrow j\) is
encoded as \(E_{ij}=2\) and \(E_{ji}=3\), while a bidirected edge
\(i\leftrightarrow j\) is encoded as \(E_{ij}=2\) and \(E_{ji}=2\).
The graph output is then evaluated using the
structural Hamming distance (SHD), 
\( 
\operatorname{SHD}(E, O) = \sum_{i=1}^d\sum_{j=1}^d \1(E_{ij} \neq O_{ij}), 
\) 
$k$-separation distance ($k$-SEP), 
\[ 
k\text{-}\operatorname{SEP}(E, O) = \sum_{i,j \in [d] }\sum_{C \subseteq [d] : i, j \notin C, \lvert C \rvert \leq k}
\1(\1_{\mconD{i}{j}{C}{E}} \neq \1_{\mconD{i}{j}{C}{O}}),
\]
and a head- and tail-specific $F_1$-score, $F_1(E, O) = \tfrac{2(1 -
\operatorname{FDR}(E, O))(1 - \operatorname{FNR}(E, O))}{2 -
\operatorname{FDR}(E, O) - \operatorname{FNR}(E, O)}$, where FDR denotes the
false discovery rate,
\[ 
\operatorname{FDR}(E, O) = \frac{1}{d}\sum_{i=1}^d \frac{ \sum_{j=1}^d
\1_{E_{ij} = 1 \land
O_{ij} = 0} }{ \max\{\sum_{j=1}^d \1_{E_{ij} = 1 \land O_{ij} = 1} + \1_{E_{ij} =
1 \land O_{ij} = 0}, 1\} }, 
\] 
and FNR the false negative rate,
\[ 
\operatorname{FNR}(E, O) = \frac{1}{d} \sum_{i=1}^d \frac{ \sum_{j=1}^d
\1_{E_{ij} = 0 \land O_{ij} = 1} }{ \max\{\sum_{j=1}^d \1_{E_{ij} = 1 \land O_{ij}
= 1} + \1_{E_{ij} = 0 \land O_{ij} = 1}, 1\} }. 
\] 
FDR and FNR are computed for both arrow heads (as above) and tails (as
above, but using the transposed $E$ and $O$) and we report $1 - F_1$, so that
for all metrics smaller values indicate better performance. For graph learning
with $k = d- 2$, we denote the separation distance by SEP.

\subsection{Implementation} 

The proposed graph learning framework is implemented in the \proglang{R}~package
\pkg{glip} which is openly available at \url{https://github.com/LucasKook/glip}.
The repository also contains code to reproduce all results in this manuscript.
To solve mixed-integer programs, we rely on \pkg{gurobi} \citep{pkg:gurobi}.
Partial correlation tests are performed using \texttt{gaussCItest()} from
\pkg{pcalg} \citep{pkg:pcalg}. For discrete variables, the $\chi^2$-test
implemented in \pkg{bnlearn} \citep{pkg:bnlearn} was used. PC and FCI were run
using \pkg{pcalg}. NOTEARS was run in \proglang{R} \citep{pkg:base} using
\pkg{reticulate} \citep{pkg:reticulate} and the \pkg{dagma} \citep{pkg:dagma}
\proglang{Python} \citep{python} library. R2SORT was run in \proglang{R} using
\pkg{reticulate} and the \pkg{CausalDisco} \citep{reisach2023r2sort}
\proglang{Python} library. The \pkg{pcalg} package provides the function
\texttt{dag2essgraph()} for computing the essential graph from a DAG. The
\pkg{glip}~package provides functions for computing representatives of
equivalence classes of ADMGs (\texttt{compute\_pag()}) and CGs
(\texttt{compute\_largest\_cg()}).

\section{Empirical results}\label{sec:results}

\subsection{Timing comparisons against other exact methods}
\label{sec:results:asp}

\begin{figure}[t!]
\begin{subfigure}{\textwidth}
\centering
\includegraphics[width=0.75\linewidth]{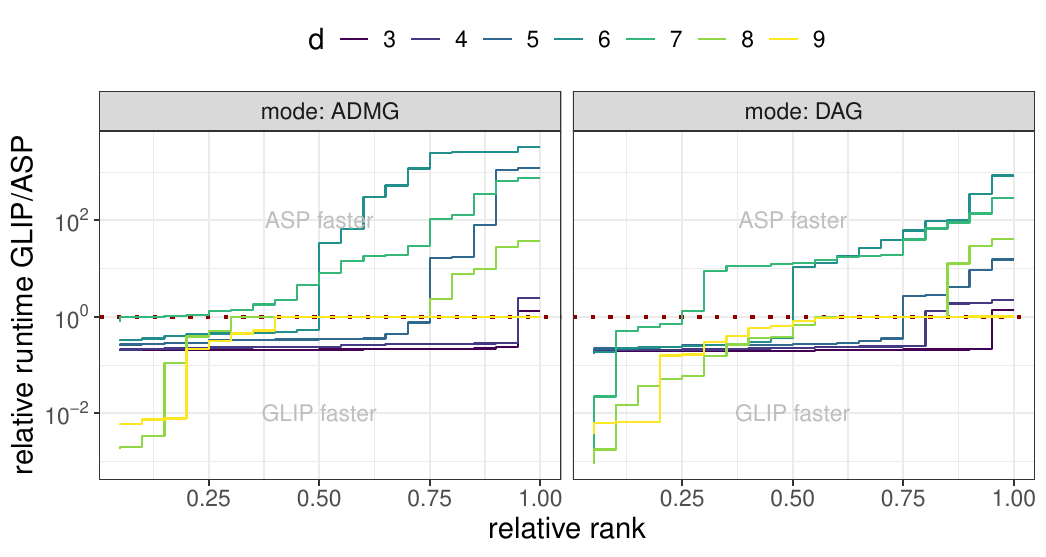}
\subcaption[]{%
Graph learning with $k = d - 2$.
}
\end{subfigure}
\begin{subfigure}{\textwidth}
\centering
\includegraphics[width=0.75\linewidth]{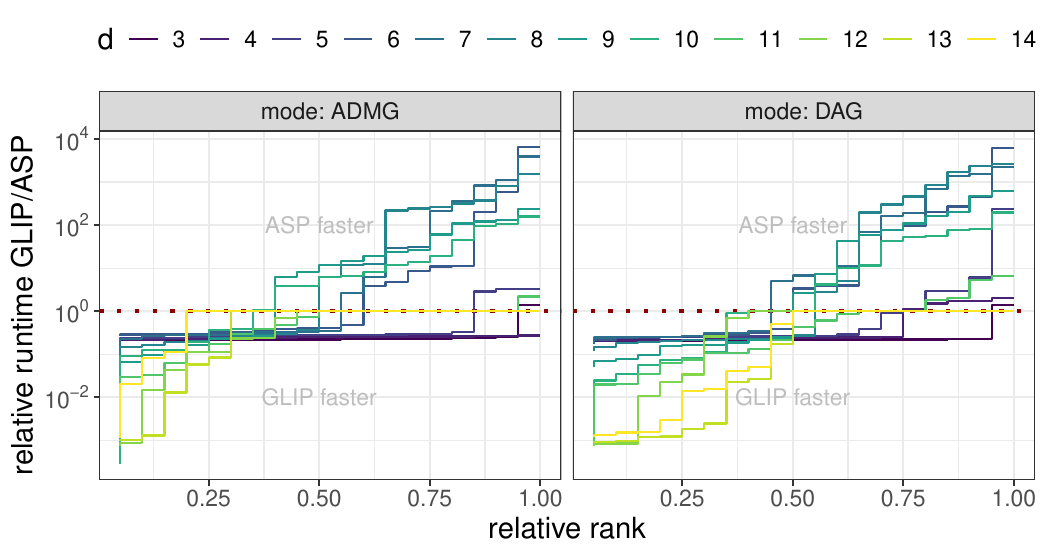}
\subcaption[]{%
Weak graph learning with $k = 1$.
}
\end{subfigure}
\caption{%
    Relative runtime comparison of GLIP and ASP. For the comparison, graphs of
    different sizes were generated according to the details in
    Section~\ref{sec:comp:exact}. To more clearly distinguish the timings for
    larger graphs, Figure~\ref{fig:asp:larged} in
    Appendix~\ref{app:empirical:asp} shows the same results but restricted to $d
    \geq 6$.
}\label{fig:asp}
\end{figure}

We first compare the relative runtime for GLIP and ASP using the empirical
quantile function of the ratios of runtimes of GLIP and ASP with different graph
sizes indicated by different colors (Figure~\ref{fig:asp}). That is, if the
quantile function falls below $1$ at a given relative rank $r \in (0, 1)$, GLIP
is faster than ASP in $r \times 100\%$ of simulation runs. Further, since the
experiment is run with a walltime at 600~seconds, if the curve flattens out at
1, this indicates both methods hit the walltime and did not necessarily arrive
at an optimal solution. Since the search space in graph learning is extremely
large, we do not expect one method to uniformly outperform the other, especially
for medium to large size graphs. 

Across both DAG and ADMG learning with $k = d - 2$ and $k = 1$, GLIP is about an
order of magnitude faster for small graphs ($d \in \{3, 4\}$). For larger
graphs, GLIP is faster in about 50\% of all instances with speed-ups of up to 3
orders of magnitude, while in the remaining 50\% of instances ASP achieves
speed-ups of up to four orders of magnitude. Therefore, neither method uniformly
dominates the other in terms of runtime and, in practice, it may be worthwhile
to run both methods in parallel. For the largest graphs ($d = 9$ for $k = d - 2$
and $d\in \{12,13,14\}$ for weak learning with $k = 1$), GLIP is able to solve
20--40\% of all instances within the 600~second walltime and all of those
instances are solved faster than ASP, whereas both methods hit walltime for the
remaining instances. For $d \in \{13, 14\}$, ASP did not solve any of the
20 instances within walltime. The absolute runtimes are reported in
Appendix~\ref{app:empirical:asp}. This comparison again highlights the hardness
of (weak) graph learning and suggests that, in practice, it may be beneficial to
run both GLIP and ASP.

\subsection{Performance comparisons against approximate methods}

\subsubsection{Simulated data}\label{sec:results:sim}

\begin{figure}[t!]
\begin{subfigure}{\textwidth}
\centering
\includegraphics[width=0.99\linewidth]{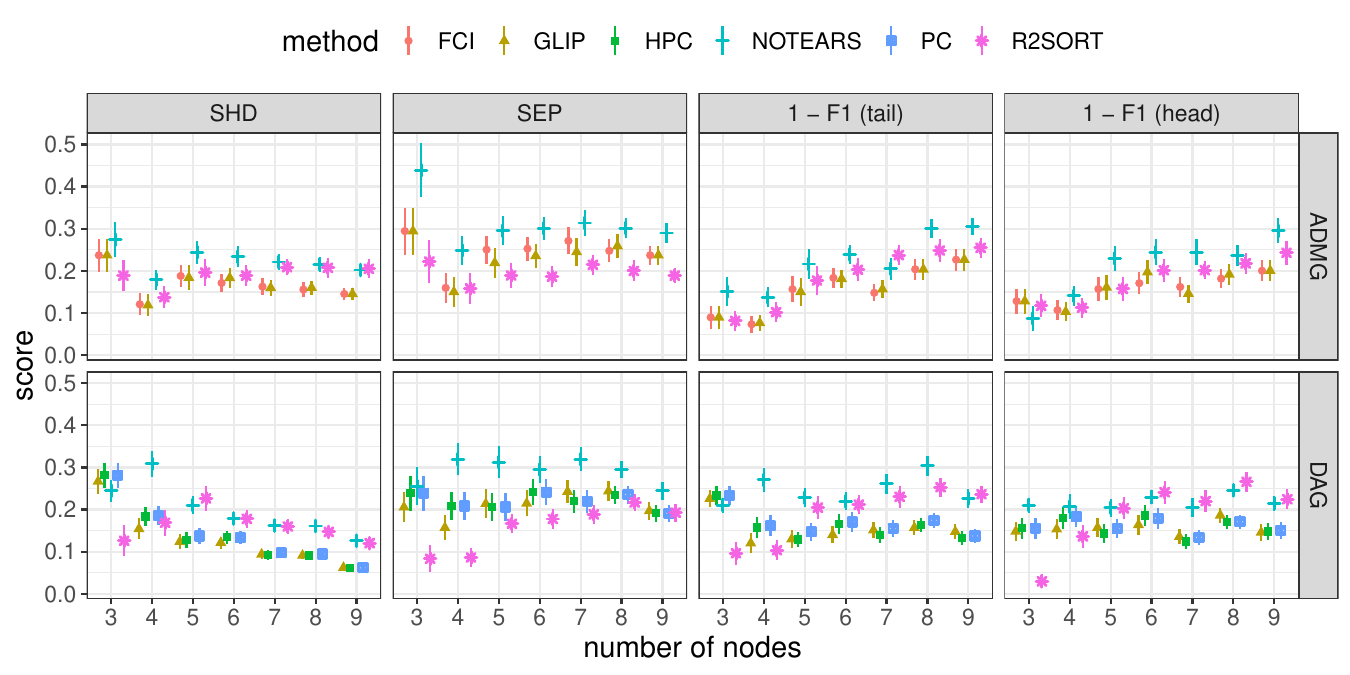}
\subcaption[]{Results for $n = 400$.}
\end{subfigure}
\begin{subfigure}{\textwidth}
\centering
\includegraphics[width=0.99\linewidth, trim=0 0 0 1cm, clip]{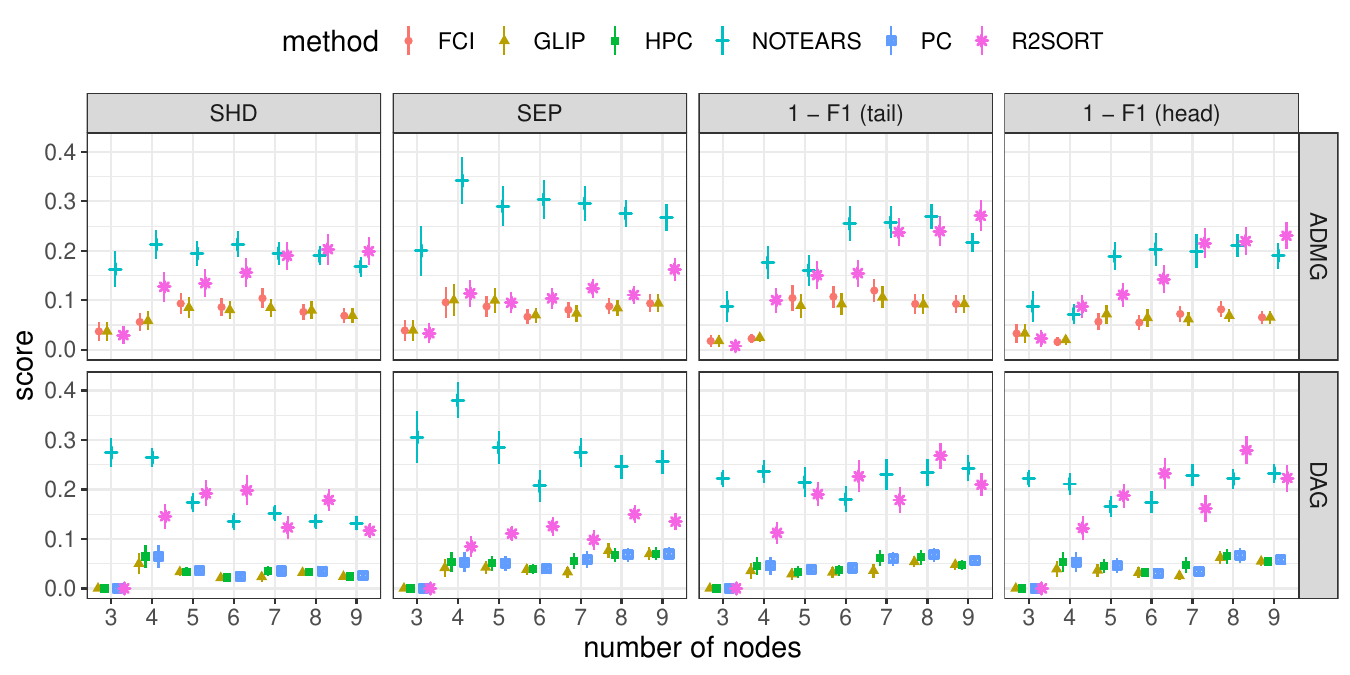}
\subcaption[]{Results for $n = 10'000$.}
\end{subfigure}
\caption{%
Performance comparison against approximate methods for graph learning with $k =
d - 2$.
}\label{fig:sim:full}
\end{figure}

\begin{figure}[t!]
\centering
\includegraphics[width=0.7\linewidth]{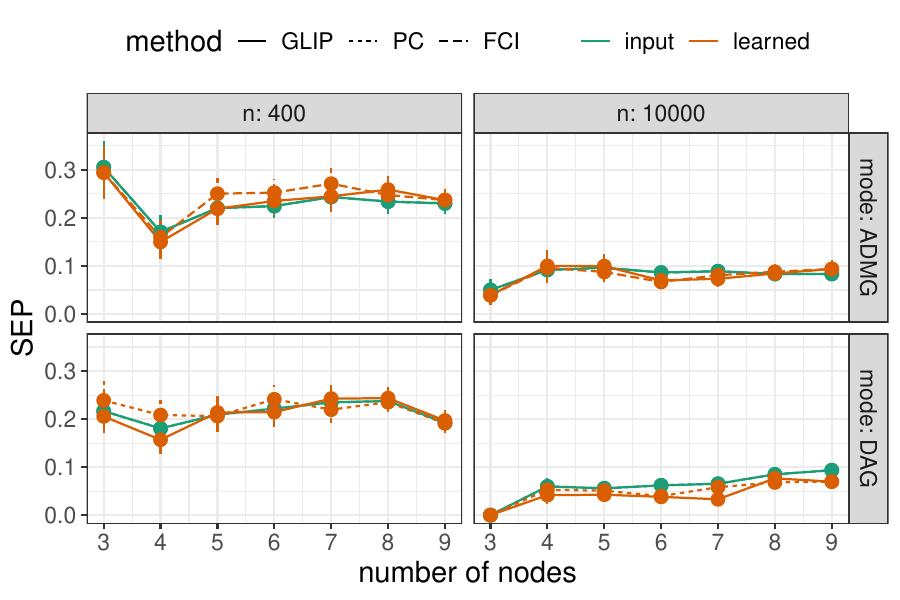}
\caption{%
Average separation distance of the thresholded input $p$-values and the output
of the graph learning methods (GLIP, PCI, FCI) against the oracle essential
graph (for DAGs) or PAG (for ADMGs).
}\label{fig:sim:full:sep}
\end{figure}

We now turn to the results for DAG and ADMG learning with $k = d - 2$ and
compare GLIP to approximate graph learning methods according to the simulation
setup in Section~\ref{sec:comp:sim}. Figure~\ref{fig:sim:full} shows that,
across both sample sizes and all graph sizes, GLIP performs at least on par with
FCI for ADMG learning and PC for DAG learning in terms of SHD, SEP, tail $F_1$
and head $F_1$ score. For $n = 400$, R2SORT achieves the best SEP scores for
graphs of size three to eight. This, however, no longer is the case for the
larger sample size, as the conditional independence tests are more powerful and
yield more consistent $p$-values (see Figure~\ref{fig:sim:full:sep}). 

Table~\ref{tab:sim:full:completion} shows the fraction of simulation runs in
which GLIP could prove optimality of the solution within the walltime limit of
300~seconds. For graphs with more than 5 nodes, this fraction drops from about
100\% to 50\% and reaches approximately 10\% for $d = 9$.

\begin{table}[!ht]
\centering
\begin{tabular}{lrrrrrrrr}
\toprule
\multicolumn{2}{c}{ } & \multicolumn{7}{c}{$d$} \\
\cmidrule(l{3pt}r{3pt}){3-9}
$n$ & mode & 3 & 4 & 5 & 6 & 7 & 8 & 9\\
\midrule
400 & ADMG & 1.000 & 1.000 & 0.933 & 0.600 & 0.533 & 0.267 & 0.067\\
400 & DAG & 1.000 & 1.000 & 1.000 & 0.567 & 0.267 & 0.367 & 0.167\\
10000 & ADMG & 1.000 & 1.000 & 0.933 & 0.633 & 0.467 & 0.267 & 0.133\\
10000 & DAG & 1.000 & 1.000 & 1.000 & 0.767 & 0.400 & 0.267 & 0.133\\
\bottomrule
\end{tabular}
\caption{Fraction of the 30 simulation iterations in which GLIP confirmed
optimality of the solution for DAG and ADMG learning in the simulation from
Section~\ref{sec:comp:sim}.}\label{tab:sim:full:completion}
\end{table}

Figure~\ref{fig:sim:full:sep} additionally shows the average separation distance
between the input $p$-values and the separations implied by the ground-truth
graph (green line), as well as the separation distance between the output of
GLIP (solid), PC (dotted), or FCI (dashed) and the separations implied by the
ground-truth graph. For the larger sample size, the learned output is closer to
the ground-truth graph than the input $p$-values, hinting at an error-correcting
property of graph learning which has been observed previously in the literature
\citep{hyttinen2014constraint,eberhardt2025discovering}. We discuss this further
in Section~\ref{sec:discussion}.

Conclusions similar to the above hold for weak learning with $k = 1$ and $k =
2$, for which all results can be found in Appendix~\ref{app:empirical:sim}.

\subsubsection{Benchmark data}\label{sec:results:bench}

Table~\ref{tab:datasets:d6} contains the results for ADMG learning with discrete
features on the benchmark datasets described in Section~\ref{sec:comp:bench}.
The ground-truth DAGs were marginalized to six nodes chosen at random, resulting
in a ground-truth ADMG, from which a PAG can be computed, and against which we
evaluate the output of the graph learning algorithms. GLIP with $k = 4$ performs
best in terms of SHD for three out of the five datasets and GLIP with $k = 1$
performs best in terms of SEP for all datasets. In Table~\ref{tab:datasets:d8}
in Appendix~\ref{app:empirical:bench}, we report the results for graphs of size
$d = 8$, in which GLIP for weak ADMG learning with $k = 1$ performs best in
terms of SHD for the majority of datasets. 

\begin{table}[t!]
\centering
\resizebox{\textwidth}{!}{\input{tables/d6}}
\caption{%
Performance of GLIP and competitors on benchmark datasets based on the PAG
output of each method at $\alpha = 0.001$. The
ground-truth DAG was marginalized to an ADMG over 6 nodes chosen uniformly at
random and only the data corresponding to those 6 variables were used for graph
learning. The analysis was repeated for $\min\{\binom{d}{6}, 50\}$ distinct
subsets; standard deviations are reported in parentheses. The best method per
dataset and metric is highlighted in bold. SHD: absolute structural Hamming
distance; SEP: separation distance. $F_1$-scores are averaged over head,
tail and circle edge marks.
}\label{tab:datasets:d6}
\end{table}

\section{Discussion}\label{sec:discussion}

In this paper we propose GLIP, a framework for exact graph learning based on
scoring graphs using $p$-values from (nonparametric) conditional independence
tests. GLIP supports graph learning and weak graph learning for directed mixed
graphs and chain graphs and for subclasses thereof. GLIP relies on solving a
mixed-integer program using an efficient encoding of graphical constraints based
on minimal lengths, resulting in linear growth of the number of variables in the
integer program as a function of the number of input $p$-values (as argued in
Section~\ref{sec:ip}). Empirically, we show that the minimal-length encodings
yield state-of-the-art performance in terms of graph size, computation time, and
quality of the learned graph output. We first discuss theoretical aspects of
GLIP, and then continue with a discussion of how GLIP can be used in
applications.

\subsection{Theoretical aspects}\label{ssec:discussTheo}

As mentioned in Section~\ref{sec:ip}, the idea of minimal-length encodings of
graphical separations can be extended to other types of graphs and separations,
such as $\delta$-separation \citep{Didelez:2008}, $\mu$-separation
\citep{Mogensen:2020}, $\sigma$-separation \citep{forre2018constraint}, and
E-separation \citep{manten2025asymmetric} by following the steps outlined in
Section~\ref{ssec:DGs}.

The objective function we target is based on a weighted sum of disagreements
between graphical separations and thresholded $p$-values of conditional
independence tests. That is, our objective is measuring distance between the
data and the graph in terms of the independence model that it encodes. Our
objective function can also be motivated as the implicit objective that
classical constraint-based algorithms such as PC and FCI are greedily
optimizing: Given an independence oracle, GLIP and PC/FCI would output the same
graph. The weights that go into the objective can be chosen in different ways,
and \citet{hyttinen2014constraint} discuss weighting schemes in the context of
ASP-based graph learning. In the context of chain graph learning,
\cite{sonntag2015learning} discuss different learning objectives, including an
objective which penalizes the cardinality of the edge set and misrepresented
conditional dependencies. To the best of our knowledge, theoretical work on the
optimality of graph learning objectives does not exist thus far, and it poses an
interesting direction for future research. Importantly, GLIP allows the choice
of any objective function that can written in a form that is linear in the
variables of the integer program, and auxiliary variables can be added to the
integer program to linearize an objective.

Using simulations for which the ground-truth graph is known, it has been
empirically demonstrated several times
\citep{hyttinen2014constraint,eberhardt2025discovering} that there are typically
fewer discrepancies between the graphical separations in the learned versus the
ground-truth graph than between the thresholded input $p$-values and the
ground-truth graph. This can be seen especially clearly in our experiments with
chain graphs (Figure~\ref{fig:chain:sep} in Appendix~\ref{app:empirical:chain}).
The input $p$-values are derived from independently conducted conditional
independence tests on finitely many samples and, therefore, they may not
correspond to a valid independence model. Graph learning methods such as GLIP
and ASP, on the other hand, are guaranteed to output a graph from the specified
class, which, in turn, will always correspond to a valid independence model,
which can be understood as a form of implicit regularization and may explain the
observed discrepancy. 

\subsection{Practical aspects}

In our experiments, we use a fast method to warmstart GLIP which then improves
upon this initial solution given a user-specified time budget. This becomes
particularly relevant for larger graphs as the search space of graphs grows
super-exponentially in the number of nodes. In our empirical results, we have
seen that this approach allows us to learn graphs of sizes where other exact
methods based on ASP \citep{hyttinen2014constraint} do not return an exact
solution within walltime. In practice, and especially for medium sized graphs,
it is still advisable to run both GLIP and ASP simultaneously, as ASP solved
around 50\% of instances faster than GLIP.

We have demonstrated empirically that GLIP can solve graph learning problems
with up to 10 nodes when considering full learning and 14 nodes when considering
weak learning with conditioning sets of up to size 1 exactly within a 600~second
walltime. It is also possible to encode large graphs and improve upon a
warmstart solution, even if an exact solution is too costly to compute. For
instance, for weak learning with $k = 1$, it is possible to encode DAGs with 20
nodes in 30~minutes on a standard laptop (intel core i5 and 16~gigabytes of
RAM). This encoding only has to be done once, as \pkg{glip} offers
infrastructure for caching and loading constraints.

Our experiments and comparisons apply a fairly restrictive walltime. In many
real applications, additional computational effort can be applied to reach an
exact solution of the graph learning problem. If the computational budget does
not allow this, GLIP may still offer an improvement over the warmstart.


\vskip 0.2in
\bibliography{references}

\clearpage

\appendix

\section*{Exact Graph Learning via Integer Programming: Supplementary Material}
\begin{center}
Lucas Kook\textsuperscript{1}, S{\o}ren Wengel Mogensen\textsuperscript{2}

{
\small
\textsuperscript{1}Vienna University of Economics and Business, Vienna, Austria\\
\textsuperscript{2}Copenhagen Business School, Copenhagen, Denmark
}
\end{center}

The appendix is structured as follows: Appendix~\ref{ssec:CGs} presents an
encoding for chain graph learning. Appendix~\ref{app:glipdemo} introduces the
\pkg{glip} package and gives a usage example. Appendix~\ref{app:encodings} gives
alternative encodings based on $d_c$- and $m_c$-separation.
Appendix~\ref{app:empirical} presents additional empirical results for the
comparison against other exact methods, approximate methods, and results for
chain graph learning. Finally, Appendix~\ref{app:proofs} contains auxiliary
theoretical results, proofs thereof, and proofs of all theoretical results in
the main text.

\section{Chain graphs}\label{ssec:CGs}

There are several interpretations of chain graphs, associated with different
Markov properties. We use the LWF-interpretation (Lauritzen, Wermuth,
Frydenberg) in this paper
\citep{lauritzen1984mixed,lauritzen1989graphical,frydenberg1990chain}. A chain
graph (Definition \ref{def:cg}) without undirected edges is a DAG, and chain
graphs are therefore a generalization of DAGs.

For ease of presentation, we define chain graphs as a subclass of \emph{hybrid
graphs} (Definition \ref{def:hybrid}). We represent a hybrid graph on the node
set $[d]$ using a single adjacency matrix corresponding to variables
$\mathbf{x}^\rightarrow = \{\xd{i}{j}\}_{i\neq j}$. A hybrid graph is, by
definition, a \emph{simple} graph, i.e.,~there is at most one edge between any
pair of distinct nodes. When $\xd{i}{j} = \xd{j}{i} = 1$, this represents an
undirected edge between $i$ and $j$. When $\xd{i}{j} = 1$ and $\xd{j}{i} = 0$,
the edge $i\rightarrow j$ is in the graph. Finally, when $\xd{i}{j} = \xd{j}{i}
= 0$, there is no edge between $i$ and $j$. We use $i - j$ as shorthand for
$\xd{i}{j} = \xd{j}{i} = 1$ and $i\rightarrow j$ as shorthand for $\xd{i}{j} =
1, \xd{j}{i} = 0$. The \emph{chain components} of a hybrid graph are the
connected components of the graph after removing all directed edges. We let
$\mathcal{T}(G)$ denote the set of chain components of $G$ (we think of each
$\tau \in \mathcal{T}(G)$ as a set of nodes, not as an undirected graph). The
\emph{boundary} of a set $A \subseteq  [d]$ is the set $\bd(A) = \{i \in [d]
\setminus A: \exists j \in A : i \rightarrow j \}$. 

We say that a graph, $G=([d],E)$, is \emph{undirected} if all edges in $E$ are
undirected. The \emph{moral graph} of a hybrid graph $G = ([d],E)$, denoted by
$G^m$, is the undirected graph on nodes $[d]$ where $i,j\in [d]$, $i\neq j$, are
adjacent if they are adjacent in $G$ or if there exists a chain component,
$\tau\in\mathcal{T}(G)$, such that $i,j \in \bd(\tau)$. When $A \subseteq [d]$
and $G =([d], E)$, we let $G_A$ denote the \emph{induced subgraph of $G$ on
$A$}, i.e., $G_A = (A, E_A)$ where $E_A$ is the subset of edges in $E$ for which
both endpoints are in $A$.

We will use moral graphs to state the global Markov property in chain graphs
(Definition~\ref{def:globalCG}), and for this purpose we also define a notion of
separation in an undirected graph. Let $i,j\in [d], C\subseteq [d]\setminus
\{i,j\}$. In an undirected graph, $G =([d],E)$, we say that a path, $\pi$,
between $i$ and $j$ is \emph{connecting given $C$} if no node on $\pi$ is in
$C$. We say that $i$ and $j$ are \emph{separated given $C$} in the undirected
graph, $G$, if there are no connecting paths between $i$ and $j$ given $C$.  For
disjoint node sets $A$, $B$, and $C$, we will say that $A$ and $B$ are separated
by $C$ in a moral graph if for each $i\in A$ and $j\in B$, nodes $i$ and $j$ are
separated given $C$. We write $\sepD{A}{B}{C}{G}$ if $A$ and $B$ are separated
by $C$ in the undirected graph $G$.

\begin{defn}[Global Markov property, chain graphs]\label{def:globalCG}
Let $X = (X_1,\ldots,X_d)$ be a random vector, and let $G = ([d], E)$ be
a chain graph. We say that the distribution of $X$ satisfies \emph{the
global Markov property} with respect to $G$ if for all disjoint sets
$A,B,C \subseteq [d]$,
\begin{align*}
    \sepD{A}{B}{C}{(G_{\an(A\cup B\cup C)})^m} \implies
    \condIndep{X_A}{X_B}{X_C}.
\end{align*}
\end{defn}

In the context of DMGs, we use $m$-separation (or $m_c$-separation) to encode
the conditional independence statements implied by a graph. These criteria are
\emph{walk-based} in the sense that they characterize separation using
existence/non-existence of paths (walks) of a certain type, and when we restrict
$m$-separation ($m_c$-separation) to DGs, we use the terms $d$-separation
($d_c$-separation). In chain graphs, \emph{$c$-separation} is an analogous
walk-based criterion, extending $d$-separation to the class of chain graphs
\citep[see][for a definition of $c$-separation]{studeny1998chain}. It is
well-known that $d$-separation in DAGs is equivalent to the so-called
\emph{moral graph criterion} \citep{lauritzen1990independence}. This equivalence
generalizes to chain graphs as $c$-separation is equivalent to the moral graph
criterion in Definition \ref{def:globalCG}: For disjoint node sets $A$, $B$, and
$C$ and a chain graph $G$, there is $c$-separation between $A$ and $B$ given $C$
in $G$ if and only if $A$ and $B$ are not separated given $C$ in the moral graph of
$G_{\an(A\cup B\cup C)}$. This means that $c$-separation can form the basis of a
minimal-length encoding, however, we use an encoding based directly on the moral
graph as will be evident from the constraints.

\subsection{Chain graph constraint}

We start from a constraint that ensures that a hybrid graph is a chain graph,
i.e.,~that it has no partially directed cycles.
\begin{align}
    & & \forall i \forall j: i\neq j: & &
   0 \leq (\ld{i}{j} -  \ld{j}{i}) + (d-1)\cdot \di{i}{j}  + (d-1)\cdot
   \di{j}{i}  \tag{CHa}\label{tag:CH1a} \\
     & & & &
   0 \leq (\ld{j}{i} - \ld{i}{j}) + (d-1)\cdot \di{i}{j}  + (d-1)\cdot \di{j}{i}
   \tag{CHb}\label{tag:CH1b}
\end{align}
The following lemmas show that the above constraints are necessary and
sufficient for a hybrid graph to be a chain graph.

\begin{lem}\label{lem:charDirCycle}
    Let $G = ([d], E)$ be a hybrid graph. For each $i,j\in [d]$, let $\ld{i}{j}$
    denote the anterior distance from $i$ to $j$ in $G$, and let $\ld{j}{i}$
    denote the anterior distance from $j$ to $i$ in $G$. There is a partially
    directed cycle in $G$ if and
    only if there exist nodes $i$ and $j$ such that $\ld{i}{j} \neq
    l_{ji}^\rightarrow$ and $\ld{i}{j}, l_{ji}^\rightarrow \neq d$. 
\end{lem}

\begin{lem}\label{lem:isCG}
    Let $G_\mathbf{x}$ be a hybrid graph. Assume that \eqref{tag:C2}--\eqref{tag:C3}
    and \eqref{tag:N1} are satisfied. Constraints~\eqref{tag:CH1a}
    and~\eqref{tag:CH1b} are satisfied if and only if $G_\mathbf{x}$ is a chain
    graph.
\end{lem}

An \emph{undirected path} in a hybrid graph is a path that consists of
undirected edges only. The \emph{undirected distance
between $i$ and $j$} in a hybrid graph, $G$, is the length of a shortest
undirected path between $i$ and $j$ if such a path exists, and otherwise it is
equal to $d$. The variables $\lu{i}{j}$ will represent the  undirected
distance between $i$ and $j$ using the below constraints, and we let
$\mathbf{l}^-$ denote the set $\{\lu{i}{j}\}_{i\neq j}$.
\begin{align*}
    \forall i \forall j: i < j :  & \\ & \lu{i}{j} \leq u_{ij}^{\text{U1}} \coloneqq 1 -
    (d-1)(\xd{i}{j} - 1) - (d-1)(\xd{j}{i} - 1) \tag{U1}\label{tag:U1} \\
     \forall i \forall j: i< j & \forall k: k\neq i,j:  \\ & \lu{i}{j} \leq
    u_{ijk}^{\text{U2}} \coloneqq 1 + \lu{i}{k} - (d-2)(\xd{j}{k}- 1)  -
    (d-2)(\xd{k}{j}- 1) \tag{U2}\label{tag:U2} \\
     \forall i \forall j: i< j:  & \\ &\lu{i}{j}  = \min_k(u_{ij}^{\text{U1}},
    u_{ijk}^{\text{U2}}, d)  \tag{W1}\label{tag:W1}
\end{align*}
The variables $\zu{i}{j}$ will indicate whether $i$ and $j$ are in the same chain
component using the below constraints, and we denote $\{\zu{i}{j}\}_{i\neq j}$
by $\mathbf{z}^-$.
\begin{align}
& \forall i , j: i< j: &  \zu{i}{j} &\leq d - \lu{i}{j} \tag{C12}\label{tag:C12} \\
& \forall i , j: i< j: &  d - \lu{i}{j} &\leq (d-1) \zu{i}{j} \tag{C13}\label{tag:C13}
\end{align}

\begin{lem}\label{lem:unDist}
    Let $G_\mathbf{x}$ be a fixed hybrid graph which corresponds to
    $\mathbf{x}^\rightarrow$. Constraint \eqref{tag:W1} is satisfied
    if and only if $\lu{i}{j}$ is the undirected distance between $i$
    and $j$ in $G_\mathbf{x}$ for all $i,j \in [d]$.
\end{lem}

The next corollary follows directly from Lemma \ref{lem:unDist}.

\begin{cor}\label{cor:chaincompindic}
For each $i,j\in [d]$, $i\neq j$, we let $\lu{i}{j}$ denote the undirected
distance between $i$ and $j$ in $G_\mathbf{x}$. Let $\{\zu{i}{j}\}$ be such that
$\zu{i}{j} = 1$ if and only if there exists an undirected path between $i$ and
$j$ in $G$. The collection $\{\zu{i}{j} \}$ is the unique solution to
Constraints \eqref{tag:C12} and \eqref{tag:C13} when $\{\lu{i}{j}\}$ are the
undirected distances of $G_\mathbf{x}$.
\end{cor}

A \emph{slide from $i$ to $j$} is a path on nodes $i = i_0, i_1,\ldots i_n = j$,
$n \geq 1$ such that $i_0 \rightarrow i_1$ and such that either $n=1$ or for all
$l = 1,\ldots, n-1$, we have $i_l - i_{l+1}$ \citep{studeny1998chain}. The path
$i \rightarrow j$, $i\neq j$, is a slide. In the context of hybrid
graphs, we use the variable $\xs{i}{j}$ to indicate whether there is a slide
from $i$ to $j$ using the following set of constraints. Therefore, the meaning
of the variable $\xs{i}{j}$ in chain graphs is different from its meaning in
directed mixed graphs. 
\begin{align}
    & \forall i \forall j: i \neq j:  & & \xs{i}{j} \geq u_{ij}^{\text{Y1}} \coloneqq
    \xd{i}{j} - \xd{j}{i}, \tag{Y1}\label{tag:Y1} \\
    & \forall i \forall j: i\neq j \forall k: i,j,k \text{ distinct }: & &
    \xs{i}{j} \geq u_{ijk}^{\text{Y2}} \coloneqq \xd{i}{k} - \xd{k}{i} + \zu{k}{j} - 1,
    \tag{Y2}\label{tag:Y2} \\
    & \forall i \forall j: i\neq j:  & & \xs{i}{j}  = \max_k(u_{ij}^{\text{Y1}},
    u_{ijk}^{\text{Y2}},0).  \tag{Z1}\label{tag:Z1}
\end{align}

\begin{prop}\label{prop:slides}
    Let $G_\mathbf{x}$ be a fixed hybrid graph such that
    $\mathbf{x}^\rightarrow$ corresponds to $G_\mathbf{x}$. We
    define $\mathbf{x}^\ast = \{\xs{i}{j} \}_{i\neq j} $ such that $\xs{i}{j} = 1$ if and
    only if there is a slide from $i$ to $j$ in $G$. Assume that Constraints
    \eqref{tag:W1} and \eqref{tag:C12}--\eqref{tag:C13} are satisfied. The set
    $\mathbf{x}^\ast$ is the unique solution of \eqref{tag:Z1}
    when $\mathbf{x}^\rightarrow$ corresponds to $G_\mathbf{x}$. 
\end{prop}

In Definition~\ref{def:globalCG}, the global Markov property in chain graphs is
described in terms of connecting paths in a moral graph. We restrict our
attention to \emph{decomposable} paths (Definition \ref{def:decomp}), and show
that these characterize separation in the relevant moral graphs (Lemma
\ref{lem:decomp}).

\begin{defn}\label{def:decomp}
    Let $G = ([d], E)$ be a hybrid graph, and let $i,j\in [d]$ and $C\subseteq
    [d]$. We say that a walk, $\pi$, between $i$ and $j$ in $(G_{\an(\{i,j\}\cup
    C)})^m$ is \emph{decomposable relative to $C$} if every subwalk of $\pi$,
    $\pi(l,k)$, is in $(G_{\an(\{k,l\}\cup C)})^m$.
\end{defn}

When $\omega$ is decomposable relative to $C$, it follows from the definition
that every subwalk of $\omega$ is also decomposable relative  to $C$. If
$\omega$ is decomposable relative to $C$ and connecting between $i$ and $j$
given $C$ in $(G_{\an(\{i,j\}\cup C)})^m$, then every subwalk $\omega(k,l)$ is
also connecting between $k$ and $l$ given $C$ in $(G_{\an(\{k,l\}\cup C)})^m$.
Finally, we see that every decomposable walk can be reduced to a decomposable
path: If we remove a loop from $k$ to $k$, then a subwalk on the new walk is
either on the original walk, or it is part of a longer subwalk (with the same
endpoints) on the original walk. The longer walk is in $(G_{\an(\{k,l\}\cup
C)})^m$ and so is the shorter walk. By repeatedly removing loops, we can obtain
a decomposable path.

The decomposable paths characterize connectivity in chain graphs in the
following sense.

\begin{lem}\label{lem:decomp}
    If $\pi$ is a connecting path between $i$ and $j$ given $C$ in
    $(G_{\an(\{i,j\}\cup C)})^m$, then there exists a connecting path between
    $i$ and $j$ given $C$ which is decomposable relative to $C$.
\end{lem}

Our proof of Lemma~\ref{lem:decomp} is very similar to the proof of Lemma~1 in
\citet{Richardson:2003} which shows a related result in DMGs equipped with
$m$-separation.

For a chain graph, $G$, we define the \emph{decomposable distance between $i$
and $j$ given $C$ in $G$} as the length of a shortest connecting and
decomposable path between $i$ and $j$ given $C$ in the moral graph of
$G_{\an(\{i,j\} \cup C)}$ if such path exists, and otherwise it is equal to $d$.
We use $\mathbf{l}^d = \{\lc{i}{j}{C}\}$ to denote the collection of
decomposable distances.

We say that the edge $l - k$ in $(G_{\an(\{i,j\} \cup C)})^m$ is \emph{moral} if
$l$ and $k$ are not adjacent in $G$.

We now list the minimal-length constraints that we need for chain graph
learning.
\begin{align}
      \forall i , j: i < j & \forall C: i,j\notin C:    \notag \\  &
      \lc{i}{j}{C}  \leq u_{ijC}^{\text{I1a}} \coloneqq 1 - (d-1)(\xd{i}{j} - 1)
      \tag{I1a}\label{tag:I1a} \\
      &   \lc{i}{j}{C}  \leq u_{ijC}^{\text{I1b}} \coloneqq 1 - (d-1)(\xd{j}{i} - 1)
      \tag{I1b}\label{tag:I1b} \\
 \forall i , j: i < j  &   \forall k: i,j,k \text{ distinct }  \forall C: i,j,k
 \notin C:  \notag \\ &     \lc{i}{j}{C} \leq u_{ijkC}^{\text{I2a}} \coloneqq 1 +
 \lc{i}{k}{C}  - (d - 2)(\xd{k}{j}  - 1)   \tag{I2a}\label{tag:I2a} \\
  &    \lc{i}{j}{C} \leq u_{ijkC}^{\text{I2b}} \coloneqq 1 + \lc{j}{k}{C}  - (d -
  2)(\xd{k}{i}  - 1)  \tag{I2b}\label{tag:I2b} \\
    \forall i, j: i < j &  \forall k: i,j,k \text{ distinct }  \forall C:
    i,j\notin C,k\in C:   \notag \\  &   \lc{i}{j}{C}  \leq u_{ijkC}^{\text{I3}}
    \coloneqq 1 - (d - 1 )(\xs{i}{k} - 1)- (d-1 )(\xs{j}{k} - 1) + (d-1 )\di{k}{C}
    \tag{I3}\label{tag:I3} \\
 \forall i, j: i< j &  \forall k: i,j,k \text{ distinct }  \forall C:
 i,j,k\notin C:   \notag \\  &    \lc{i}{j}{C}   \leq u_{ijkC}^{\text{I4}} \coloneqq
 \lc{i}{k}{C} + \lc{k}{j}{C} + (d-2)\di{k}{C} \tag{I4}\label{tag:I4} \\
   \forall i,  j: i< j  & \forall C:i,j\notin C    \notag \\   &    \lc{i}{j}{C}
   = \min_{k}(u_{ijC}^{\text{I1a}},u_{ijC}^{\text{I1b}},u_{ijkC}^{\text{I2a}},
   u_{ijkC}^{\text{I2b}},u_{ijkC}^{\text{I3}},u_{ijkC}^{\text{I4}})
   \tag{Q1}\label{tag:Q1}
\end{align}

Recall that when $G_\mathbf{x}$ is a hybrid graph and $\mathbf{x} =
(\mathbf{x}^\rightarrow,\mathbf{x}^\leftrightarrow)$ corresponds to
$G_\mathbf{x}$, we have $\mathbf{x}^\leftrightarrow = 0$. The set
$\{\xs{i}{j}\}_{i\neq j}$ is denoted by $\mathbf{x}^\ast$.

\begin{thm}[CGs]\label{thm:CGs}
    Let $G_\mathbf{x}$ be a fixed hybrid graph, and assume that $\mathbf{x}$
    corresponds to $G_\mathbf{x}$. Assume that $\mathbf{x}^\rightarrow$,
    $\mathbf{x}^\ast$,
    $\mathbf{l}^\rightarrow$, $\mathbf{d}^{\not\rightarrow}$,
    $\mathbf{d}_C^{\not\rightarrow}$,  $\mathbf{l}^-$, and $\mathbf{z}^-$
    satisfy
    \eqref{tag:C2}--\eqref{tag:C3}, \eqref{tag:C12}--\eqref{tag:C13},
    \eqref{tag:N1}, \eqref{tag:R1a}--\eqref{tag:R1b}, \eqref{tag:W1}, and
    \eqref{tag:Z1}. Let $a_\mathbb{G} = d - 1$.
    \begin{enumerate}[label=(\roman*)]
        \item Constraints~\eqref{tag:CH1a} and~\eqref{tag:CH1b} are satisfied if
        and only if $G_\mathbf{x}$ is a chain graph. \label{bul:thmCG1}
        \item The set
        \begin{align*}
            \{\lc{i}{j}{C}: i,j\in [d], C\subseteq [d]\setminus \{i,j\}, i\neq j\}
        \end{align*}
        is a solution to \eqref{tag:Q1} if and only if $\lc{i}{j}{C}$ equals the
        decomposable distance between $i$ and $j$ given $C$ in $G_\mathbf{x}$
        for each triple $(i,j,C)$ such that $i\neq j$ and $i,j\notin C$. That
        is, the decomposable distances in $G_\mathbf{x}$ are the unique solution
        of \eqref{tag:Q1} when $\mathbf{x}$ corresponds to $G_\mathbf{x}$.
        \label{bul:thmCG2}
    \end{enumerate}
\end{thm}

\section{Software implementation}\label{app:glipdemo}

We illustrate how to use the \pkg{glip} \proglang{R}~package with a simple
simulated example with three variables. We begin by loading the package,
setting a seed for reproducibility purposes and simulating data from a
distribution that is Markov and faithful w.r.t.\ $Y \leftarrow X \to Z$.
\begin{verbatim}
R> library("glip")
R> set.seed(1)
R> n <- 500
R> X <- rnorm(n)
R> Y <- X + rnorm(n)
R> Z <- X + rnorm(n)
R> data <- data.frame(X = X, Y = Y, Z = Z)
\end{verbatim}
The main function for graph learning in \pkg{glip} is \texttt{learn\_graph()},
which takes arguments \texttt{data} (a data frame), \texttt{max\_size} (the value
for $k$, which defaults to \texttt{ncol(data) - 2}, and an argument \texttt{mode},
which can be \texttt{"dg"}, \texttt{"dag"}, \texttt{"dmg"}, \texttt{"admg"}, or
\texttt{"chain"}. In the code chunk below, we perform ADMG learning with $k =
1$.
\begin{verbatim}
R> glip <- learn_graph(data, max_size = 1, mode = "admg", 
+    test_args = list(reg_YonZ = "lrm", reg_XonZ = "lrm"))
\end{verbatim}
The progress of the optimization, performed via \pkg{gurobi}, is printed to the
console and arguments to gurobi can be supplied to \texttt{learn\_graph()} via
the \texttt{gurobi\_args} argument. By default, \pkg{glip} performs a
nonparametric CI test (the Generalised Covariance Measure test,
\citealp{shah2020hardness}, with random forest regressions, as
implemented in \pkg{comets}, \citealp{kook2024algorithm}). We choose this as the
default because the implementation supports discrete, continuous, count, and
survival time variables \citep{kook2024tramgcm}. Arguments for the
test can be supplied via the \texttt{test\_args} argument. In the code chunk
above, we perform the test with linear regressions, so that the test corresponds
to a partial correlation test. The package also supports tests from
\pkg{bnlearn} and \pkg{pcalg} by setting, for instance, \texttt{test =
"gaussCItest"} and \texttt{use\_comets = FALSE}. The output \texttt{glip}
contains a data frame with test results, the learned graph, and the computed
graphical representation of the equivalence class.
\begin{verbatim}
R> glip$graph 

$M1
  X Y Z
X 0 1 1
Y 0 0 0
Z 0 0 0

$M2
  X Y Z
X 0 0 0
Y 0 0 0
Z 0 0 0
\end{verbatim}
In this case, the learned graph is an ADMG with two adjacency matrices
(\texttt{M1}: directed, \texttt{M2}: bidirected). The corresponding PAG, using
the same edge coding as \pkg{pcalg}, is given below.
\begin{verbatim}
R> glip$computed

  X Y Z
X 0 1 1
Y 1 0 0
Z 1 0 0
attr(,"class")
[1] "pag"    "matrix" "array" 
\end{verbatim}
Indeed, the learned graph is Markov equivalent with the true graph. The
\pkg{glip} package is hosted on GitHub at
\url{https://github.com/LucasKook/glip}
and can be installed from within \proglang{R} using
\texttt{remotes::install\_github("LucasKook/glip")}.

\section{Other encodings}\label{app:encodings}

Naturally, there are many possible ways to translate a graph learning problem
into a integer-programming problem. In this section, we discuss some
alternatives to the encodings used in the main paper.

\subsection{Constraint reduction}

It is possible to reduce the number of constraints in the DG-encoding by using
Constraints~\eqref{tag:L1am}--\eqref{tag:L4m} (below) instead of Constraints
\eqref{tag:L1am}--\eqref{tag:L4m} (Section~\ref{ssec:DGs}). There are two
differences between the two encodings: First, the below encoding has fewer
constraints as \eqref{tag:L2a}--\eqref{tag:L2b} have simply been removed.
Second, Constraint \eqref{tag:L4m} uses the variable $\di{k}{(C\cup\{i,j\})}$,
and Constraint \eqref{tag:L4} uses $\di{k}{C}$. This means that the below
encoding uses more variables; in the weak equivalence setting in particular.
Moreover, Constraints~\eqref{tag:L1am}--\eqref{tag:L4m} mix variables
corresponding to different conditioning sets whereas
\eqref{tag:L1a}--\eqref{tag:L4} do not.

\begin{align}
      \forall i , j: i < j & \forall C: i,j\notin C:    \notag \\  &
      \lc{i}{j}{C}  \leq u_{ijC}^{\text{L1a$^\text{m}$}} = 1 - (d-1)(\xd{i}{j} -
      1)    \tag{L1a$^\text{m}$}\label{tag:L1am} \\
      &   \lc{i}{j}{C}  \leq u_{ijC}^{\text{L1b$^\text{m}$}} = 1 -
      (d-1)(\xd{j}{i} - 1)  \tag{L1b$^\text{m}$}\label{tag:L1bm} \\
    \forall i, j: i < j &  \forall k: i,j,k \text{ distinct }  \forall C:
    i,j\notin C, k\in C:   \notag \\  &   \lc{i}{j}{C}  \leq
    u_{ijkC}^{\text{L3$^\text{m}$}} = 2 - (d - 2 )(\xd{i}{k} - 1)- (d-2
    )(\xd{j}{k}  - 1) \tag{L3$^\text{m}$}\label{tag:L3m} \\
 \forall i, j: i< j &  \forall k: i,j,k \text{ distinct }  \forall C:
 i,j,k\notin C:   \notag \\  &    \lc{i}{j}{C}   \leq
 u_{ijkC}^{\text{L4$^\text{m}$}} = \lc{i}{k}{C} + \lc{k}{j}{C} + (d-2)\di{k}{(C
 \cup \{i,j\})}   \tag{L4$^\text{m}$}\label{tag:L4m} \\
   \forall i,  j: i< j  & \forall C:i,j\notin C :   \notag \\   &
   \lc{i}{j}{C}   =
   \min_{k}(u_{ijC}^{\text{L1a$^\text{m}$}},u_{ijC}^{\text{L1b$^\text{m}$}},
   u_{ijkC}^{\text{L3$^\text{m}$}},u_{ijkC}^{\text{L4$^\text{m}$}})
   \tag{M1$^\text{m}$}\label{tag:M1m}
\end{align}

An analogous reduction in the number of constraints is possible for DMGs and CGs
by making the same changes.

\subsection[dc connecting constraints]{$d_c$-connecting constraints}
\label{sec:dcConstraints}

This section describes a set of constraints that use minimal-length
$d_c$-connecting paths. This avoids the use of ancestor indicators. This
approach uses more constraints, but fewer variables than the approach described
in the main paper. Empirically, the approach in the main paper is far more
computationally efficient because only triples (instead of quintuplets)
of nodes are considered. We include the $d_c$-connectivity constraints as they
encode the length of $d_c$-distances which can be used in the objective to
weight paths of different lengths. We conjecture that they are better
for this purpose than the $d$-distances.

\subsubsection{DGs}

The value of $\tilde{n}$ depends on $d$, and it can be found in Section
\ref{sssec:dcsep}. The following are the minimal-length constraints for DGs with
$d_c$-separation.

\begin{align}
      \forall i , j: i < j & \forall C: i,j\notin C:    \notag \\  &
      \lc{i}{j}{C}  \leq u_{ijC}^{\text{L1a$^\text{c}$}} \coloneqq 1 -
      \tilde{n}(\xd{i}{j} - 1)    \tag{L1a$^\text{c}$}\label{tag:L1ac} \\
      &   \lc{i}{j}{C}  \leq u_{ijC}^{\text{L1b$^\text{c}$}} \coloneqq 1 -
      \tilde{n}(\xd{j}{i} - 1)  \tag{L1b$^\text{c}$}\label{tag:L1bc} \\
 \forall i , j: i < j  &   \forall k: i,j,k \text{ distinct }  \forall C: i,j,k
 \notin C:  \notag \\ &     \lc{i}{j}{C} \leq u_{ijkC}^{\text{L2a$^\text{c}$}} \coloneqq
 1 + \lc{i}{k}{C}  - (\tilde{n} - 1)(\xd{k}{j} - 1)
 \tag{L2a$^\text{c}$}\label{tag:L2ac} \\
  &    \lc{i}{j}{C} \leq u_{ijkC}^{\text{L2b$^\text{c}$}} \coloneqq 1 + \lc{j}{k}{C}  -
  (\tilde{n} - 1)(\xd{k}{i} - 1)   \tag{L2b$^\text{c}$}\label{tag:L2bc}
  \\
    \forall i, j: i < j &  \forall k: i,j,k \text{ distinct }  \forall C:
    i,j\notin C, k\in C:   \notag \\  &   \lc{i}{j}{C}  \leq
    u_{ijkC}^{\text{L3$^\text{c}$}} \coloneqq 2 - (\tilde{n} -1 )(\xd{i}{k} - 1)-
    (\tilde{n} -1 )(\xd{j}{k} - 1) \tag{L3$^\text{c}$}\label{tag:L3c}\\
 \forall i, j: i< j &  \forall k, l: i,j,k,l \text{ distinct }  \forall C:
 i,j,l\notin C , k\in C:   \notag \\  &    \lc{i}{j}{C}   \leq
 u_{ijklC}^{\text{L4a$^\text{c}$}} \coloneqq 2 + \lc{j}{l}{C} - (\tilde{n} -2
 )(\xd{i}{k} - 1) - (\tilde{n} -2 )(\xd{l}{k}  - 1)
 \tag{L4a$^\text{c}$}\label{tag:L4ac} \\
  &   \lc{i}{j}{C}   \leq u_{ijklC}^{\text{L4b$^\text{c}$}} \coloneqq 2 + \lc{i}{l}{C} -
  (\tilde{n} -2 )(\xd{j}{k} - 1) - (\tilde{n} -2 )(\xd{l}{k}  - 1)
  \tag{L4b$^\text{c}$}\label{tag:L4bc} \\
   \forall i , j: i < j &  \forall k, l, m: i,j,k,l \text{ distinct  and }
   i,j,k,m \text{ distinct} \forall C: i,j,l,m\notin C, k\in C:  \notag \\  &
   \lc{i}{j}{C}   \leq  u_{ijklmC}^{\text{L5$^\text{c}$}} \coloneqq 2 + \lc{i}{l}{C} +
   \lc{m}{j}{C} - (\tilde{n} -3 )(\xd{l}{k}  - 1) - (\tilde{n} -3
   )(\xd{m}{k}  - 1)   \tag{L5$^\text{c}$}\label{tag:L5c} \\
   \forall i,  j: i< j  & \forall C:i,j\notin C    \notag \\   &    \lc{i}{j}{C}
   = \min_{k,l,m}(u_{ijC}^{\text{L1a$^\text{c}$}},u_{ijC}^{\text{L1b$^\text{c}$}},
   u_{ijkC}^{\text{L2a$^\text{c}$}},u_{ijkC}^{\text{L2b$^\text{c}$}},
   u_{ijkC}^{\text{L3$^\text{c}$}},u_{ijklC}^{\text{L4a$^\text{c}$}},
   u_{ijklC}^{\text{L4b$^\text{c}$}},u_{ijklmC}^{\text{L5$^\text{c}$}})
   \tag{M1$^\text{c}$}\label{tag:M1c}
\end{align}

\begin{thm}[DGs and DAGs, $d_c$-separation]\label{thm:DAGsdc}
    Let $G_\mathbf{x}$ be a fixed DG, assume that $\mathbf{x}$ corresponds
    to $G_\mathbf{x}$, and let $a_\mathbb{G} = \tilde{n}$.
    \begin{enumerate}[label=(\roman*)]
        \item If $\mathbf{x}^\rightarrow$, $\mathbf{l}^\rightarrow$, and
        $\mathbf{d}^{\not\rightarrow}$ satisfy \eqref{tag:C2}--\eqref{tag:C3}
        and \eqref{tag:N1}, then Constraint \eqref{tag:AC} is satisfied if and
        only if $G_\mathbf{x}$ is a directed acyclic graph.
        \label{bul:thmDAGdc1}
        \item The set
        \begin{align*}
            \{\lc{i}{j}{C}: i,j\in [d], C\subseteq [d]\setminus \{i,j\}, i\neq j\}
        \end{align*}
        is a solution to \eqref{tag:M1c} if and only if $\lc{i}{j}{C}$ equals the
        $d_c$-distance between $i$ and $j$ given $C$ in $G_\mathbf{x}$ for each
        triple $(i,j,C)$ such that $i\neq j$ and $i,j\notin C$. That is, the 
        $d_c$-distances in $G_\mathbf{x}$ are the unique solution of
        \eqref{tag:M1c} when $\mathbf{x}$ corresponds to $G_\mathbf{x}$.
        \label{bul:thmDAGdc2}
    \end{enumerate}
\end{thm}

\subsubsection{DMGs}

Minimal-length constraints for DMGs with $m_c$-separation are found below (see
Section \ref{sssec:dcsep} for a definition of $\tilde{n}$).

\begin{align}
  \forall i , j: i < j & \forall C: i,j\notin C:     \notag \\  &  \lc{i}{j}{C}
  \leq u_{ijC}^{\text{K1a$^\text{c}$}} \coloneqq 1 - \tilde{n}(\xs{i}{j} - 1)
  \tag{K1a$^\text{c}$}\label{tag:K1ac} \\
 & \lc{i}{j}{C}  \leq u_{ijC}^{\text{K1b$^\text{c}$}} \coloneqq 1 - \tilde{n}(\xs{j}{i}
 - 1)  \tag{K1b$^\text{c}$}\label{tag:K1bc} \\
 \forall i , j: i < j   & \forall k: i,j,k \text{ distinct }  \forall C: i,j,k
 \notin C:  \notag \\  
 &     \lc{i}{j}{C}  \leq u_{ijkC}^{\text{K2a$^\text{c}$}} \coloneqq  1 + \lc{i}{k}{C}
 - (\tilde{n} - 1)(\xd{k}{j}  - 1)  \tag{K2a$^\text{c}$}\label{tag:K2ac}
 \\
  &  \lc{i}{j}{C}  \leq u_{ijkC}^{\text{K2b$^\text{c}$}} \coloneqq  1 + \lc{j}{k}{C}  -
  (\tilde{n} - 1)(\xd{k}{i} - 1)   \tag{K2b$^\text{c}$}\label{tag:K2bc}
  \\
  \forall i, j: i < j &  \forall k: i,j,k \text{ distinct } \forall C: i,j\notin
  C, k\in C:   \notag \\  &   \lc{i}{j}{C}  \leq
  u_{ijkC}^{\text{K3a$^\text{c}$}} \coloneqq 2 - (\tilde{n} -1 )(\xs{i}{k} - 1)-
  (\tilde{n} -1 )(\xs{j}{k} - 1) \tag{K3a$^\text{c}$}\label{tag:K3ac} \\
   \forall i, j: i < j &  \forall k_1,k_2: i,j,k_1,k_2 \text{ distinct }
   \forall C: i,j\notin C, k_1,k_2\in C:   \notag 
   \\  
   &   \lc{i}{j}{C}  \leq u_{ijk_1k_2C}^{\text{K3b$^\text{c}$}} \coloneqq 2 + 
   \lcb{k_1}{k_2}{C} - (\tilde{n} -2
   )(\xs{i}{k_1} - 1) \ -  \notag \\ & \quad\quad\quad\quad\quad\quad\quad\quad\quad
   (\tilde{n} -2 )(\zcb{k_1}{k_2}{C} - 1) -
   (\tilde{n} -2 )(\xs{j}{k_2} - 1) \tag{K3b$^\text{c}$}\label{tag:K3bc} \\
  \forall i, j: i< j &  \forall k, l: i,j,k,l \text{ distinct }  \forall C:
  i,j,l\notin C , k\in C:   \notag \\   
  &  \lc{i}{j}{C}   \leq
  u_{ijklC}^{\text{K4a$^\text{c}$}} \coloneqq 2 + \lc{j}{l}{C} - (\tilde{n} -2
  )(\xs{i}{k} - 1) - (\tilde{n} -2 )(\xd{l}{k} - 1)
  \tag{K4a$^\text{c}$}\label{tag:K4ac} 
  \\
 &    \lc{i}{j}{C}   \leq u_{ijklC}^{\text{K4b$^\text{c}$}} \coloneqq 2 + \lc{i}{l}{C} -
 (\tilde{n} -2 )(\xs{j}{k} - 1) - (\tilde{n} -2 )(\xd{l}{k} - 1)
 \tag{K4b$^\text{c}$}\label{tag:K4bc} \\
    \forall i, j: i< j &  \forall k_1,k_2, l: i,j,k_1,k_2,l \text{ distinct }
    \forall C: i,j,l\notin C , k_1,k_2\in C:   \notag \\  &     \lc{i}{j}{C}
    \leq u_{ijk_1k_2lC}^{\text{K4c$^\text{c}$}} \coloneqq  2 + \lc{j}{l}{C} +
    \lcb{k_1}{k_2}{C} - (\tilde{n} -3 )(\xs{i}{k} - 1) \ -  
      \notag \\ & \quad\quad\quad\quad\quad\quad\quad\quad\quad
    (\tilde{n} -3
    )(\zcb{k_1}{k_2}{C} - 1) - (\tilde{n} -3 )(\xd{l}{k} - 1)
    \tag{K4c$^\text{c}$}\label{tag:K4cc} 
\end{align}
\begin{align}
 &   \lc{i}{j}{C}   \leq u_{ijk_1k_2lC}^{\text{K4d$^\text{c}$}} \coloneqq 2 +
 \lc{i}{l}{C} + \lcb{k_1}{k_2}{C} - (\tilde{n} -3 )(\xs{j}{k} - 1)  \ - 
      \notag \\ & \quad\quad\quad\quad\quad\quad\quad\quad\quad
 (\tilde{n}
 -3 )(\zcb{k_1}{k_2}{C} - 1) - (\tilde{n} -3 )(\xd{l}{k} - 1)
 \tag{K4d$^\text{c}$}\label{tag:K4dc} \\
   \forall i , j: i < j &  \forall k, l, m: i,j,k,l \text{ dist. and }
   i,j,k,m \text{ dist. }  \forall C: i,j,l,m\notin C, k\in C:  \notag \\  &
   \lc{i}{j}{C}   \leq u_{ijklmC}^{\text{K5a$^\text{c}$}} \coloneqq 2 + \lc{i}{l}{C} +
   \lc{m}{j}{C} - (\tilde{n} -3 )(\xd{l}{k} - 1) - (\tilde{n} -3 )(\xd{m}{k} -
   1)   \tag{K5a$^\text{c}$}\label{tag:K5ac} \\
   \forall i , j: i < j &  \forall k_1,k_2, l, m: i,j,k_1,k_2,l \text{ dist., }  i,j,k_1,k_2,m \text{ dist. }  \forall C: i,j,l,m\notin C,
   k_1,k_2\in C:  \notag \\   &  \lc{i}{j}{C}   \leq
   u_{ijk_1k_2lmC}^{\text{K5b$^\text{c}$}} \coloneqq 2 + \lc{i}{l}{C} +
   \lcb{k_1}{k_2}{C} + \lc{m}{j}{C} - (\tilde{n} -4 )(\xd{l}{k} - 1) \ -
      \notag \\ & \quad\quad\quad\quad\quad\quad\quad\quad\quad
   (\tilde{n} -4 )(\zcb{k_1}{k_2}{C} - 1)  - (\tilde{n} -4 )(\xd{m}{k} - 1)
   \tag{K5b$^\text{c}$}\label{tag:K5bc} \\
   \forall i,  j: i< j &  \forall C:i,j\notin C    \notag \\   &    \lc{i}{j}{C}
   = \min_{k,k_1,k_2,l,m}(u_{ijC}^{\text{K1a$^\text{c}$}},
   u_{ijC}^{\text{K1b$^\text{c}$}},
   u_{ijkC}^{\text{K2a$^\text{c}$}},
   u_{ijkC}^{\text{K2b$^\text{c}$}},
   u_{ijkC}^{\text{K3a$^\text{c}$}},u_{ijk_1k_2C}^{\text{K3b$^\text{c}$}},
   u_{ijklC}^{\text{K4a$^\text{c}$}},
  \notag \\ & \quad\quad\quad\quad\quad\quad\quad\quad\quad
   u_{ijklC}^{\text{K4b$^\text{c}$}},u_{ijk_1k_2lC}^{\text{K4c$^\text{c}$}},
   u_{ijk_1k_2lC}^{\text{K4d$^\text{c}$}},u_{ijklmC}^{\text{K5a$^\text{c}$}},
   u_{ijk_1k_2lmC}^{\text{K5b$^\text{c}$}})
   \tag{P1$^\text{c}$}\label{tag:P1c}
 \end{align}

 \begin{thm}[DMGs and ADMGs, $m_c$-separation]\label{thm:DMGsdc}
    Let $G_\mathbf{x}$ be a fixed DMG, and assume that $\mathbf{x}$ corresponds
    to $G_\mathbf{x}$. Assume that $\mathbf{x}^\rightarrow$, $\mathbf{x}^\leftrightarrow$, $\mathbf{x}^\ast$,
    $\mathbf{l}^\leftrightarrow$, and $\mathbf{z}^\leftrightarrow$ satisfy  \eqref{tag:C6}--\eqref{tag:C11}, and \eqref{tag:O1}, and let $a_\mathbb{G} = \tilde{n}$.
    \begin{enumerate}[label=(\roman*)]
        \item If $\mathbf{x}$,
    $\mathbf{l}^\rightarrow$, and $\mathbf{d}^{\not\rightarrow}$ satisfy \eqref{tag:C2}--\eqref{tag:C3} and \eqref{tag:N1}, then Constraint \eqref{tag:AC} is satisfied if and only if
        $G_\mathbf{x}$ is an acyclic directed mixed graph. \label{bul:thmDMGdc1}
        \item The set
        \begin{align*}
            \{\lc{i}{j}{C}: i,j\in [d], C\subseteq [d], i\neq j, i,j\notin C\}
        \end{align*}
        is a solution to \eqref{tag:P1c} if and only if $\lc{i}{j}{C}$ equals the
        $m_c$-distance between $i$ and $j$ given $C$ in $G_\mathbf{x}$ for each
        triple $(i,j,C)$ such that $i\neq j$ and $i,j\notin C$. That is, the 
        $m_c$-distances in $G_\mathbf{x}$ are the unique solution of
        \eqref{tag:P1c} when $\mathbf{x}$ corresponds to $G_\mathbf{x}$.
        \label{bul:thmDMGdc2}
    \end{enumerate}
\end{thm}

\section{Additional empirical results}\label{app:empirical}

\subsection{Runtimes for GLIP and ASP}\label{app:empirical:asp}

\begin{figure}[t!]
\begin{subfigure}{\textwidth}
\centering
\includegraphics[width=0.75\linewidth]{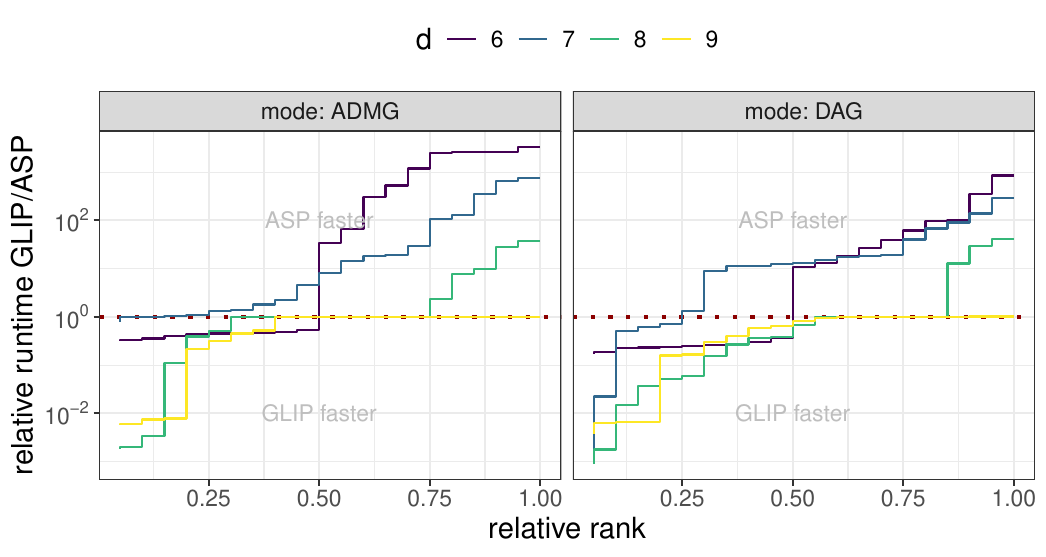}
\subcaption[]{Graph learning with $k = d - 2$.}
\end{subfigure}
\begin{subfigure}{\textwidth}
\centering
\includegraphics[width=0.75\linewidth]{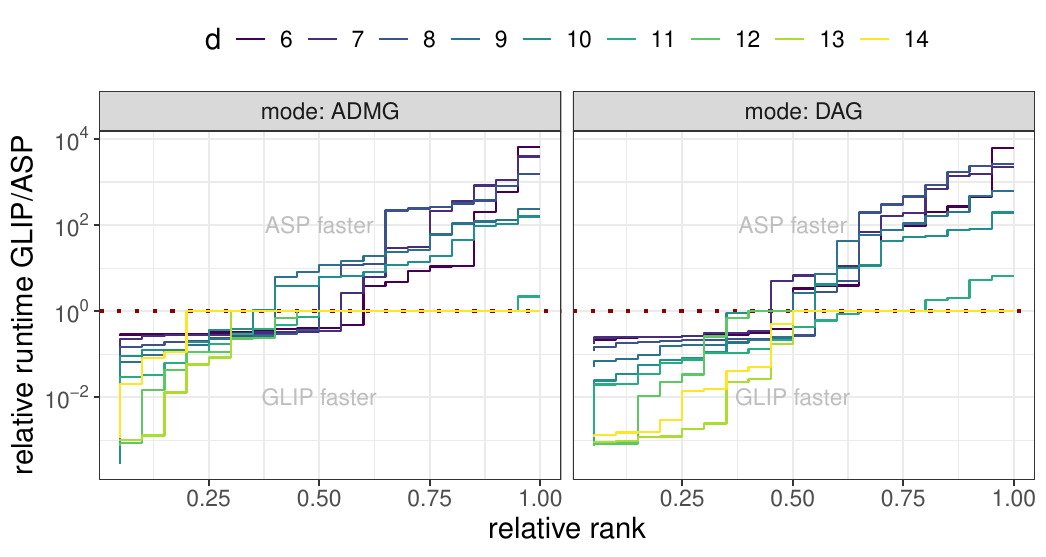}
\subcaption[]{Weak graph learning with $k = 1$.}
\end{subfigure}
\caption{%
    Relative runtime comparison of GLIP and ASP restricted to larger graphs.
}\label{fig:asp:larged}
\end{figure}

\begin{figure}[t!]
\begin{subfigure}{\textwidth}
\centering
\includegraphics[width=0.75\linewidth]{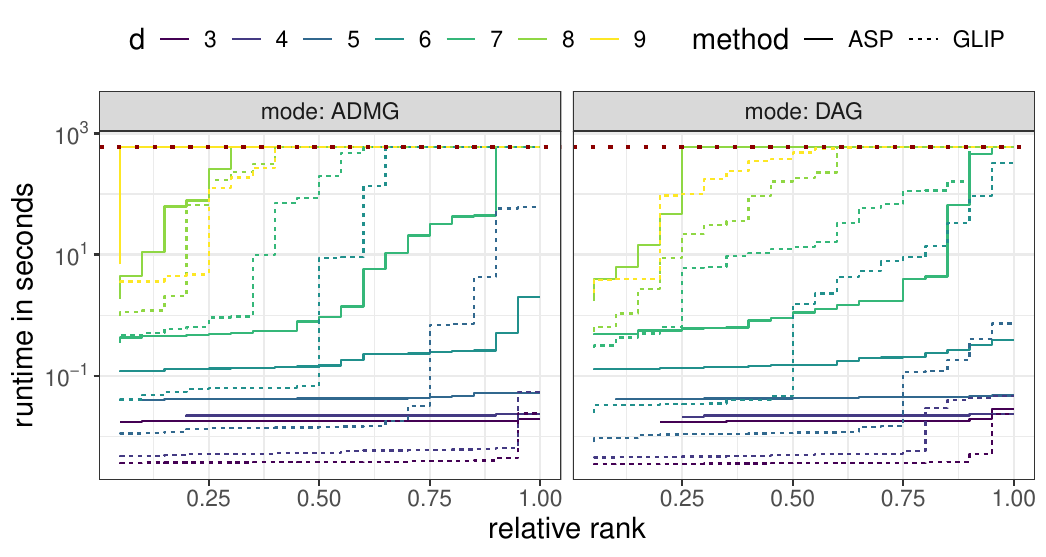}
\subcaption[]{Graph learning with $k = d - 2$.}
\end{subfigure}
\begin{subfigure}{\textwidth}
\centering
\includegraphics[width=0.75\linewidth]{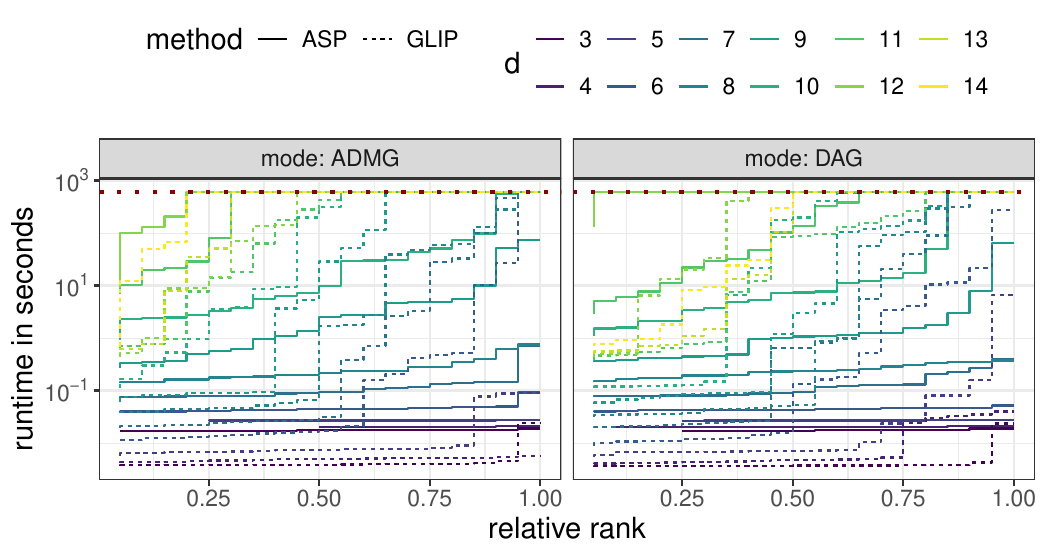}
\subcaption[]{Weak graph learning with $k = 1$.}
\end{subfigure}
\caption{%
    Absolute runtime comparison of GLIP and ASP. For the comparison, graphs of
    different sizes were generated according to the details in
    Section~\ref{sec:comp:exact}. The red dotted line indicates the walltime of
    600~seconds.
}\label{fig:asp:abs}
\end{figure}

Figure~\ref{fig:asp:larged} displays the same runtimes as Figure~\ref{fig:asp}
in the main text, but restricted to graphs with at least 6 nodes.
Figure~\ref{fig:asp:abs} shows the absolute runtimes for GLIP and ASP.

\subsection{Additional results using simulated data}\label{app:empirical:sim}

\begin{figure}[t!]
\begin{subfigure}{\textwidth}
\centering
\includegraphics[width=0.99\linewidth]{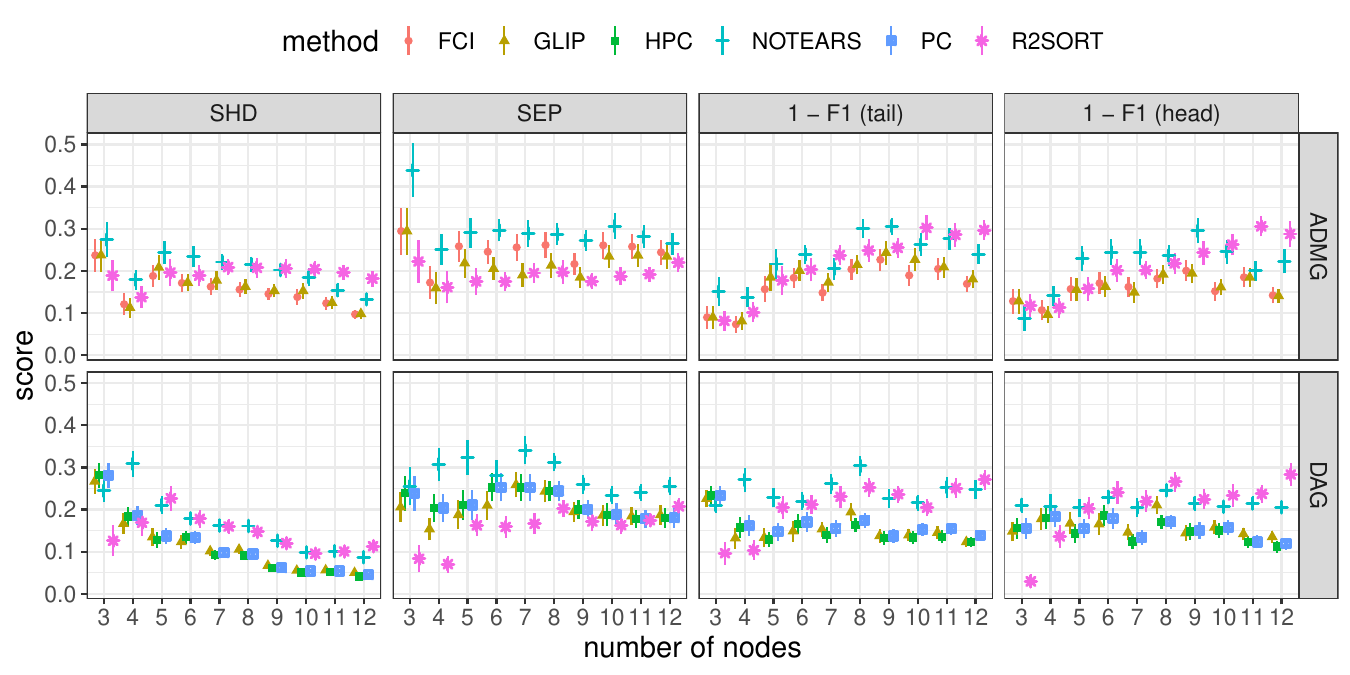}
\subcaption[]{Results for $n = 400$.}
\end{subfigure}
\begin{subfigure}{\textwidth}
\centering
\includegraphics[width=0.99\linewidth, trim=0 0 0 1cm, clip]{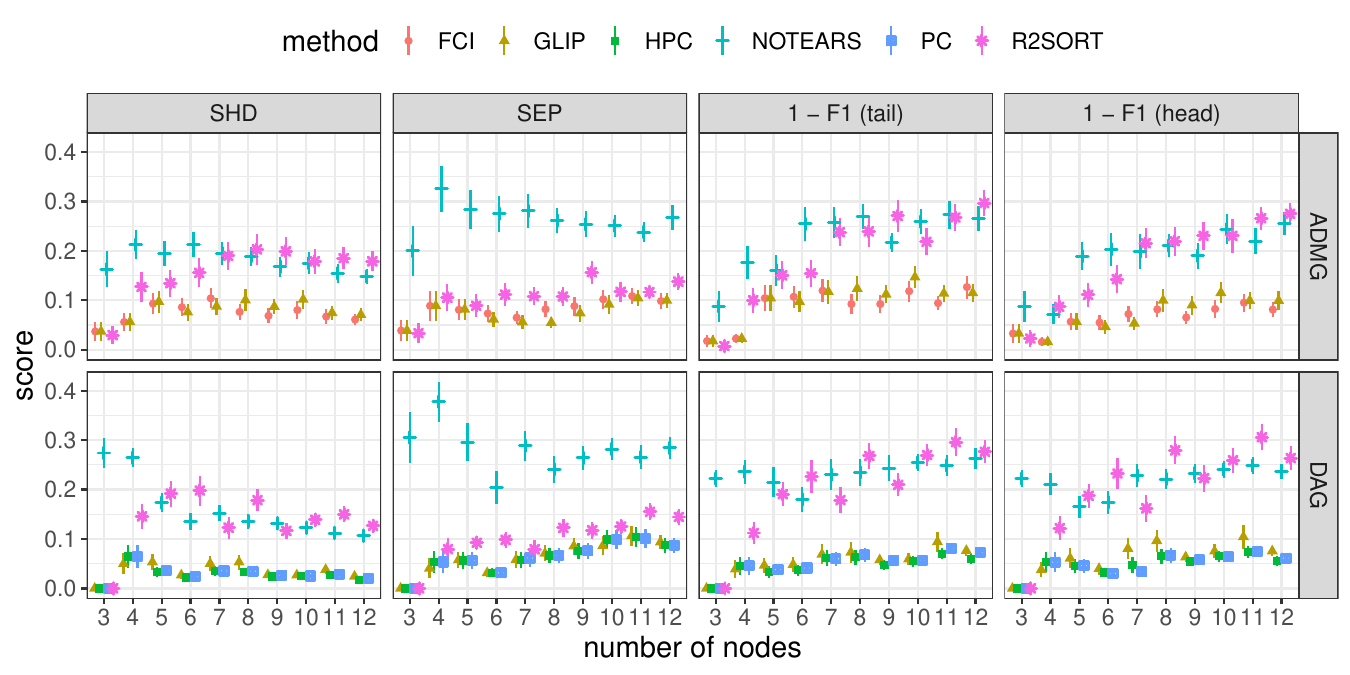}
\subcaption[]{Results for $n = 10'000$.}
\end{subfigure}
\caption{%
    Performance comparison against approximate methods for weak graph learning
    with $k = 1$.
}\label{fig:sim:weak}
\end{figure}

\begin{figure}[t!]
\begin{subfigure}{\textwidth}
\centering
\includegraphics[width=0.99\linewidth]{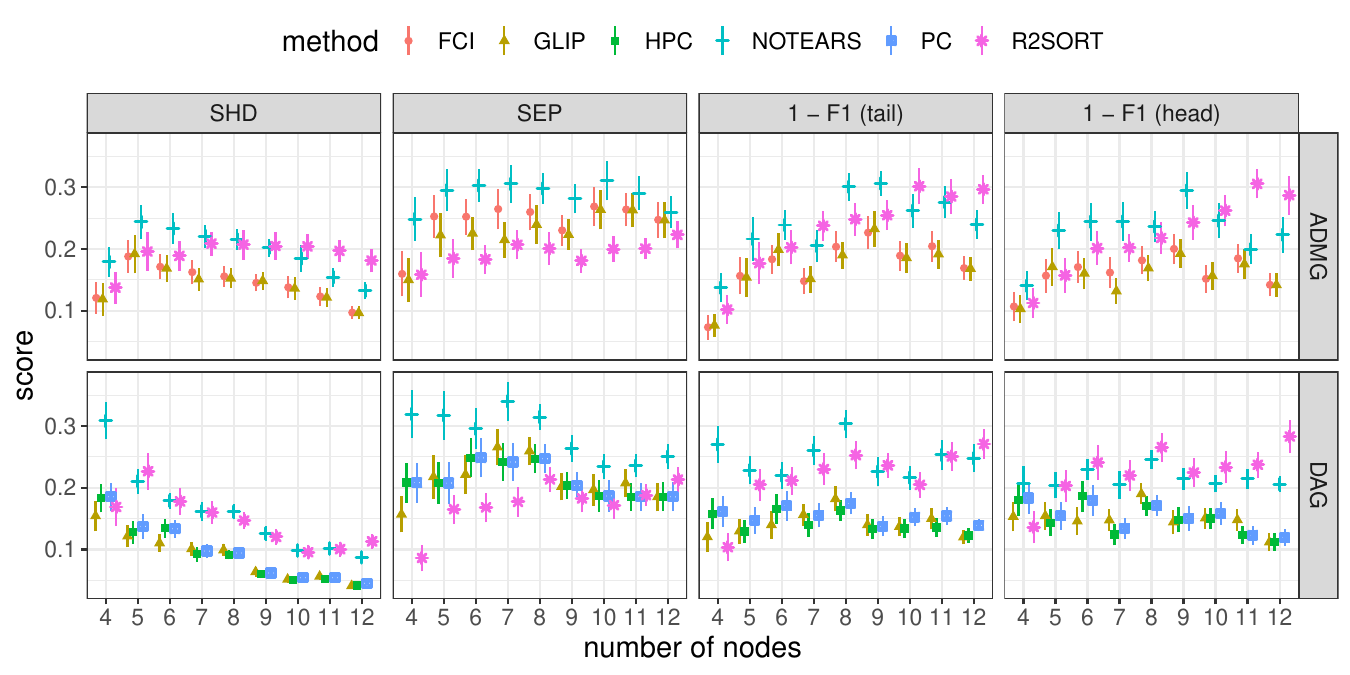}
\subcaption[]{Results for $n = 400$.}
\end{subfigure}
\begin{subfigure}{\textwidth}
\centering
\includegraphics[width=0.99\linewidth, trim=0 0 0 1cm, clip]{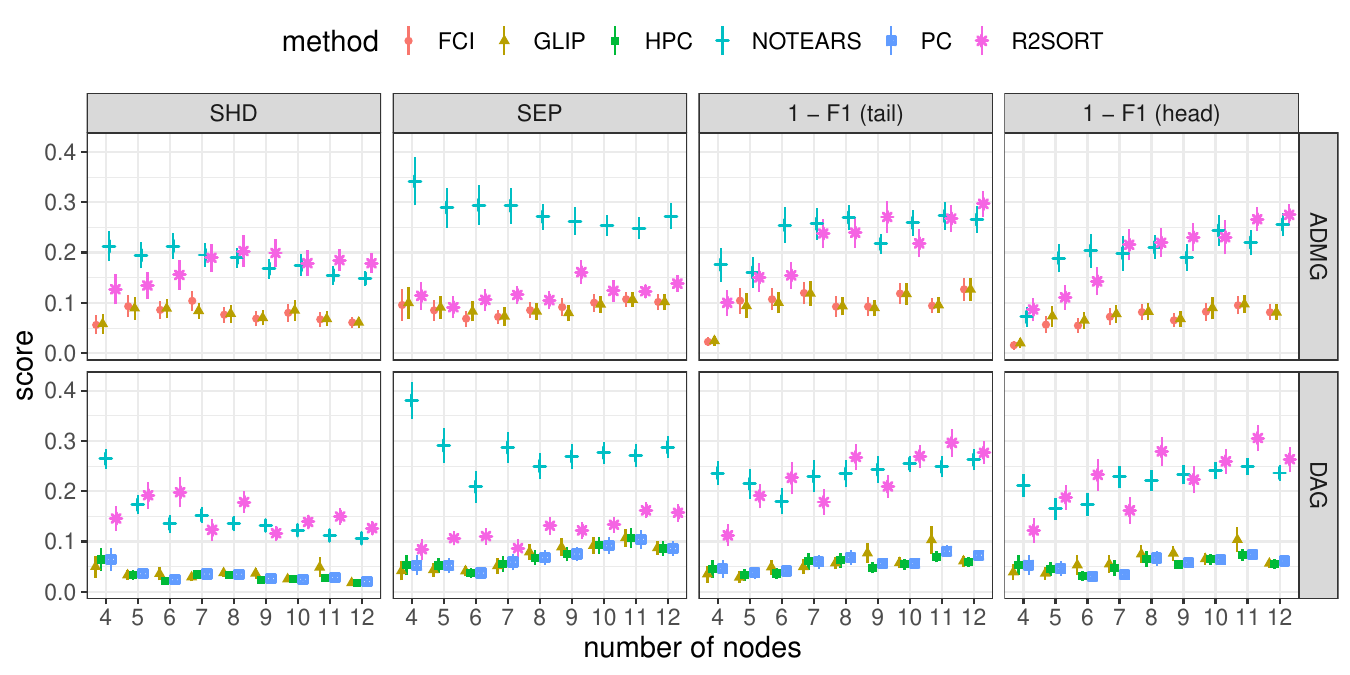}
\subcaption[]{Results for $n = 10'000$.}
\end{subfigure}
\caption{%
Performance comparison against approximate methods for weak graph learning with
$k = 2$.
}\label{fig:sim:weak2}
\end{figure}

\begin{figure}[t!]
\begin{subfigure}{0.49\textwidth}
\centering
\includegraphics[width=0.99\linewidth]{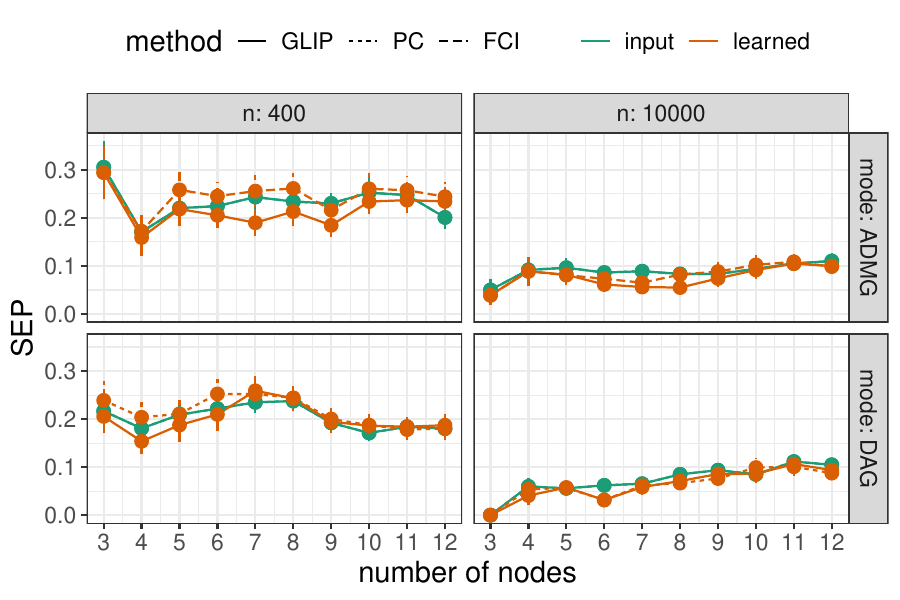}
\subcaption[]{Results for $k = 1$.}
\end{subfigure}
\begin{subfigure}{0.49\textwidth}
\centering
\includegraphics[width=0.99\linewidth]{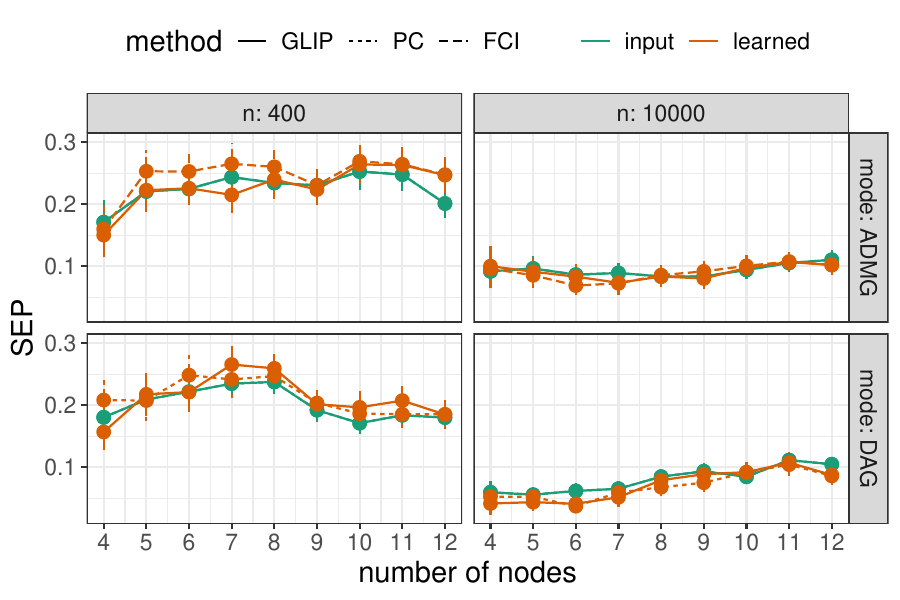}
\subcaption[]{Results for $k = 2$.}
\end{subfigure}
\caption{%
Average separation distance against the oracle essential graph (DAG) or PAG (ADMG).
}\label{fig:sim:weak:sep}
\end{figure}

We present results on the performance of GLIP for weak graph learning under the
same simulation settings as described in Section~\ref{sec:comp:sim}. This
comparison puts GLIP at a disadvantage over the other competitors, as GLIP only
has access to test results for conditioning sets of up to size $k \in \{1, 2\}$.
Together with the short walltime of 300~seconds, this leads to a conservative
evaluation of GLIP's performance for weak learning. Figures~\ref{fig:sim:weak}
and~\ref{fig:sim:weak2} contain the results, which show that, especially for the
larger sample size, GLIP performs almost on par in terms of SHD and head/tail
F1, despite having access to less information. The fractions of runs in which
GLIP could confirm optimality of the solution are shown in
Table~\ref{tab:sim:weak:completion}. Separation distance between input
$p$-values and the oracle graph, as well as between the learned output and the
oracle graph is shown in Figure~\ref{fig:sim:weak:sep}.

\begin{table}[t!]
\centering
\resizebox{\textwidth}{!}{%
\begin{tabular}{lrrrrrrrrrrrr}
\toprule
\multicolumn{2}{c}{ } & \multicolumn{11}{c}{$d$} \\
\cmidrule(l{3pt}r{3pt}){3-13}
$n$ & mode & $k$ & 3 & 4 & 5 & 6 & 7 & 8 & 9 & 10 & 11 & 12\\
\midrule
400 & ADMG & 1 & 1.000 & 1.000 & 1.000 & 0.933 & 0.667 & 0.567 & 0.433 & 0.433 & 0.367 & 0.367\\
400 & DAG & 1 & 1.000 & 1.000 & 1.000 & 0.767 & 0.400 & 0.467 & 0.333 & 0.333 & 0.233 & 0.133\\
400 & ADMG & 2 & - & 1.000 & 0.933 & 0.733 & 0.567 & 0.333 & 0.167 & 0.233 & 0.233 & 0.200\\
400 & DAG & 2 & - & 1.000 & 1.000 & 0.667 & 0.300 & 0.333 & 0.267 & 0.233 & 0.167 & 0.000\\
10000 & ADMG & 1 & 1.000 & 1.000 & 1.000 & 0.833 & 0.733 & 0.500 & 0.467 & 0.400 & 0.433 & 0.133\\
10000 & DAG & 1 & 1.000 & 1.000 & 1.000 & 0.867 & 0.600 & 0.467 & 0.400 & 0.167 & 0.167 & 0.100\\
10000 & ADMG & 2 & - & 1.000 & 0.967 & 0.667 & 0.567 & 0.300 & 0.300 & 0.167 & 0.200 & 0.033\\
10000 & DAG & 2 & - & 1.000 & 0.967 & 0.733 & 0.367 & 0.300 & 0.233 & 0.067 & 0.067 & 0.067\\
\bottomrule
\end{tabular}}
\caption{Fraction of the 30 simulation iterations in which GLIP confirmed
optimality of the solution for weak DAG and ADMG learning in the simulation from
Section~\ref{sec:comp:sim}.}\label{tab:sim:weak:completion}
\end{table}

\subsection{Additional results on the benchmark datasets}\label{app:empirical:bench}

Table~\ref{tab:datasets:d8} contains the results for ADMG learning with discrete
features on the benchmark datasets described in Section~\ref{sec:comp:bench}.
The ground-truth DAGs were marginalized to eight nodes chosen at random,
resulting in a ground-truth ADMG, from which a PAG can be computed, and against
which we evaluate the output of the graph learning algorithms. GLIP performs
best in terms of SHD for four out of the five datasets. 

\begin{table}[!t]
\centering
\resizebox{\textwidth}{!}{\input{tables/d8}}
\caption{%
Performance of GLIP and competitors on benchmark datasets based on the PAG
output of each method at $\alpha = 0.001$. The
ground-truth DAG was marginalized to an ADMG over 8 nodes chosen uniformly at
random and only the data corresponding to those 8 variables were used for graph
learning. The analysis was repeated for $\min\{\binom{d}{8}, 50\}$ distinct
subsets; standard deviations are reported in parentheses. The best method per
dataset and metric is highlighted in bold. SHD: absolute structural Hamming
distance; SEP: separation distance; FDR: false discovery rate; FNR: false
negative rate. FDR and FNR are averaged over arrow heads and tails and
circle-edges.
}\label{tab:datasets:d8}
\end{table}

\subsection{Empirical results on chain graph learning}\label{app:empirical:chain}

We generate random chain graphs and data from a corresponding multivariate
normal distribution according to the algorithms presented in
\citet{ma2008structural} with degree 4, chain component length 2, and edge
probability 0.8. GLIP is warmstarted with a DAG computed from the output of PC
and run with a walltime of 300~seconds. For the simulation, we consider $d$ up
to 10 and sample sizes $1000$ and $10^5$. Partial correlation tests are
performed for all tests at $\alpha = 0.001$. For each combination of $d$ and
$n$, the simulation is repeated 30~times.

\begin{figure}[t!]
\begin{subfigure}{\textwidth}
\centering
\includegraphics[width=0.85\linewidth]{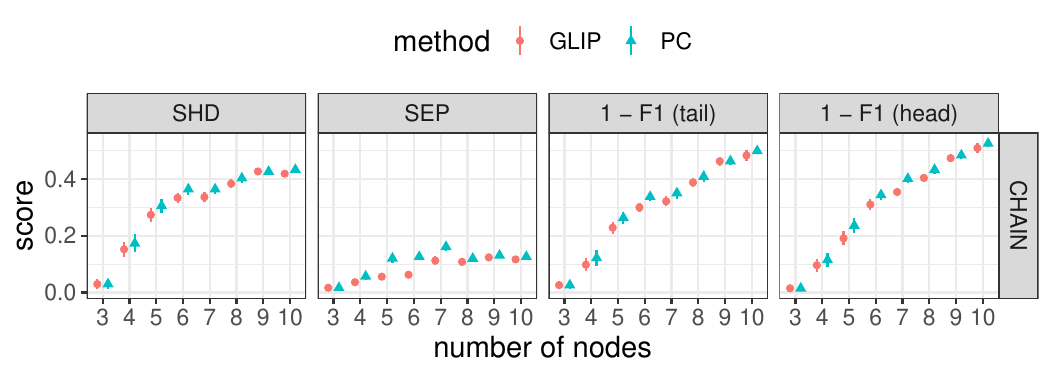}
\subcaption[]{Results with $n = 1000$.}
\end{subfigure}
\begin{subfigure}{\textwidth}
\centering
\includegraphics[width=0.85\linewidth, trim=0 0 0 1cm, clip]{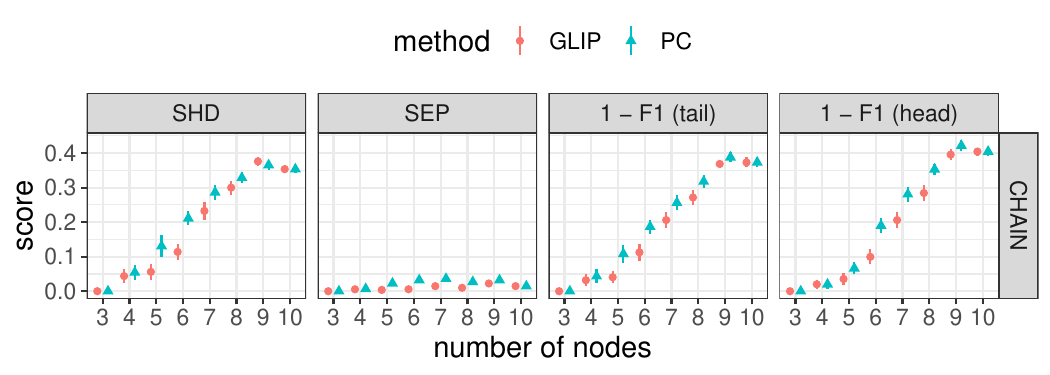}
\subcaption[]{Results with $n = 100'000$.}
\end{subfigure}
\caption{%
    Empirical evaluation of GLIP for chain graph learning.
}\label{fig:sim:chain}
\end{figure}

\begin{figure}[t!]
\centering
\includegraphics[width=0.85\linewidth]{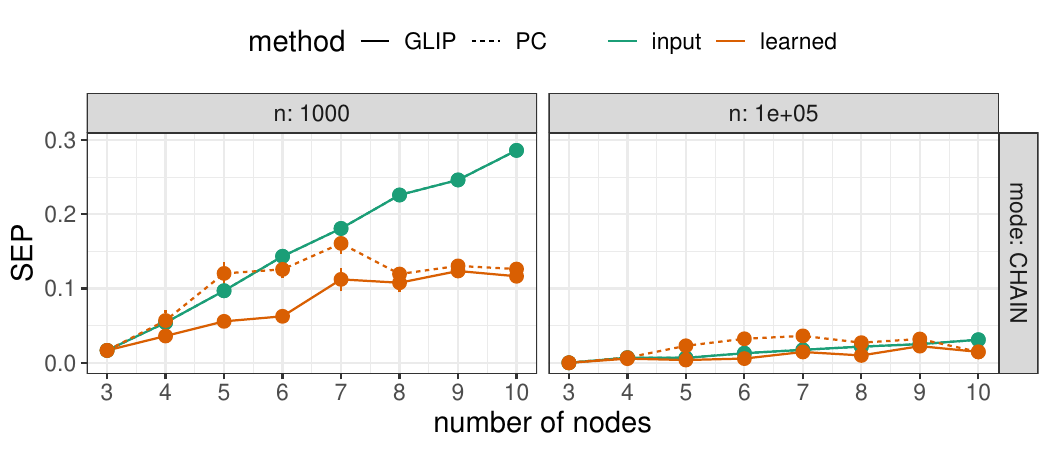}
\caption{%
Average separation distance against the oracle largest chain graph.
}\label{fig:chain:sep}
\end{figure}

Figure~\ref{fig:sim:chain} shows the results for chain graph learning for GLIP
versus PC. The PC algorithm is not intended for chain
graph learning, only for the subclass of DAGs. We use it as a
competitor method as the only publicly available implementations of chain graph
learning methods we found were outdated, and it would require considerable effort to
re-write them due to deprecated dependencies, etc. GLIP is able
to improve on the PC solution for all metrics, especially at the larger sample
size. Figure~\ref{fig:chain:sep} also shows that GLIP learns chain graphs that
are more consistent with the underlying ground-truth than the input $p$-values
suggest, especially for $n = 1000$. The effect is less pronounced for $n =
10^5$, which may be due to near-oracle performance of the partial correlation
test at such large sample sizes.

Table~\ref{tab:chain:completion} gives the fraction of runs in which GLIP could
show optimality of the solution within the walltime limit, which drops rather
abruptly for graphs of size 6. Nonetheless, GLIP is able to considerably improve
upon the warmstart solution within the walltime limit for graphs with $d \leq
8$.

\begin{table}[t!]
\centering
\begin{tabular}{lrrrrrrrr}
\toprule
\multicolumn{1}{c}{ } & \multicolumn{8}{c}{$d$} \\
\cmidrule(l{3pt}r{3pt}){2-9}
$n$ & 3 & 4 & 5 & 6 & 7 & 8 & 9 & 10\\
\midrule
$10^3$ & 1.000 & 1.000 & 0.800 & 0.000 & 0.000 & 0.000 & 0.000 & 0.000\\
$10^5$ & 1.000 & 1.000 & 1.000 & 0.333 & 0.033 & 0.000 & 0.000 & 0.000\\
\bottomrule
\end{tabular}
\caption{Fraction of the 30 simulation iterations in which GLIP confirmed
optimality of the solution for chain graph
learning.}\label{tab:chain:completion}
\end{table}

\clearpage

\section{Proofs}\label{app:proofs}

\subsection{Proof of Theorem~\ref{thm:dc}}

\begin{proof}
If there is an $m$-connecting walk between $i$ and $j$ given $C$, then every
collider $k$ which is not in $C$ is an ancestor of some $c \in C$, $k \neq c$.
By expanding the original walk by $k \rightarrow \ldots \rightarrow c \leftarrow
\ldots \leftarrow k$  for each collider, we can obtain an $m_c$-connecting walk
between $i$ and $j$ given $C$.

We assume now that there is an $m_c$-connecting walk between $i$ and $j$ given
$C$ and wish to show that we can find such a walk which is of length at most
$\tilde{n}$. For $d < 4$, this follows immediately. Assume $d \geq 4$, and
assume that $\omega$ is an $m_c$-connecting walk between $i$ and $j$ given $C$.

If $i$ or $j$ appears more than once on $\omega$, then we can find a
strictly shorter $m_c$-connecting walk, and we assume that $i$ and $j$ each
appear only once. If there are no colliders, then we can find a
$m_c$-connecting walk of length at most $d - 1$.

If $k$ is a collider on $\omega$, then $k \in C$. If it appears more than
once, then it is a collider in each instance, and we can find a strictly
shorter $m_c$-connecting walk.

If $l$ is a noncollider, then it is a noncollider in each instance on the
walk, and in each instance it must have a tail to the left or to the right
(having fixed the orientation of the walk such that $i$ is the left-most
node and $j$ is the right-most node). If $l$ appears three times or more,
then at least two instances both have tails on the left or on the right.
Concatenating the subwalks between $i$ and the first such $l$-instance and
the subwalk between the second such $l$-instance and $j$ will create a
strictly shorter $m_c$-connecting walk between $i$ and $j$ given $C$.

This means that we can find an $m_c$-connecting walk such that $i$, $j$, and
the collider $k$ appear at most once, and every other node appears at most
twice giving a total of $2(d- 3) + 3 = 2d - 3$ nodes, i.e., $2d - 4$ edges.
\end{proof}

\subsection{Proof of Lemma~\ref{lem:mconn}}

\begin{proof}
    Assume that $\omega$ is an $m$-connecting walk between $i$ and $j$ given
    $C$, $i \neq j$. If $\omega$ is not a path, then there must be a node $k$
    which appears at least twice on $\omega$. If $k$ is an endpoint of $\omega$,
    then there is a strictly shorter subwalk of $\omega$ which is $m$-connecting
    between $i$ and $j$ given $C$. Otherwise, we consider the concatenation of
    two subwalks of $\omega$: We let $\omega_1$ denote the subwalk between $i$
    and the first occurrence of $k$, and we let $\omega_2$ denote the subwalk
    between the last occurrence of $k$ and $j$. If $k$ is a collider on the
    concatenation of $\omega_1$ and $\omega_2$, then $k$ was either collider or
    an ancestor of a collider on $\omega$, and therefore $k\in \an(C)$. If $k$
    is a noncollider on the concatenation of $\omega_1$ and $\omega_2$, then it
    was a noncollider in at least one instance on $\omega$, and therefore
    $k\notin C$. In either case, we see that the concatenation of $\omega_1$ and
    $\omega_2$ is an $m$-connecting walk between $i$ and $j$ given $C$ which is
    strictly shorter than $\omega$. Repeating this argument, we can construct an
    $m$-connecting path between $i$ and $j$ given $C$.
\end{proof}

\subsection{Proof of Corollary~\ref{cor:dConndcConn}}

\begin{proof}
    If there is an $m$-connecting path between $i$ and $j$ given $C$, then there
    is also an $m_c$-connecting walk by Theorem \ref{thm:dc}. On the other hand,
    an $m_c$-connecting walk is also $m$-connecting, and it can be reduced to an
    $m$-connecting path (Lemma \ref{lem:mconn}). 
\end{proof}

\subsection{Proof of Lemma~\ref{lem:antDist}}

\begin{proof}
    \ref{bul:lDirdDir1} We assume that for all $i$ and $j$, $i\neq j$,
    $\ld{i}{j}$ equals the anterior distance from $i$ to $j$ in $G_\mathbf{x}$,
    and we show that $\{\ld{i}{j}\}$ in this case are a unique solution to
    \eqref{tag:N1} when $\mathbf{x}$ corresponds to $G_\mathbf{x}$. We do this
    by induction on the anterior distance, $\phi$. If $\phi = 1$ (that is, there
    is a descending path from $i$ to $j$ of length 1), we have $\xd{i}{j} = 1$.
    We see that $u_{ij}^{\text{D1}} = 1$, and that $\ld{i}{j} = 1$ equals the
    minimum in \eqref{tag:N1} such that \eqref{tag:N1} is satisfied for the pair
    $(i,j)$, and that this is the only possible solution for this pair. If
    $\phi>1$, there is a minimal-length descending path from $i$ to $j$ of
    length $\phi$, and we consider the penultimate node on this path, $k$, $k
    \neq i,j$. The subpath from $i$ to $k$ is descending and of length $\phi-1$,
    and $\xd{k}{j}  = 1$. Moreover, the subpath is of minimal length, and
    therefore $\ld{i}{k} = \phi - 1$ (by the induction assumption, this is the
    unique solution for the pair $(i,k)$). There can be no
    $u_{il}^{\text{D2}}<\phi - 1$ as this would create a strictly shorter
    descending path between $i$ and $j$ using the induction assumption. This
    means that $\ld{i}{j} = \phi$ satisfies \eqref{tag:N1}, and that it is the
    only value to do so. If $\phi = d$, then there is no descending path from
    $i$ to $j$, and $u_{ij}^{\text{D1}}=d$. By the induction assumption,
    $\lc{i}{k}{C}$ is the anterior distance for each $k$, and therefore
    $u_{ijk}^{\text{D2}}\geq d$ for each $k$. This means that $\ld{i}{j} = d$ is
    the only solution to \eqref{tag:N1}. 

    \ref{bul:lDirdDir2} By definition, we have $\ld{i}{j}= d$ if and only if
    there is no descending path from $i$ to $j$ in $G_\mathbf{x}$. We see that
    the stated solution satisfies \eqref{tag:C2}--\eqref{tag:C3}. Moreover, for
    any solution we see that if $\di{i}{j}$ satisfies
    \eqref{tag:C2}--\eqref{tag:C3}, then $\di{i}{j} = 0$ if and only if
    $\ld{i}{j} = d$ which means that the stated solution is the unique solution.

    \ref{bul:lDirdDir3} When $\mathbf{l}^\rightarrow$ are the anterior distances
    and \eqref{tag:C2}--\eqref{tag:C3} hold, then $\di{i}{j} = 1$ if and only if
    there is a descending path from $i$ to $j$ in $G_\mathbf{x}$. From this we
    see that $\di{i}{C}$ are the unique solution to
    \eqref{tag:R1a}--\eqref{tag:R1b}.
\end{proof}

\subsection{Proof of Theorem~\ref{thm:DAGs}}

\begin{proof}
\ref{bul:thmDAG1} By Lemma \ref{lem:antDist}, $\ld{i}{j}$ equals the anterior
distance from $i$ to $j$ in $G_{\mathbf{x}}$. Moreover, $\di{i}{j} = 1$
if no descending path exists from $i$ to $j$, and $\di{i}{j} = 0$ otherwise. 
Recall that, in
DGs, a path is descending if and only if it is directed.

If there is a directed cycle, there must exist $i$ and $j$, $i \neq j$, such
that $i \rightarrow \ldots \rightarrow j$ and $j \rightarrow \ldots
\rightarrow i$ in $G_\mathbf{x}$. Therefore, $\di{i}{j}  = \di{j}{i} = 0$,
and  \eqref{tag:AC} is violated for $i$ and $j$. If \eqref{tag:AC} is violated,
then there exist $i$ and $j$, $i\neq j$, such that $\di{i}{j} = \di{j}{i} =
0$. By concatenating the corresponding directed paths, we obtain a walk
starting and ending in $i$. We can reduce this to a directed cycle from $i$
to $i$, that is, $G_\mathbf{x}$ is not a DAG.

\ref{bul:thmDAG2}
We assume that $\mathbf{x}$ corresponds to the graph $G_\mathbf{x}$. For all
$i\in [d]$ and $C \subseteq [d]$, we have that $i$ is an ancestor of $C$ in $G$
if and only if $\di{i}{C} = 0$ (Lemma \ref{lem:antDist}). In this proof, we let
$\phic{i}{j}{C}$ denote the $d$-distance between $i$ and $j$ given $C$
in the graph $G_\mathbf{x}$. We define $\mathbf{l} = \{\lc{i}{j}{C}\}$ such
that $\lc{i}{j}{C} = \phic{i}{j}{C}$ for all $i,j,C$. We show that this is the
unique solution to \eqref{tag:M1} when $\mathbf{x}$ corresponds to
$G_\mathbf{x}$, and we show this by induction on the $d$-distance,
$\phi$.

If $\phi = 1$ (i.e, the $d$-distance between $i$ and $j$ given $C$ is
1), then $i$ and $j$ are adjacent. Therefore $\xd{i}{j} = 1$ or $\xd{j}{i} = 1$,
and $u_{ijC}^{\text{L1a$^\ast$}} = 1$ or $u_{ijC}^{\text{L1b$^\ast$}} = 1$. We
see that $\phic{i}{j}{C} = 1$ is the unique solution of \eqref{tag:M1} for the
triple $(i,j,C)$.

We now assume that $1 < \phi < d$, and we assume that whenever $\lc{i}{j}{C} <
\phi$, the $d$-distance between $i$ and $j$ given $C$ is the unique solution to
\eqref{tag:M1} for the triple $(i,j,C)$. This means that there is a
minimal-length
$d$-connecting path, $\pi$, between $i$ and $j$ given $C$ of length $\phi$, and
we first assume that this path has a head at $j$. In this case, there is a
$k\neq i,j$ such that there is a minimal-length $d$-connecting path of length
$\phi - 1$ between $i$ and $k$, $k\notin C$, and $\xd{k}{j} = 1$. Therefore
$\lc{i}{k}{C} = \phi - 1$, and $u_{ijkC}^{\text{L2a$^\ast$}} = \phi$. If there
is a head at $i$, we can argue similarly. Assume there is a tail at both
endpoints, $i \rightarrow \ldots \sim l \sim k \leftarrow j$. If $i = l$, then
$\phi = 2$, and we see that $u_{ijkC}^{\text{L3$^\ast$}} = 2$. Assume instead $i
\neq l$. If $k$ is a collider, then the path between $i$ and $l$ and the path
between $l$ and $j$ are both $d$-connecting, and $l \in \an(C)\setminus C$. Both
paths are minimal-length as $l\in\an(C)\setminus C$.  By Lemma
\ref{lem:antDist}, $\di{l}{C} = 0$. We see that $\lc{i}{l}{C} + \lc{l}{j}{C} =
\phi$, and therefore $u_{ijC}^{\text{L4a$^\ast$}} = \phi$. If $k$ is a
noncollider, then $k$ is an ancestor of some other collider on the path since
there are tails at both endpoints of the path. This collider is in $\an(C)$, and
therefore $k \in \an(C) \setminus C$. The subpaths that meet at $k$ are
minimal-length as $k\in\an(C)\setminus C$, and by Lemma~\ref{lem:antDist},
$\di{k}{C} = 0$. We can use that $\lc{i}{k}{C} + \lc{k}{j}{C} = \phi$ such that
$u_{ijC}^{\text{L4a$^\ast$}} = \phi$. In either case, we have shown that for the
triple $(i,j,C)$, there exists a $u$-variable in the minimum in \eqref{tag:M1}
which has the value $\phi$.

We will now argue that if any $u$-variable takes a value, $\psi$, such that
$\psi<\phi$, then there is a $d$-connecting path between $i$ and $j$ given $C$
of length $\psi$, leading to a contradiction. 
\begin{itemize}
    \item If $u_{ijC}^{\text{L2a$^\ast$}} = \psi$, then $\lc{i}{k}{C} = \psi -
    1$, and by the induction assumption $\lc{i}{k}{C} < \phi$ equals the
    $d$-distance between $i$ and $k$ given $C$, i.e., there is a $d$-connecting
    path between $i$ and $k$ given $C$ in $G_\mathbf{x}$ of length $\psi - 1$.
    Concatenating this with the edge $k \rightarrow j$ gives us a $d$-connecting
    walk between $i$ and $j$ given $C$ of length $\psi$ in $G_\mathbf{x}$. If
    $u_{ijC}^{\text{L2a$^\ast$}} = \psi$, we can argue similarly.
    \item If $u_{ijC}^{\text{L3$^\ast$}} = \psi$, then $\psi = 2$ and edges $i
    \rightarrow k$ and $j \rightarrow k$ are in $G_\mathbf{x}$. Moreover,
    $\di{k}{C} = 0$. Using Lemma \ref{lem:antDist}, this means that $k\in
    \an_{G_\mathbf{x}}(C)$, and therefore $i \rightarrow k \leftarrow j$ is
    $d$-connecting given $C$ and of length $\psi = 2$.
    \item If $u_{ijC}^{\text{L4a$^\ast$}} = \psi$, then $k\notin C$ and
    $\di{k}{C} = 0$. We have $\lc{i}{k}{C}, \lc{k}{j}{C} < \psi$, and
    $\lc{i}{k}{C} + \lc{k}{j}{C} = \psi$. By the induction assumption, there is
    a $d$-connecting path between $i$ and $k$ given $C$ of length $\lc{i}{k}{C}$
    and there is a $d$-connecting path between $k$ and $j$ given $C$ of length
    $\lc{k}{j}{C}$. Their concatenation is a $d$-connecting walk between $i$ and
    $j$ given $C$ of length $\psi$ as $\di{k}{C} = 0$ implies $k\in
    \an_{G_\mathbf{x}}(C)$ (Lemma \ref{lem:antDist}).
\end{itemize}
In each case above, we have argued that we can find a $d$-connecting
walk of length $\psi$, and these can be reduced to $d$-connecting paths of
length at most $\psi$. This contradicts the assumption that $\pi$ is of minimal
length, and we see that for the triple $(i,j,C)$, the minimum in \eqref{tag:M1}
takes the value $\phi$.

Finally, if $\phi = d$, there is no $d$-connecting path between $i$ and $j$
given $C$. From the above, it follows that there is no $u$-variable which is
strictly less than $d$ as this would imply the existence of a $d$-connecting
path. Therefore, $\phi = d$ satisfies \eqref{tag:M1} for the triple $(i,j,C)$.
\end{proof}

\subsection{Proof of Lemma~\ref{lem:biDist}}

\begin{proof}
We assume that $\mathbf{x} \coloneqq (\mathbf{x}^\rightarrow,
\mathbf{x}^\leftrightarrow)$ corresponds to $G_\mathbf{x}$. We show that the
bidirected distances are the unique solution to \eqref{tag:O1}, and we show this
by induction on $\phi$ (the  bidirected distance between $i$ and $j$ relative to
$C$ in $G_\mathbf{x}$). 

If $\phi = 1$, then $\xb{i}{j} = 1$ and $i,j\in C$. This means that the 
bidirected distance between $i$ and $j$ relative to $C$ is 1.  If the 
bidirected distance between $i$ and $j$ relative to $C$ is $1<\phi<d$, then
there is a $k\in C$ such that there the  bidirected distance between $i$
and $k$ relative to $C$ is $\phi - 1$ and such that $\xb{j}{k} = 1$. By the
induction assumption, $\lcb{i}{j}{C}$ equals the  bidirected distance
between $i$ and $k$ relative to $C$. Moreover, there cannot be an $l\in C$ such
that $\lcb{i}{l}{C} < \phi - 1$ and $\xb{l}{j} = 1$ as this would create a
contradiction to the fact that $\phi$ is the 
bidirected distance between $i$ and $j$ relative to $C$ (using the induction
assumption). This means that for $i,j,C$, the minimum in \eqref{tag:O1} equals
$\phi$, and that this is the unique solution for the triple $(i,j,C)$. 

If $\phi = d$, then $\xb{i}{j} = 0$. We have shown that the  bidirected
distances solve \eqref{tag:O1} for $\phi < d$. If there is a $k\in C$ such that
$\lcb{i}{k}{C} < d-1$ and $\xb{j}{k} = 1$, there is a bidirected path between
$i$ and $k$ such that all nodes are in $C$ of length strictly less than $d-1$ and
there is an edge $j\leftarrow k$ which creates a contradiction. This means that
the minimum in \eqref{tag:O1} is $\phi = d$, and this is the unique solution for
the triple $(i,j,C)$. 
\end{proof}

\subsection{Proof of Lemma~\ref{lem:semibiDist}}

\begin{proof}
If the  semi-bidirected distance from $i$ to $j$ relative to $C$ is 1,
then there is $\xd{i}{j} = 1$ or $\xb{i}{j} = 1$, and $i\notin C, j \in C$.
Using Lemma \ref{lem:consistDMG}, we have $\xs{i}{j} = 1$, and
$u_{ij}^{\text{E1}} = 1$. This means that $\lcd{i}{j}{C} = 1$ is the unique
solution of \eqref{tag:G1} for the triple $(i,j,C)$.

If the  semi-bidirected distance from $i$ to $j$ relative to $C$ is $1 <
\phi < d$, then $i \notin C, j \in C$, and there exists $k\in C$ such that
$\xs{i}{k} = 1$, and such that the bidirected distance between $i$ and $j$
relative to $C$ is $\phi - 1$. By Lemma \ref{lem:biDist}, we have $\lcb{k}{j}{C}
= \phi - 1$. We see that there can be no $l\in C$ such that $\lcd{l}{j}{C} <
\phi -1$ and $\xs{i}{l} = 1$ as this would this would make the 
semi-bidirected distance from $i$ to $j$ relative to $C$ strictly less than
$\phi$ (using Lemma \ref{lem:biDist} again).

If the  semi-bidirected distance from $i$ to $j$ relative to $C$ is $\phi = d$,
then $u_{ij}^{\text{E1}} = d$. By Lemma \ref{lem:biDist}, there is no $k\in C$
such that $\lcb{k}{j}{C} < d-1$ and $\xs{i}{k} = 1$, and therefore,
$\lcd{i}{j}{C} =d $ is the unique solution of \eqref{tag:G1} for the triple
$(i,j,C)$. 
\end{proof}

\subsection{Proof of Theorem~\ref{thm:DMGs}}

\begin{proof}
    \ref{bul:thmDMG1} This is proven by using Lemma \ref{lem:antDist} and
    arguing as in the proof of Theorem \ref{thm:DAGs}\ref{bul:thmDAG1}.

    \ref{bul:thmDMG2} We assume that $\mathbf{x} = (\mathbf{x}^\rightarrow,
\mathbf{x}^\leftrightarrow)$ corresponds to $G_\mathbf{x}$.
We will use the following facts. By Lemma \ref{lem:antDist}, we have that
$\di{i}{C} = 0$ if and only if $i \in \an_{G_\mathbf{x}}(C)$. From Lemma
\ref{lem:consistDMG}, we have that $\xs{i}{j} = 0$ if and only if $\xd{i}{j}=0$
and $\xb{i}{j}= 0$. By Lemma \ref{lem:biDist}, we have that $\lcb{i}{j}{C}$
equals the  bidirected distance between $i$ and $j$ relative to
$C$. Finally, $\lcd{i}{j}{C}$ equals the  semi-bidirected
distance from $i$ to $j$ relative to $C$ (Lemmas \ref{lem:consistDMG} and \ref{lem:semibiDist}).

We will use induction on $\phi$, the $m$-distance between $i$ and $j$
given $C$, to show that the $m$-distances are the unique solution. If
$\phi = 1$, then there is an $m$-connecting path between $i$ and $j$ given $C$ of
length $\phi = 1$, and $\xs{i}{j} = 1$ or $\xs{j}{i} = 1$. We see that
$u_{ijC}^{\text{K1a}} = 1$ or $u_{ijC}^{\text{K1b}} = 1$, and we see that $\phi$
is the unique solution of \eqref{tag:P1} for the triple $(i,j,C)$.

Assume now that $1 < \phi < d$, and that $\pi$ is an $m$-connecting path between
$i$ and $j$ given $C$ of length $\phi$. 
\begin{itemize}
    \item If there is a directed edge with a head at $j$,  $i \sim \ldots \sim k
    \to j$, then $k \notin C$, and there is a minimal-length $m$-connecting path
    between $i$ and $k$ given $C$ of length $\phi -1 $ and $\xd{k}{j} = 1$. By
    the induction assumption, $\lc{i}{k}{C} = \phi- 1$, and we see that
    $u_{ijkC}^{\text{K2a}} = \phi$. 
    \item Otherwise, if there are no noncolliders on $\pi$ and every collider on
    $\pi$ is in $C$, there exists a node $k\neq i,j$ on $\pi$ such that $k \in
    C$. The subpath of $\pi$ between $i$ and $k$ is semi-bidirected and its
    length equals $\lcd{i}{j}{C}$. Similarly, there is a semi-bidirected path
    from $j$ to $k$ of length $\lcd{j}{k}{C}$, and $\lcd{i}{j}{C} +
    \lcd{j}{k}{C} = \phi$. This means that $u_{ijC}^{\text{K3}} = \phi$ for the
    triple $(i,j,C)$. 
    \item If the path has no noncolliders, and there is a collider $k$ on the
    path such that $k\notin C$, then the $m$-distance between $i$ and $k$ given
    $C$, $\phi_1$, and the $m$-distance between $k$ and $j$ given $C$,
    $\phi_2$, are such that $\phi_1 + \phi_2 = \phi$. By  the induction
    assumption $\lc{i}{k}{C}= \phi_1$ and $\lc{k}{j}{C}=\phi_2$. We see that
    $u_{ijC}^{\text{K4}} = \phi$ as $k \in \an(C)$ and therefore $\di{k}{C} =
    0$. 
    \item If $\pi$ has a noncollider, $k$, then there is a directed subpath from
    $k$ to some $l$, $k \neq l$, and we consider such a subpath of maximal
    length. The path $\pi$ has neither a directed edge with a head at $i$ nor a
    directed edge with a head at $j$, and therefore $l\neq i,j$. The directed
    subpath from $k$ to $l$ is of maximal length, and $l$ is therefore a
    collider. We have that $l\in \an(C)$, and therefore $k$ must be an ancestor
    of $C$ in $G_\mathbf{x}$. As before $k\notin \an(C)\setminus C$, and we can
    therefore argue that $u_{ijC}^{\text{K4}}= \phi$.
\end{itemize}
In every case above, we have argued that there is as $u$-variable which
takes the value $\phi$ when the $m$-distance between $i$ and $j$ given $C$ is
$\phi$, $1 < \phi < d$.

We will now argue that no $u$-variable can take a smaller value, $\psi < \phi$.
\begin{itemize}
    \item If $u_{ijC}^{\text{K1a}} = \psi$ or $u_{ijC}^{\text{K1b}} = \psi$,
    then $\psi = 1$ and $\xs{i}{j} = 1$ or $\xs{j}{i} = 1$. This implies that
    $\xd{i}{j} = 1$, $\xd{j}{i}  =1$, or $\xb{i}{j}= 1$ such that $i$ and $j$
    are adjacent.
    \item If $u_{ijkC}^{\text{K2a}} = \psi$, then by the induction assumption
    $\lc{i}{k}{C}$ is the $m$-distance between $i$ and $k$ given $C$ as
    $\lc{i}{k}{C} = \psi-1<\phi$, and furthermore $\xd{k}{j}=1$. The
    concatenation of the $m$-connecting path of length $\psi-1$ between $i$ and
    $k$ given $C$ and the edge $k\rightarrow j$ is an $m$-connecting walk
    between $i$ and $j$ given $C$ in $G_\mathbf{x}$. If $u_{ijkC}^{\text{K2b}} =
    \psi$, then a similar argument works.
    \item If $u_{ijkC}^{\text{K3}} = \psi$, then $\lcd{i}{k}{C} + \lcd{j}{k}{C}
    = \psi$. Therefore, there are semi-bidirected paths from $i$ to $k$ relative
    to $C$ and from $j$ to $k$ relative $C$, and their concatenation is an
    $m$-connecting walk between $i$ and $j$ given $C$ of length $\psi$.
    \item If $u_{ijkC}^{\text{K4}} = \psi$, we that $\di{k}{C} = 0$ and
    $\lc{i}{k}{C} + \lc{k}{j}{C}  = \psi$. By the induction assumption
    $\lc{i}{k}{C} < \psi$ is the $m$-distance between $i$ and $k$ given $C$ in
    $G_\mathbf{x}$ and $\lc{k}{j}{C} <\psi$ is the $m$-distance between $k$ and
    $j$ given $C$ in $G_\mathbf{x}$, and there exist an $m$-connecting path
    between $i$ and $k$ given $C$ of length $\lc{i}{k}{C}$ and an $m$-connecting
    path between $k$ and $j$ given $C$ of length $\lc{k}{j}{C}$, respectively.
    We have $\di{k}{C} = 0$, i.e., $k\in\an_{G_\mathbf{x}}(C)$. Therefore, the
    concatenation of the paths is an $m$-connecting walk between $i$ and $j$
    given $C$ of length $\psi$.
\end{itemize}
In each case above, we have found an $m$-connecting walk between $i$ and
$j$ given $C$ of length at most $\psi$. Such a walk can be reduced to an
$m$-connecting path between $i$ and $j$ given $C$ of length at most $\psi$. This
contradicts the assumption that $\pi$ is of minimal length, and we see that
$\phi$ is the unique value which satisfies \eqref{tag:P1}.

If $\phi = d$, there is no $m$-connecting path between $i$ and $j$ given $C$.
The above argument shows there is no $u$-variable which is strictly less than
$d$, and therefore $\lc{i}{j}{C} = d$ is the unique solution of \eqref{tag:P1}
for the triple $(i,j,C)$.
\end{proof}

\subsection{Proof of Lemma~\ref{lem:charDirCycle}}

\begin{proof}
If there is a partially directed cycle, there exist $i,j$ on this cycle such
that $i \rightarrow j$, and therefore $\ld{i}{j} = 1$. Edges $j \rightarrow i$
and $i \mathdash j$ are not in the graph, thus $l_{ji}^\rightarrow > 1$. On the
other hand, there is a descending path from $j$ to $i$, i.e.,
$l_{ji}^\rightarrow < d$.

If $i$ and $j$ are such that $\ld{i}{j} \neq l_{ji}^\rightarrow$ and $\ld{i}{j},
l_{ji}^\rightarrow \neq d$, then there exist descending paths from $i$ to $j$
and from $j$ to $i$. The shorter of the two cannot be undirected as this would
make it descending in the opposite direction as well, contradicting $\ld{i}{j}
\neq \ld{i}{j}$. If we concatenate the two descending paths, we can obtain a
descending walk starting and ending in $i$. As argued above, there is a directed
edge on this walk, say $k \rightarrow l$, and we can consider the walk as a
concatenation of this edge and a descending walk from $l$ to $k$. The descending
walk from $l$ to $k$ can be reduced to a descending path. The concatenation of
this path with the edge $k\rightarrow l$ creates a partially directed cycle.
\end{proof}

\subsection{Proof of Lemma~\ref{lem:isCG}}

\begin{proof}
    By Lemma \ref{lem:antDist}, we have that $\ld{i}{j}$ is the anterior
    distance in $G_\mathbf{x}$ for each $i,j\in [d]$. By the same lemma, we also
    have that $\di{i}{j} = 1$ if and only if $\ld{i}{j} = d$. Assume first that
    $G_\mathbf{x}$ is not a chain graph. By Lemma \ref{lem:charDirCycle}, there
    exists $i,j$ such that $\ld{i}{j} \neq \ld{j}{i}$ and $\ld{i}{j},\ld{j}{i}
    \neq d$. We see that $\di{i}{j} =\di{j}{i} = 0$, and therefore one of
    \eqref{tag:CH1a} or \eqref{tag:CH1b} must be violated. On the other hand if
    one of  \eqref{tag:CH1a} or \eqref{tag:CH1b} are violated, we see that for
    some $i,j$, $\ld{i}{j} \neq \ld{j}{i}$ and $\di{i}{j}  = \di{j}{i} = 0$, and
    the result follows from Lemma \ref{lem:charDirCycle}.
\end{proof}

\subsection{Proof of Lemma~\ref{lem:unDist}}

\begin{proof}
We show that the undirected distances are the unique solution of \eqref{tag:W1}
by induction on the undirected distance, $\phi$.

If $\phi = 1$, then $\xd{i}{j} = \xd{j}{i} = 1$, and $u_{ij}^{\text{U1}} = 1$.
We see that $\ld{i}{j}$ is the unique solution. We assume now that $1<\phi<d$,
and that the undirected distances are the unique solution for each pair $(i,j)$
such that the undirected distance between $i$ and $j$ is strictly less than
$\phi$. As $\phi > 1$, there exists a node $k$ on the path such
that $k\neq i,j$, and the  undirected distance between $i$ and
$k$ is $\phi - 1$, and 
$\xd{k}{j} = \xd{j}{k} = 1$. Using the induction
assumption, $\ld{i}{k} = \phi -1$, and $u_{ijk}^{\text{U2}} = \phi$. Moreover,
no $u$-variable can be smaller as this would imply the existence of an
undirected path between $i$ and $j$ which is strictly shorter than $\lu{i}{j}$
(using the induction assumption). If $\lu{i}{j} = d$, then there is no
undirected path between $i$ and $j$, and we see that
$u_{ij}^{\text{U1}},u_{ijk}^{\text{U2}}\geq d$. 
\end{proof}

\subsection{Proof of Proposition~\ref{prop:slides}}

\begin{proof}
From Lemma~\ref{lem:unDist}, $\lu{i}{j}$ is the undirected distance between $i$
and $j$ for each $i<j$, and from Corollary~\ref{cor:chaincompindic}, $\zu{i}{j}
= 1$ if and only if there is an undirected path between $i$ and $j$ in
$G_\mathbf{x}$.

We show first that a solution exists. This is obvious as $\xs{i}{j}$ does not
appear in the expressions that define $u_{ij}^{\text{Y1}}$ and
$u_{ijk}^{\text{Y2}}$. Assume that $\mathbf{x}^\ast$ is any solution.    If
$\xs{i}{j} = 1$, then $u_{ij}^{\text{Y}1} = 1$ or $u_{ijk}^{\text{Y}2} = 1$ for
some $k \neq i,j$. If $u_{ij}^{\text{Y}1} = 1$, then $\xd{i}{j} = 1$ and
$\xd{j}{i} = 0$ and there is a slide from $i$ to $j$ in $G$. If
$u_{ijk}^{\text{Y}2} = 1$ for some $k \neq i,j$, then $\xd{i}{k} = 1$,
$\xd{k}{i} = 0$, and $\zu{k}{j} = 1$. By Lemma~\ref{lem:unDist} and
Corollary~\ref{cor:chaincompindic}, $\zu{k}{j} = 1$ implies that there is an
undirected path between $k$ and $j$, and therefore there is a slide from $i$ to
$j$.

If there is a slide from $i$ to $j$ in $G$, then it is either a single edge
in which case \eqref{tag:Z1} implies that $\xs{i}{j} = 1$. Otherwise there
exists a $k$, $k\neq i,j$ such that $i \rightarrow k$ and $k$ and $j$ are in
the same chain component. Lemma \ref{lem:unDist} and Corollary
\ref{cor:chaincompindic} imply that $\zu{k}{j} = 1$, and $\xs{i}{j} = 1$. We
see that $\xs{i}{j} = 0$ implies that there is no slide from $i$ to $j$ in
$G$. In conclusion, the set $\mathbf{x}^\ast$ described in the
proposition is the unique solution to \eqref{tag:Z1}.
\end{proof}

\subsection{Proof of Lemma~\ref{lem:decomp}}

\begin{proof}
Let $\pi$ be a connecting path between $i$ and $j$ given $C$ in $(G_{\an(\{i,j\}\cup
C)})^m$. We order the vertices such that $i$ is the first, and $j$ is the last.
We let $k$ be the last vertex on the path such that $k\in \an(i)\setminus
\an(C)$ if such a vertex exists, and otherwise we let $k=i$. On the subpath
between $k$ and $j$, we let $l$ be the first vertex on the subpath such that
$l\in \an(j)\setminus \an(C)$ if such a vertex exists, and otherwise we let $l =
j$.

The subpath of $\pi$ between $k$ and $l$ is in $(G_{\an(\{i,j\}\cup C)})^m$, and
every nonendpoint node on this path is in $\an(C)$. Moreover, $k \in \an(i)$ and
$l \in \an(j)$. The nodes $k$ and $l$ are not in $\an(C)$, and this means that,
in $G$, there is a descending path from $k$ to $i$ and a descending path from
$l$ to $j$ such that no node is in $C$. We can concatenate the three paths to
create a connecting walk in $(G_{\an(\{i,j\}\cup C)})^m$. This walk is
decomposable, and we can reduce this to a connecting and decomposable path.
\end{proof}

\subsection{Proof of Theorem~\ref{thm:CGs}}

Recall that $\mathcal{T}(G)$ denotes the set of chain components of a hybrid
graph $G$, and that $\tau\in \mathcal{T}(G)$ is a node set, not an undirected
graph.

\begin{proof}
    \ref{bul:thmCG1} Constraints \eqref{tag:C2}--\eqref{tag:C3}, and \eqref{tag:N1}
    are satisfied by assumption. Lemma \ref{lem:isCG} gives the result.

    \ref{bul:thmCG2} We will show by induction that the decomposable distances
    are the unique solution of~\eqref{tag:Q1}. We use induction on the
    decomposable distance, $\phi$.

If $\phi = 1$, $i$ and $j$ are either adjacent or parents of the same chain
component $\tau$ such that $\tau \subseteq \an(C)$. In the first case,
$u_{ijC}^{\text{I1a}} = 1$ or $u_{ijC}^{\text{I1b}} = 1$. In the second case,
Proposition~\ref{prop:slides} shows that there exists a $k$ such that $\xs{i}{k}
= \xs{j}{k} = 1$. By Lemma~\ref{lem:antDist}, $\di{k}{C} = 0$, and we see that
$u_{ijkC}^{\text{I3}} = 1$. 

Assume now that $1 <\phi<d$, and we assume that there is a connecting and
decomposable path $\pi$ of length $\phi$ between $i$ and $j$ in the moral graph
of $G_{\an( \{i,j\}\cup C)}$. Assume that $\pi$ is of the form $i \sim \ldots
\sim k \sim j$ and that $\xd{k}{j} = 1$. The subpath between $i$ and $k$ is
connecting in $(G_{\an( \{i,k\}\cup C)})^m$ as $\pi$ is decomposable. The
decomposable distance between $i$ and $k$ relative to $C$ cannot be less than
$\phi - 1$ as this would create a contradiction to $\pi$ being minimal. This
means that the decomposable distance between $i$ and $k$ relative to $C$ must be
$\phi - 1$, and by the induction assumption $\lc{i}{k}{C} = \phi - 1$. It
follows that $u_{ijkC}^{\text{I2a}} = \phi$. If instead $i \sim k \sim \ldots
\sim j$ and $\xd{k}{i} = 1$, a similar argument applies.

Assume that that the edges that are adjacent to $i$ and to $j$ both correspond
to directed edges with tails at $i$ and at $j$ or that they are both moral
edges. In this case, there must exist a moral edge on the path. Assume that $l -
h$ is a moral edge. The path $\pi$ is decomposable, and therefore its subpath
$\pi(l,h)$ is in $(G_{\an(\{l,h\}\cup C)})^m$. This subpath equals the edge $l -
h$, and there must exists $\tau \in \mathcal{T}(G)$ such that $l,h \in
\bd(\tau)$ and such that $\tau \in \an(C)$. We see that $l,h\in \an(C)$, and one
of them must be different from $i$ and $j$. We let $k$ denote $l$ or $h$ such
that $k\neq i,j$. We see that $k \in \an(C)$. The subpaths of $\pi$ between $i$
and $k$ and between $k$ and $j$ are connecting  in $(G_{\an(\{i,k\}\cup C)})^m$
and $(G_{\an(\{k,j\}\cup C)})^m$ respectively, and by the induction assumption
$\lc{i}{k}{C} + \lc{k}{j}{C} = \phi$.

In each case above, we have found a $u$-variable which takes the value $\phi$
for the triple $(i,j,C)$.

We will now argue that no $u$-variable can take a strictly smaller value. 
\begin{itemize}
    \item Assume that $\psi < \phi$. If $\psi = 1$, then $\xd{i}{j} = 1$,
    $\xd{j}{i} = 1$ or $u_{ijkC}^{\text{I3}} = 1$. If $\xd{i}{j} = 1$ or
    $\xd{j}{i} = 1$, we see that $i - j$ is in $((G_\mathbf{x})_{\an(\{i,j\}\cup
    C)})^m$. If $u_{ijkC}^{\text{I3}} = 1$, then $\xs{i}{k} = \xs{j}{k} = 1$ and
    $\di{k}{C} = 0$. Proposition~\ref{prop:slides} shows that there exists a
    slide from $i$ to $k$ and a slide from $j$ to $k$. By
    Lemma~\ref{lem:antDist}, $\di{k}{C} = 0$ implies $k \in \an(C)$. This means that 
    $i,j\in \bd(\tau_k)$ where $\tau_k$ is the chain component of $k$. The node
    $k$ is anterior to $C$, and this means that $i$ and $j$ are adjacent in
    $((G_\mathbf{x})_{\an(\{i,j\}\cup C)})^m$.
    \item If $u_{ijkC}^{\text{I2a}} = \psi$, then $\lc{i}{k}{C} = \psi - 1$ and by the
    induction hypothesis, this means that there is a connecting and decomposable
    path between $i$ and $k$ given $C$ in $(G_\mathbf{x})_{\an(\{i,k\}\cup C)}^m$.
    We have $\xd{k}{j} = 1$ and therefore $\an(\{i,k\}\cup C) \subseteq
    \an(\{i,j\}\cup C)$. This means that the path between $i$ and $k$ is also in
    $((G_\mathbf{x})_{\an(\{i,j\}\cup C)})^m$. Concatenating it with the edge $k -
    j$ gives a connecting and decomposable path between $i$ and $j$ given $C$ in
    $((G_\mathbf{x})_{\an(\{i,j\}\cup C)})^m$. If $u_{ijkC}^{\text{I2b}} = \psi$, a
    similar argument applies.
    \item If $u_{ijkC}^{\text{I4}} = \psi$, then $\lc{i}{k}{C} + \lc{k}{j}{C} =
    \psi$ and $\di{k}{C} = 0$. By the induction assumption, there is a connecting and
    decomposable path between $i$ and $k$ in $((G_\mathbf{x})_{\an(\{i,k\}\cup
    C)})^m$ of length $\psi_1$ and a connecting and decomposable path between $k$
    and $j$ in $((G_\mathbf{x})_{\an(\{k,j\}\cup C)})^m$ of length $\psi_2$ such
    that $\psi_1 + \psi_2 = \psi$. Their concatenation is a connecting and
    decomposable path in $((G_\mathbf{x})_{\an(\{i,j\}\cup C)})^m$.
\end{itemize}
In each case, we have shown that the existence of a $u$-variable which takes the
value $\psi < \phi$ creates a contradiction to $\pi$ being of minimal length.
Therefore, $\phi$ is the unique solution to \eqref{tag:Q1} for the triple
$(i,j,C)$.

Finally, assume that $\phi = d$. The above shows that any $u$-variable strictly
less than $d$ would imply the existence of a connecting path, and therefore $d$
is the unique solution to \eqref{tag:Q1}.
\end{proof}

\subsection{Proof of Theorem~\ref{thm:optim}}

\begin{proof}
We assume throughout that $G_{\hat{\mathbf{x}}}$ corresponds to
$\hat{\mathbf{x}}$. We will give a complete argument in the case where
$\mathbb{G}_d$ is the set of DAGs on $d$ nodes and using $d$-separation. The
other cases follow from the analogous results (see the end of this proof).

Let $\mathbb{G}_d$ be the set of DAGs on $d$ nodes. We need to show that 1)
every DAG is represented by a vector $\mathbf{y}_{\mathbb{G}_d}$ in the feasible
region of~\eqref{tag:IPG}, 2) every $\mathbf{y}_{\mathbb{G}_d}$ in the feasible
region of~\eqref{tag:IPG} corresponds to a DAG, and 3) each DAG $G$ is
scored correctly in the sense that the $\zc{i}{j}{C}$-variables correspond to the
$d$-separations in $G$. We start by considering a specific DAG, $G$,
to show that it is feasible and scored correctly. The graph $G$ defines a vector
of edge variables, $\mathbf{x}$, that we consider fixed. 

1) We see from Lemma \ref{lem:antDist}\ref{bul:lDirdDir1}, that \eqref{tag:N1}
is satisfied if and only if $\mathbf{l}^\rightarrow$ are the anterior distances
in $G$. When $\mathbf{l}^\rightarrow$ are the anterior distances, Lemma
\ref{lem:antDist} also gives that there is a unique solution
$\mathbf{d}^{\not\rightarrow}$ to \eqref{tag:C2}--\eqref{tag:C3} (Lemma
\ref{lem:antDist}\ref{bul:lDirdDir2}) and to \eqref{tag:R1a}--\eqref{tag:R1b}
(Lemma \ref{lem:antDist}\ref{bul:lDirdDir3}). From Theorem \ref{thm:DAGs}, we
see that \eqref{tag:AC} is satisfied when $\mathbf{x}$ defines a DAG. We let
$\mathbf{l}^d$ denote the $d$-distances in $G$. By Theorem \ref{thm:DAGs},
$\mathbf{l}^d$ is the unique solution to \eqref{tag:M1}.  Define now
$\{\zc{i}{j}{C} \}$ such that $\zc{i}{j}{C} = 0$ if and only if $\lc{i}{j}{C} =
d$ for each $i,j,C$. The set $\{\zc{i}{j}{C}\}$  is the unique solution to
\eqref{tag:C4}--\eqref{tag:C5} when $\mathbf{l}^d$ are the $d$-distances
(Lemma \ref{lem:consist}). Jointly, this implies that there is a feasible vector
$\mathbf{y}_{\mathbb{G}_d}$ such that its edge variables correspond to $G$, and
such that the subvector $\{\zc{i}{j}{C}\}$ encodes the $d$-separations and the
$d$-connections of $G$. This proves 1) and 3).

2) We assume that $G$ is a DG, but not a DAG, and we assume that $\mathbf{x}$
corresponds to $G$. If there are no values of $\mathbf{l}_{\mathbb{G}_d}$,
$\mathbf{z}_{\mathbb{G}_d}$, and $\mathbf{w}_{\mathbb{G}_d}$ such that
constraints \eqref{tag:C2}--\eqref{tag:C3}, \eqref{tag:N1}, and
\eqref{tag:R1a}--\eqref{tag:R1b} are satisfied, then this DG cannot be
represented in the feasible region. Assume instead that these constraints are
satisfied. It now follows from Theorem~\ref{thm:DAGs} that the~\eqref{tag:AC}
constraint is violated.

When $\mathbb{G}_d$ is the set of DGs on $d$ nodes using $d$-separation, the
proof is analogous to the above when omitting the \eqref{tag:AC} constraint.

When $\mathbb{G}_d$ is the set of ADMGs on $d$ nodes using $m$-separation, a
similar proof applies, using Theorem \ref{thm:DMGs}. This also applies to DMGs
using $m$-separation when omitting \eqref{tag:AC}.

When $\mathbb{G}_d$ is the set of chain graphs on $d$ nodes, an analogous proof
holds (using Theorem~\ref{thm:CGs}).

When $\mathbb{G}_d$ is the set of DGs/DAGs on $d$ using $d_c$-separation, we can
apply Theorem~\ref{thm:DAGsdc}, and when $\mathbb{G}_d$ is the set of DMGs/ADMGs
on $d$ using $m_c$-separation, we can apply Theorem~\ref{thm:DMGsdc} (see
Appendix~\ref{app:encodings}).
\end{proof}

\subsection{Proof of Theorem~\ref{thm:DAGsdc}}

\begin{proof}
\ref{bul:thmDAGdc1} The proof of Theorem \ref{thm:DAGs}\ref{bul:thmDAG1} also
applies in this case.

\ref{bul:thmDAGdc2}
We assume that $\mathbf{x}$ corresponds to the graph $G_\mathbf{x}$. We let
$\phic{i}{j}{C}$ denote the $d_c$-distance between $i$ and $j$ given $C$
in the graph $G_\mathbf{x}$.  We define $\mathbf{l} = \{\lc{i}{j}{C}\}$ such
that $\lc{i}{j}{C} = \phic{i}{j}{C}$ for all $i,j,C$. We show that this is the
unique solution to \eqref{tag:M1c} by induction on the $d_c$-distance,
$\phi$.

If $\phi = 1$, then $i$ and $j$ are adjacent. Therefore $\xd{i}{j} = 1$ or $\xd{j}{i} = 1$,
and $u_{ijC}^{\text{L1a$^\text{c}$}} = 1$ or $u_{ijC}^{\text{L1b$^\text{c}$}} = 1$. We
see that $\phic{i}{j}{C} = 1$ is the unique solution of \eqref{tag:M1c} for the
triple $(i,j,C)$.

We now assume that $1 < \phi < \tilde{n} + 1$. This means that there is a
minimal-length $d_c$-connecting walk, $\omega$, between $i$ and $j$ given $C$ of
length $\phi$, and we first assume that this walk has a head at $j$. In this
case, there is a $k\neq i,j$ such that there is a minimal-length
$d_c$-connecting walk of length $\phi - 1$ between $i$ and $k$, $k\notin C$, and
$\xd{k}{j} = 1$. Therefore $\lc{i}{k}{C} = \phi - 1$, and
$u_{ijkC}^{\text{L2a$^\text{c}$}} = \phi$. If there is a head at $i$, we can
argue similarly. Assume there is a tail at both endpoints, $i \rightarrow \ldots
\sim l \sim k \leftarrow j$. If $i = l$, then $\phi = 2$, and we see that
$u_{ijkC}^{\text{L3$^\text{c}$}} = 2$. Assume instead $i \neq l$. If $k$ is a
collider, then the path between $i$ and $l$ and the path between $l$ and $j$ are
both $d_c$-connecting. We see that $\lc{i}{l}{C} + \lc{l}{j}{C} \leq \phi$, and
therefore $u_{ijC}^{\text{L4a$^\text{c}$}} \leq \phi$. If $k$ is a noncollider,
and $i \rightarrow h \sim \ldots l \sim k \leftarrow j$ such that $h$ is a
collider, then we can argue analogously using \ref{tag:L4ac}. If neither $h$ nor
$k$ is a collider, then we can find some other collider, and argue using
\ref{tag:L5c}. In either case, we have shown that for the triple $(i,j,C)$,
there exists a $u$-variable in the minimum in \eqref{tag:M1c} which has the
value $\phi$.

We will now argue that if any $u$-variable takes a value, $\psi$, such that
$\psi<\phi$, then there is a $d_c$-connecting walk between $i$ and $j$ given $C$
of length $\psi$, leading to a contradiction. 
\begin{itemize}
    \item If $u_{ijC}^{\text{L2a$^\text{c}$}} = \psi$, then $\lc{i}{k}{C} = \psi
    - 1$, and by the induction assumption $\lc{i}{k}{C} < \phi$ equals the
    $d_c$-distance between $i$ and $k$ given $C$, i.e., there is a
    $d_c$-connecting walk between $i$ and $k$ given $C$ in $G_\mathbf{x}$ of
    length $\psi - 1$. Concatenating this with the edge $k \rightarrow j$ gives
    us a $d_c$-connecting walk between $i$ and $j$ given $C$ of length $\psi$ in
    $G_\mathbf{x}$. If $u_{ijC}^{\text{L2a$^\ast$}} = \psi$, we can argue
    similarly.
    \item If $u_{ijC}^{\text{L3$^\text{c}$}} = \psi$, then $\psi = 2$ and edges
    $i \rightarrow k$ and $j \rightarrow k$ are in $G_\mathbf{x}$. Moreover,
    $\di{k}{C} = 0$. Using Lemma \ref{lem:antDist}, this means that $k\in
    \an_{G_\mathbf{x}}(C)$, and therefore $i \rightarrow k \leftarrow j$ is
    $d_c$-connecting given $C$ and of length $\psi = 2$.
    \item If $u_{ijC}^{\text{L4a$^\text{c}$}} = \psi$, then $k\in C$ and
    $l\notin C$. We have $\lc{j}{l}{C} < \psi$, and $\lc{j}{l}{C} + 2 = \psi$.
    By the induction assumption, there is a $d_c$-connecting walk between $j$
    and $l$ given $C$ of length $\lc{j}{l}{C}$, and furthermore $\xd{i}{k} =
    \xd{l}{k} = 1$. Their concatenation is a $d_c$-connecting walk between $i$
    and $j$ given $C$ of length $\psi$. If $u_{ijC}^{\text{L4b$^\text{c}$}} =
    \psi$, we can argue analogously.
    \item If $u_{ijC}^{\text{L5$^\text{c}$}} = \psi$, then
    $\lc{i}{l}{C},\lc{m}{j}{C} < \phi$ and $\xd{l}{k} = \xd{m}{k} = 1$ where
    $l,m\notin C$ and $k\in C$. By the induction assumption there are
    $d_c$-connecting walks between $i$ and $l$ given $C$ and between $m$ and $j$
    given $C$. Their concatenation with the edges $l\rightarrow k$ and
    $m\rightarrow k$ is a $d_c$-connecting walk between $i$ and $j$ given $C$ in
    $G_\mathbf{x}$. 
\end{itemize}
In each case above, we have argued that we can find a $d_c$-connecting
walk of length $\psi$. This contradicts the assumption that $\omega$ is of minimal
length, and we see that for the triple $(i,j,C)$, the minimum in \eqref{tag:M1}
takes the value $\phi$.

Finally, if $\phi = \tilde{n} + 1$, there is no $d_c$-connecting path between
$i$ and $j$ given $C$. From the above, it follows that there is no $u$-variable
which is strictly less than $\tilde{n} + 1$ as this would imply the existence of
a $d_c$-connecting path. Therefore, $\phi = \tilde{n} + 1$ satisfies
\eqref{tag:M1} for the triple $(i,j,C)$.
\end{proof}

\subsection{Proof of Theorem~\ref{thm:DMGsdc}}

\begin{proof} \ref{bul:thmDMGdc1} The proof of Theorem
\ref{thm:DMGs}\ref{bul:thmDMG1} also applies here.

\ref{bul:thmDMGdc2} We assume that $\mathbf{x} = (\mathbf{x}^\rightarrow,
\mathbf{x}^\leftrightarrow)$ corresponds to $G_\mathbf{x}$. We have $\xs{i}{j} =
0$ if and only if $\xd{i}{j}=0$ and $\xb{i}{j}= 0$ (Lemma \ref{lem:consistDMG}),
and $\lcb{i}{j}{C}$ equals the bidirected distance between $i$ and $j$ relative
to $C$ (Lemma \ref{lem:biDist}).

We will use induction on $\phi$, the $m_c$-distance between $i$ and $j$ given
$C$, to show that the  $m_c$-distances are the unique solution of
\eqref{tag:P1c}. If $\phi = 1$, there is an $m_c$-connecting path between $i$
and $j$ given $C$ of length $\phi = 1$, and $\xs{i}{j} = 1$ or $\xs{j}{i} = 1$.
We see that $u_{ijC}^{\text{K1a$^\text{c}$}} = 1$ or
$u_{ijC}^{\text{K1b$^\text{c}$}} = 1$, and we see that $\phi$ is the unique
solution of \eqref{tag:P1c} for the triple $(i,j,C)$.

Assume now that $1 < \phi < \tilde{n} + 1$, and that $\omega$ is an
$m_c$-connecting path between $i$ and $j$ given $C$ of length $\phi$. 
\begin{itemize}
    \item If there is a directed edge with a head at $j$,  $i \sim \ldots \sim k
    \to j$, then $k \notin C$, and there is a minimal-length $m_c$-connecting walk
    between $i$ and $k$ given $C$ of length $\phi -1 $ and $\xd{k}{j} = 1$. By
    the induction assumption, $\lc{i}{k}{C} = \phi- 1$, and we see that
    $u_{ijkC}^{\text{K2a$^\text{c}$}} = \phi$. If there is a directed edge at
    $i$, we can argue analogously.
    \item Otherwise, if there are no noncolliders on $\omega$, then every collider on
    $\omega$ must be in $C$, there exists a node $k\neq i,j$ on $\omega$ such that $k \in
    C$. We see that $u_{ijC}^{\text{K3a$^\text{c}$}} = \phi$ or
    $u_{ijC}^{\text{K3b$^\text{c}$}} \leq \phi$ for the
    triple $(i,j,C)$. 
    \item Otherwise, assume that $\omega$ has a noncollider, $k$. There must
    also be a collider on $\omega$ as the first edge is not directed and pointed
    toward $i$ ($i \leftarrow \ldots$) and the last edge is not directed and
    pointed towards $j$ ($\ldots \rightarrow j$). We consider the \emph{collider
    segment} of this collider on $\omega$ (that is, the longest possible subwalk
    of $\omega$ which contains this collider and such that every node is a
    collider. If $i \rightarrow h_1$ or
    $i\leftrightarrow h_1$ where $h_1$ is the first node on this collider
    segment, we see that $u_{ijkC}^{\text{K4a$^\text{c}$}} \leq \phi$ or
    $u_{ijkC}^{\text{K4c$^\text{c}$}} \leq \phi$. If $j \rightarrow h_2$ or
    $j\leftrightarrow h_2$ where $h_2$ is the last node on this collider
    segment, then $u_{ijkC}^{\text{K4b$^\text{c}$}} \leq \phi$ or
    $u_{ijkC}^{\text{K4d$^\text{c}$}} \leq \phi$. Finally, if the endpoint $i$
    is not adjacent to $h_1$ and the endpoint $j$ is not adjacent to $h_2$, then
    $u_{ijkC}^{\text{K5$^\text{c}$}} \leq \phi$. 
\end{itemize}
In every case above, we have argued that there is a $u$-variable which is less
than or equal to  $\phi$ when the  $m_c$-distance between $i$ and $j$ given $C$
is $\phi$, $1 < \phi < d$.

We will now argue that no $u$-variable can take a smaller value, $\psi < \phi$.
\begin{itemize}
    \item If $u_{ijC}^{\text{K1a$^\text{c}$}} = \psi$ or
    $u_{ijC}^{\text{K1b$^\text{c}$}} = \psi$, then $\psi = 1$ and $\xs{i}{j} =
    1$ or $\xs{j}{i} = 1$. This implies that $\xd{i}{j} = 1$, $\xd{j}{i}  =1$,
    or $\xb{i}{j}= 1$ such that $i$ and $j$ are adjacent.
    \item If $u_{ijkC}^{\text{K2a$^\text{c}$}} = \psi$, then by the induction
    assumption $\lc{i}{k}{C}$ is the  $m_c$-distance between $i$ and $k$ given
    $C$ as $\lc{i}{k}{C} = \psi-1<\phi$, and furthermore $\xd{k}{j}=1$. The
    concatenation of an $m_c$-connecting walk of length $\psi-1$ between $i$ and
    $k$ given $C$ and the edge $k\rightarrow j$ is an $m_c$-connecting walk
    between $i$ and $j$ given $C$ in $G_\mathbf{x}$. If
    $u_{ijkC}^{\text{K2b$^\text{c}$}} = \psi$, then a similar argument works.
    \item If $u_{ijkC}^{\text{K3a$^\text{c}$}} = \psi$, then
    $\xs{i}{k}=\xs{k}{j} = 1$ and $k\in C$, creating an $m_c$-connecting walk of
    length $2$. If $u_{ijkC}^{\text{K3a$^\text{c}$}} = \psi$, then there is a
    bidirected path between $k_1$ and $k_2$ of length $\psi - 2$ such that all
    nodes are in $C$, and $\xs{i}{k_1} = \xs{j}{k_2} = 1$. The concatenation is
    an $m_c$-connecting walk for length $\psi$. \item If
    $u_{ijkC}^{\text{K4a$^\text{c}$}} = \psi$, $u_{ijkC}^{\text{K4b$^\text{c}$}}
    = \psi$, $u_{ijkC}^{\text{K4c$^\text{c}$}} = \psi$, or
    $u_{ijkC}^{\text{K4d$^\text{c}$}} = \psi$, we can in each case use the
    induction assumption and the implied edges to create an $m_c$-connecting
    walk between $i$ and $j$ given $C$ of length $\psi$. 
\end{itemize}
In each case above, we have found an $m_c$-connecting walk between $i$ and $j$
given $C$ of length at most $\psi$.  This contradicts the assumption that
$\omega$ is of minimal length, and we see that $\phi$ is the unique value which
satisfies \eqref{tag:P1}.

If $\phi = \tilde{n} + 1$, there is no $m_c$-connecting walk between $i$ and $j$
given $C$. The above argument shows there is no $u$-variable which is strictly
less than $\tilde{n} + 1$, and therefore $\lc{i}{j}{C} = \tilde{n} + 1$ is the
unique solution of \eqref{tag:P1c} for the triple $(i,j,C)$.
\end{proof}

\end{document}

%% file: tables/d6.tex
\begin{tabular}{lrrrrrrr}
\toprule
\bf Dataset & \bf Method & \bf SHD & \bf 1-SEP & \bf 2-SEP & \bf SEP & \bf 1 -- F1 (head) & \bf 1 -- F1 (tail)\\
\midrule
 & FCI & 7.73 (5.91) & 0.20 (0.16) & 0.22 (0.16) & 0.23 (0.17) & 0.16 (0.16) & {\bf 0.14} (0.12)\\
 & GLIP ($k=1$) & 6.77 (5.49) & {\bf 0.13} (0.13) & - & {\bf 0.17} (0.15) & 0.16 (0.14) & 0.18 (0.17)\\
 & GLIP ($k=2$) & {\bf 6.47} (5.19) & - & 0.18 (0.15) & 0.19 (0.16) & {\bf 0.15} (0.12) & 0.17 (0.15)\\
 & GLIP ($k=d-2$) & {\bf 6.47} (5.19) & - & - & 0.19 (0.16) & {\bf 0.15} (0.12) & 0.17 (0.15)\\
 & NOTEARS & 12.30 (5.77) & 0.47 (0.19) & 0.49 (0.20) & 0.51 (0.21) & 0.31 (0.22) & 0.30 (0.23)\\
\multirow{-6}{*}{ALARM} 
 & R2SORT & 8.00 (4.48) & 0.16 (0.13) & 0.18 (0.13) & 0.19 (0.14) & 0.22 (0.14) & 0.22 (0.14)\\
\cmidrule{1-8}
 & FCI & 4.71 (2.88) & 0.32 (0.19) & 0.28 (0.26) & 0.26 (0.15) & 0.12 (0.13) & {\bf 0.05} (0.10)\\
 & GLIP ($k=1$) & {\bf 4.21} (2.52) & {\bf 0.19} (0.10) & - & {\bf 0.17} (0.08) & {\bf 0.05} (0.04) & 0.10 (0.09)\\
 & GLIP ($k=2$) & 8.63 (7.95) & - & {\bf 0.25} (0.26) & 0.26 (0.26) & 0.17 (0.18) & 0.20 (0.20)\\
 & GLIP ($k=d-2$) & 8.63 (7.95) & - & - & 0.26 (0.26) & 0.17 (0.18) & 0.20 (0.20)\\
 & NOTEARS & 14.92 (6.28) & 0.65 (0.08) & 0.62 (0.21) & 0.61 (0.21) & 0.39 (0.18) & 0.40 (0.18)\\
\multirow{-9}{*}{ASIA} 
 & R2SORT & 13.37 (6.94) & {\bf 0.15} (0.07) & 0.27 (0.24) & 0.29 (0.24) & 0.31 (0.17) & 0.28 (0.19)\\
\cmidrule{1-8}
 & FCI & 17.60 (7.80) & 0.24 (0.18) & {\bf 0.23} (0.16) & 0.22 (0.14) & 0.33 (0.19) & {\bf 0.26} (0.16)\\
 & GLIP ($k=1$) & 16.77 (8.15) & 0.19 (0.15) & - & {\bf 0.15} (0.12) & 0.38 (0.20) & 0.37 (0.19)\\
 & GLIP ($k=2$) & {\bf 13.63} (8.47) & - & {\bf 0.23} (0.18) & 0.23 (0.17) & {\bf 0.28} (0.18) & 0.32 (0.18)\\
 & GLIP ($k=d-2$) & {\bf 13.63} (8.47) & - & - & 0.23 (0.17) & {\bf 0.28} (0.18) & 0.32 (0.18)\\
 & NOTEARS & 19.63 (6.42) & 0.76 (0.15) & 0.73 (0.16) & 0.71 (0.17) & 0.48 (0.14) & 0.46 (0.15)\\
\multirow{-6}{*}{CHILD} 
 & R2SORT & 18.60 (4.59) & 0.22 (0.17) & 0.24 (0.15) & 0.26 (0.15) & 0.43 (0.12) & 0.39 (0.14)\\
\cmidrule{1-8}
 & FCI & 16.20 (5.94) & 0.61 (0.18) & 0.64 (0.18) & 0.66 (0.19) & {\bf 0.30} (0.20) & 0.38 (0.22)\\
 & GLIP ($k=1$) & {\bf 16.13} (5.90) & {\bf 0.60} (0.18) & - & {\bf 0.65} (0.19) & 0.32 (0.19) & 0.40 (0.21)\\
 & GLIP ($k=2$) & 16.23 (5.95) & - & 0.64 (0.18) & 0.66 (0.19) & 0.33 (0.19) & 0.39 (0.21)\\
 & GLIP ($k=d-2$) & 16.23 (5.95) & - & - & 0.66 (0.19) & 0.33 (0.19) & 0.39 (0.21)\\
 & NOTEARS & 18.40 (6.13) & 0.74 (0.19) & 0.77 (0.18) & 0.79 (0.17) & 0.25 (0.00) & {\bf 0.00} (0.00)\\
\multirow{-6}{*}{HEPAR2} 
 & R2SORT & 16.60 (5.75) & {\bf 0.60} (0.20) & {\bf 0.63} (0.20) & 0.66 (0.20) & 0.31 (0.20) & 0.40 (0.21)\\
\cmidrule{1-8}
 & FCI & 2.80 (4.94) & 0.01 (0.01) & {\bf 0.01} (0.02) & 0.02 (0.03) & {\bf 0.03} (0.06) & {\bf 0.03} (0.07)\\
 & GLIP ($k=1$) & 2.50 (4.37) & {\bf 0.00} (0.01) & - & {\bf 0.01} (0.02) & 0.05 (0.09) & 0.08 (0.13)\\
 & GLIP ($k=2$) & {\bf 2.10} (4.16) & - & {\bf 0.01} (0.01) & {\bf 0.01} (0.01) & {\bf 0.03} (0.07) & 0.08 (0.15)\\
 & GLIP ($k=d-2$) & {\bf 2.10} (4.16) & - & - & {\bf 0.01} (0.01) & {\bf 0.03} (0.07) & 0.08 (0.15)\\
 & NOTEARS & 12.40 (4.21) & 0.47 (0.16) & 0.47 (0.16) & 0.46 (0.15) & 0.41 (0.17) & 0.41 (0.17)\\
\multirow{-6}{*}{SACHS} 
 & R2SORT & 8.97 (5.77) & 0.08 (0.08) & 0.07 (0.06) & 0.07 (0.06) & 0.29 (0.18) & 0.18 (0.17)\\
\bottomrule
\end{tabular}

%% file: tables/d8.tex
\begin{tabular}{lrrrrrrr}
\toprule
\bf Dataset & \bf Method & \bf SHD & \bf 1-SEP & \bf 2-SEP & \bf SEP & \bf 1 -- F1 (head) & \bf 1 -- F1 (tail)\\
\midrule
 & FCI & 13.47 (9.28) & 0.21 (0.16) & 0.23 (0.16) & 0.24 (0.17) & {\bf 0.15} (0.15) & {\bf 0.12} (0.12)\\
 & GLIP ($k=1$) & {\bf 12.47} (8.47) & {\bf 0.13} (0.10) & - & {\bf 0.16} (0.11) & 0.19 (0.12) & 0.19 (0.14)\\
 & GLIP ($k=2$) & 13.27 (9.17) & - & {\bf 0.19} (0.15) & 0.21 (0.16) & 0.20 (0.13) & 0.20 (0.16)\\
 & GLIP ($k=d-2$) & 13.27 (9.17) & - & - & 0.21 (0.16) & 0.20 (0.13) & 0.20 (0.16)\\
 & NOTEARS & 20.33 (8.90) & 0.46 (0.19) & 0.49 (0.20) & 0.52 (0.22) & 0.35 (0.16) & 0.36 (0.19)\\
\multirow{-6}{*}{ALARM} 
 & R2SORT & 17.23 (8.77) & 0.20 (0.15) & 0.22 (0.15) & 0.23 (0.15) & 0.30 (0.13) & 0.31 (0.13)\\
\cmidrule{1-8}
 & FCI & {\bf 8.00} (0.00) & 0.43 (0.00) & 0.41 (0.00) & 0.34 (0.00) & {\bf 0.08} (0.00) & {\bf 0.00} (0.00)\\
 & GLIP ($k=1$) & 12.00 (0.00) & {\bf 0.17} (0.00) & - & {\bf 0.22} (0.00) & 0.27 (0.00) & 0.06 (0.00)\\
 & GLIP ($k=2$) & 17.00 (0.00) & - & 0.29 (0.00) & 0.30 (0.00) & 0.31 (0.00) & 0.20 (0.00)\\
 & GLIP ($k=d-2$) & 17.00 (0.00) & - & - & 0.30 (0.00) & 0.31 (0.00) & 0.20 (0.00)\\
 & NOTEARS & 15.00 (0.00) & 0.66 (0.00) & 0.61 (0.00) & 0.50 (0.00) & 0.33 (0.00) & 0.50 (0.00)\\
\multirow{-6}{*}{ASIA} 
 & R2SORT & 38.00 (0.00) & 0.21 (0.00) & {\bf 0.26} (0.00) & 0.38 (0.00) & 0.38 (0.00) & 0.21 (0.00)\\
\cmidrule{1-8}
 & FCI & 26.93 (11.65) & 0.35 (0.16) & 0.32 (0.15) & 0.29 (0.13) & {\bf 0.35} (0.22) & {\bf 0.32} (0.20)\\
 & GLIP ($k=1$) & {\bf 25.77} (10.36) & 0.24 (0.13) & - & {\bf 0.19} (0.08) & {\bf 0.35} (0.17) & 0.39 (0.16)\\
 & GLIP ($k=2$) & 26.63 (10.91) & - & 0.30 (0.15) & 0.28 (0.13) & {\bf 0.35} (0.19) & 0.45 (0.20)\\
 & GLIP ($k=d-2$) & 26.63 (10.91) & - & - & 0.28 (0.13) & {\bf 0.35} (0.19) & 0.45 (0.20)\\
 & NOTEARS & 29.00 (10.14) & 0.77 (0.12) & 0.73 (0.13) & 0.66 (0.15) & 0.42 (0.14) & 0.41 (0.13)\\
\multirow{-6}{*}{CHILD} 
 & R2SORT & 34.13 (7.92) & {\bf 0.17} (0.11) & {\bf 0.19} (0.10) & 0.25 (0.10) & 0.46 (0.14) & 0.48 (0.12)\\
\cmidrule{1-8}
 & FCI & {\bf 31.33} (9.22) & 0.66 (0.13) & 0.70 (0.12) & 0.73 (0.13) & {\bf 0.32} (0.18) & {\bf 0.39} (0.26)\\
 & GLIP ($k=1$) & 31.53 (9.00) & 0.65 (0.13) & - & {\bf 0.71} (0.15) & 0.39 (0.17) & 0.44 (0.22)\\
 & GLIP ($k=2$) & 31.73 (9.18) & - & 0.70 (0.12) & 0.73 (0.14) & 0.41 (0.15) & 0.43 (0.22)\\
 & GLIP ($k=d-2$) & 31.73 (9.18) & - & - & 0.73 (0.14) & 0.41 (0.15) & 0.43 (0.22)\\
 & NOTEARS & 34.20 (9.38) & 0.76 (0.14) & 0.80 (0.12) & 0.84 (0.12) & 0.39 (0.14) & 0.28 (0.26)\\
\multirow{-6}{*}{HEPAR2} 
 & R2SORT & 31.97 (8.90) & {\bf 0.64} (0.14) & {\bf 0.68} (0.14) & 0.72 (0.14) & 0.41 (0.15) & 0.43 (0.20)\\
\cmidrule{1-8}
 & FCI & 11.77 (8.90) & 0.02 (0.03) & 0.03 (0.04) & 0.05 (0.06) & 0.10 (0.11) & {\bf 0.09} (0.11)\\
 & GLIP ($k=1$) & 13.77 (10.89) & {\bf 0.00} (0.00) & - & 0.05 (0.04) & 0.21 (0.18) & 0.21 (0.15)\\
 & GLIP ($k=2$) & {\bf 5.50} (7.08) & - & {\bf 0.01} (0.04) & {\bf 0.02} (0.04) & {\bf 0.08} (0.12) & 0.10 (0.12)\\
 & GLIP ($k=d-2$) & {\bf 5.50} (7.08) & - & - & {\bf 0.02} (0.04) & {\bf 0.08} (0.12) & 0.10 (0.12)\\
 & NOTEARS & 19.67 (4.99) & 0.47 (0.11) & 0.47 (0.10) & 0.44 (0.09) & 0.48 (0.12) & 0.48 (0.12)\\
\multirow{-6}{*}{SACHS} 
 & R2SORT & 22.17 (7.30) & 0.05 (0.05) & 0.05 (0.03) & 0.08 (0.04) & 0.36 (0.12) & 0.23 (0.15)\\
\bottomrule
\end{tabular}